\pdfoutput=1
\documentclass[a4paper,11pt]{article}
\usepackage{heppub}
\usepackage{bookmark}
\bookmarksetup{depth=2} %

\usepackage{xspace}
\usepackage{placeins}
\usepackage{wasysym}
\usepackage{slashed}
\usepackage{empheq}
\usepackage[usenames,dvipsnames]{xcolor}
\usepackage{textcomp}
\usepackage{mathtools}
\usepackage{tikz}
\usepackage{caption}
\usepackage{subcaption}
\usepackage{amsmath}

\newcommand{\ord}[1]{\mathcal{O}(#1)}
\newcommand{\Ord}[1]{\mathcal{O}\bigl(#1\bigr)}
\newcommand{\ORd}[1]{\mathcal{O}\Bigl(#1\Bigr)}

\newcommand{\ket}[1]{\lvert#1\rangle}

\newcommand{\bra}[1]{\langle#1\rvert}

\newcommand{\nn}{\nonumber}

\newcommand{\df}{\mathrm{d}}
\newcommand{\img}{\mathrm{i}}

\renewcommand{\Re}{\operatorname{Re}}

\newcommand{\Tr}{\operatorname{Tr}}

\newcommand{\Sl}[1]{\slashed{#1}}

\newcommand{\cA}{\mathcal{A}}
\newcommand{\cB}{\mathcal{B}}

\newcommand{\cI}{\mathcal{I}}

\newcommand{\cL}{\mathcal{L}}
\newcommand{\cM}{\mathcal{M}}
\newcommand{\cO}{\mathcal{O}}
\newcommand{\cP}{\mathcal{P}}
\newcommand{\cR}{\mathcal{R}}

\newcommand{\cJ}{\mathcal{J}}

\DeclareMathOperator*{\SumInt}{%
\mathchoice%
  {\ooalign{$\displaystyle\sum$\cr\hidewidth$\displaystyle\int$\hidewidth\cr}}
  {\ooalign{\raisebox{.14\height}{\scalebox{.7}{$\textstyle\sum$}}\cr\hidewidth$\textstyle\int$\hidewidth\cr}}
  {\ooalign{\raisebox{.2\height}{\scalebox{.6}{$\scriptstyle\sum$}}\cr$\scriptstyle\int$\cr}}
  {\ooalign{\raisebox{.2\height}{\scalebox{.6}{$\scriptstyle\sum$}}\cr$\scriptstyle\int$\cr}}
}

\newcommand{\eps}{\epsilon}
\newcommand{\la}{\lambda}
\newcommand{\w}{\omega}

\newcommand{\bn}{{\bar n}}
\newcommand{\nbar}{\bar{n}}
\newcommand{\nb}{\bar{n}}

\newcommand{\as}{\alpha_s}

\newcommand{\MSbar}{\overline{\text{MS}}\xspace}

\newcommand{\GeV}{\,\mathrm{GeV}}

\newcommand{\WidthTwoSubfigs}{0.49\textwidth}

\allowdisplaybreaks[2]

\newcommand{\nnnllp}{N$^3$LL$'$}

\newcommand{\nnnlo}{N$^3$LO}
\newcommand{\loone}{LO$_1$}

\newcommand{\mloone}{\mathrm{LO}_{1}}
\newcommand{\mlozero}{\mathrm{LO}_{0}}

\title{Subleading beauty effects in Higgs production at small transverse momentum}

\author[a,b]{Rebecca von Kuk,}
\emailAdd{rebecca.vonkuk@unibe.ch}

\author[c,d]{Johannes K.~L.~Michel,}
\emailAdd{j.k.l.michel@uva.nl}

\author[e]{Iain W.~Stewart,}
\emailAdd{iains@mit.edu}

\author[e,f,g]{Zhiquan Sun,}
\emailAdd{zqsun@berkeley.edu}

\author[a]{and Frank J.~Tackmann}
\emailAdd{frank.tackmann@desy.de}

\affiliation[a]{Deutsches Elektronen-Synchrotron DESY, Notkestr. 85, 22607 Hamburg, Germany}
\affiliation[b]{Albert Einstein Center for Fundamental Physics, Institut f\"ur Theoretische Physik,\\
	Universit\"at Bern, Sidlerstrasse 5, CH-3012 Bern, Switzerland}
\affiliation[c]{Institute for Theoretical Physics Amsterdam and Delta Institute for Theoretical Physics, \\
University of Amsterdam, Science Park 904, 1098 XH Amsterdam, The Netherlands}
\affiliation[d]{Nikhef, Theory Group, Science Park 105, 1098 XG, Amsterdam, The Netherlands}
\affiliation[e]{Center for Theoretical Physics - a Leinweber Institute,\,Massachusetts Institute of Technology,\\
Cambridge,\,MA\,02139,\,USA}
\affiliation[f]{Leinweber Institute for Theoretical Physics,\,University of California,\,Berkeley,\,CA\,94720}
\affiliation[g]{Theory Group,\,Physics Division,\,Lawrence Berkeley National Laboratory,\,Berkeley,\,CA\,94720}

\abstract{%
We study $b$-quark mass effects at small Higgs transverse momentum $q_T$
where the Higgs is produced in gluon fusion via a closed $b$-quark loop.
In contrast to top-quark loops, which can be integrated out at a high scale, the
$b$-quark mass, $m_b$, induces a second small scale $m_b \sim q_T \ll m_H$ such that
standard $q_T$ resummation does not apply in this region, which includes the peak
of the distribution.
The resummation of mass effects for the $gg\to H$ form factor, which has been studied
previously, applies to the $q_T$ spectrum in the limit $q_T \ll m_b$.
Unlike this case, for $q_T \sim m_b$, real emissions can resolve the $b$-quark loop,
requiring a more complicated simultaneous $m_b$ and $q_T$ factorization. It involves
new next-to-leading-power $m_b$ and $q_T$ dependent beam and soft functions, and
the corresponding bare factorization theorem suffers -- as expected -- from
endpoint divergences.
We use soft-collinear effective theory to explicitly compute all amplitudes with
a $b$-quark loop and an additional real soft or collinear emission. Our results
provide important insight into the nontrivial structure of
the endpoint divergences and show how their expected cancellation happens
in the sum of all amplitudes separately for soft and collinear emissions.
Interestingly, we find that at this order the endpoint-divergent
contributions have a separable dependence on $q_T$ and $m_b$, providing
strong motivation that in the future their cancellations can be made manifest in the
factorized cross section.
As an important further cross check we show that the sum of all
SCET amplitudes correctly reproduces the full-QCD one-emission amplitudes expanded
in the relevant soft or collinear limit, which also requires nontrivial contributions
involving massive Glauber quarks in the effective-theory calculation.
}

\date{\today}

\preprint{\vbox{%
\hbox{MIT-CTP 6143}
\hbox{DESY-26-153}
\hbox{Nikhef 2026-017}
}
}

\definecolor{Magenta}{rgb}{0.75,0,0.75}
\definecolor{Green}{rgb}{0,0.75,0}
\begin{document}
\setcounter{tocdepth}{2}
\maketitle
\newpage

\section{Introduction}
Since the discovery of the Higgs boson at the LHC~\cite{ATLAS:2012yve, CMS:2012qbp}, precise measurements of its couplings to the other Standard Model particles have become a central part of the LHC physics program.
The dominant production mechanism of the Higgs boson at the LHC is gluon fusion, $gg \to H$, which accounts for close to $90\%$ of the total production cross section~\cite{Cepeda:2019klc}.
As the Higgs boson does not couple to gluons directly, this process is loop-induced and is mediated by virtual quarks, each of which couples to the Higgs boson proportionally to its mass.
The process is dominated by the top-quark loop which is usually integrated out, so that the Higgs boson couples to gluons through an effective local operator~\cite{Wilczek:1977zn,Shifman:1979eb,Inami:1982xt,DAWSON1991283,Djouadi:1991tka,Spira:1995rr}.
In this heavy-top limit, the inclusive cross section is known to \nnnlo{} in QCD~\cite{Harlander:2002wh, Ravindran:2003um, Anastasiou:2015ema,Anastasiou:2016cez,Mistlberger:2018etf}.
The Higgs transverse-momentum $q_T$ spectrum in $pp \to H$ is of particular interest since it serves as an important input for many experimental analyses.
For finite values of $q_T$, the spectrum is available at NNLO from $H+$jet calculations \cite{deFlorian:1999zd,Ravindran:2002dc,Glosser:2002gm,Chen:2014gva,Boughezal:2015aha,Boughezal:2015dra,Caola:2015wna,Chen:2016zka,Campbell:2019gmd,Chen:2019wxf}.
For small values of $q_T$, the spectrum contains large logarithms of $q_T/m_H$ which need to be resummed in order to obtain a meaningful prediction.
Indeed, this resummation has been achieved up to \nnnllp~\cite{Bozzi:2005wk,Becher:2012yn,Neill:2015roa, Bizon:2017rah, Chen:2018pzu,Bizon:2018foh,Gutierrez-Reyes:2019rug,Becher:2020ugp,Billis:2021ecs}.

\begin{figure}
	\centering
	\begin{subfigure}{0.33\textwidth}
		\centering
		\includegraphics[height=2.5cm]{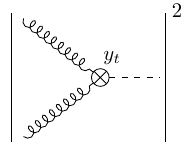}
		\subcaption{}  \label{fig:ggHsq}
	\end{subfigure}%
	\hspace{2cm}
	\begin{subfigure}{0.33\textwidth}
		\centering
		\includegraphics[height=2.5cm]{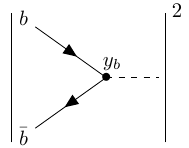}
		\subcaption{}  \label{fig:bbHsq}
	\end{subfigure}\\%
		\begin{subfigure}{0.33\textwidth}
		\centering
		\includegraphics[height=2.5cm]{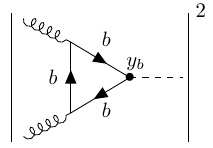}
		\subcaption{}  \label{fig:ggHbsq}
	\end{subfigure}%
		\hspace{2cm}
	\begin{subfigure}{0.33\textwidth}
		\centering
		\includegraphics[height=2.5cm]{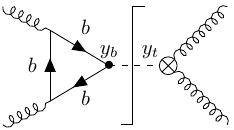}
		\subcaption{} \label{fig:ggHbbH}
	\end{subfigure}%
	\caption{(a) Top-quark contribution to the gluon fusion process, (b) quark-initiated Higgs production, (c) bottom-quark contribution to gluon fusion and (d) $y_b$-$y_t$-interference contribution.}
\end{figure}

A high-precision prediction for the $q_T$ spectrum allows one to measure the quark Yukawa couplings from fitting the shape of the spectrum, which is sensitive to the initial states \cite{Ebert:2016idf,Bishara:2016jga, Soreq:2016rae}.
The quark- and gluon-induced channels can also be well distinguished by their shape as their predictions come with different Casimir scaling.
To allow a fit to the $q_T$ spectrum it is particularly important that all contributions of the spectrum are known to sufficiently high orders.
The full $q_T$ spectrum in proton collisions reads
\begin{align}\label{eq:higgsqTxsec}
	\frac{\text{d} \sigma (pp\to H)}{\text{d} q_T}= y_t^2\frac{\text{d}  \sigma_{tt} }{\text{d} q_T} +y_t y_b \frac{\text{d}  \sigma_{tb} }{\text{d} q_T}+ y_b^2\frac{\text{d}  \sigma_{bb} }{\text{d} q_T} +( y_b \to y_c),
\end{align}
where $y_q$ denotes the respective quark Yukawa coupling.
The first term describes the top-quark contribution to the gluon fusion process and is diagrammatically shown in \fig{ggHsq}.
The third term in \eq{higgsqTxsec} contains the $b$-quark-initiated contribution, which is shown in \fig{bbHsq}.
Both contributions are known at \nnnllp+\nnnlo~\cite{Billis:2021ecs, Cal:2023mib}.
In addition, the $y_b^2$ term receives a contribution from gluon fusion through a squared bottom-quark loop, shown in \fig{ggHbsq}.
The focus of this paper is the second contribution in \eq{higgsqTxsec} which we refer to as the $y_b$-$y_t$-interference term.
This contribution is illustrated in \fig{ggHbbH}.
Usually, in the context of resumming the Higgs $q_T$ spectrum, only the top contribution in gluon fusion is considered explicitly as the top is by far the heaviest quark.
However, the contribution with the bottom-quark loop is enhanced by the interference with the top quark and contributes $\cO(5\%)$ to the gluon fusion cross section~\cite{LHCHiggsCrossSectionWorkingGroup:2016ypw,Anastasiou:2016cez,Czakon:2023kqm}.
Lighter quarks only make up a few percent of the cross section.
At the current level of precision of the experimental measurements and further significant improvements expected at the HL-LHC, the proper resummation of this contribution becomes mandatory.

At fixed order, the next-to-leading-order corrections to the top-bottom interference in Higgs plus jet production were obtained in \refcite{Lindert:2017pky}, using the two-loop amplitudes mediated by a nearly massless quark from \refscite{Melnikov:2016qoc,Melnikov:2017pgf}.
Bottom-quark effects in the resummed $q_T$ spectrum and in the jet-vetoed cross section were studied in \refscite{Grazzini:2013mca,Banfi:2013eda,Caola:2018zye}. Since emissions with $q_T \gtrsim m_b$ resolve the bottom-quark loop, the standard $q_T$ resummation does not directly apply, which leads to sizable uncertainties in the top-bottom interference contribution~\cite{Caola:2018zye}.
A reliable prediction of the interference contribution therefore requires the joint resummation of the logarithms of $q_T/m_H$ as well as of $m_b/m_H$.

Large logarithms of $m_b/m_H$ already appear at the level of the $gg\to H$ amplitude, and share common features with logarithms in the related decay $H\to \gamma\gamma$ via a $b$-quark loop.
Here, the hard scale of the problem is given by the mass of the Higgs boson $m_H$, and the low scale is given by the bottom-quark mass $m_b \ll m_H$.
The leading (double) logarithms (LL) for the $H\to \gamma\gamma$ case
were first predicted a long time ago~\cite{Kotsky:1997rq,Akhoury:2001mz}.
For $gg\to H$ the LL resummation was obtained in \refscite{Liu:2017vkm,Liu:2018czl} using the methods of perturbative QCD, building on earlier work in QED~\cite{Penin:2014msa}.
These methods were also applied to the Higgs production cross section~\cite{Anastasiou:2020vkr} and to Higgs plus jet production
for $m_b \ll q_T$~\cite{Liu:2024tkc} at next-to-leading-logarithmic (NLL) order,
where in both cases the treatment of additional real radiation
was simplified by working in the soft threshold ($z \to 1$) limit.

The factorization and resummation of $b$-mass effects for the Higgs boson decay and production amplitudes beyond LL order
were first considered systematically in \refscite{Liu:2019oav,Liu:2020wbn,Liu:2022ajh}, using methods from the soft-collinear effective theory (SCET)~\cite{Bauer:2000ew,Bauer:2000yr,Bauer:2001ct,Bauer:2001yt}.
In this framework, the bottom-quark loop contributes through subleading-power effective operators whose contributions are suppressed by $m_b/m_H$.
\Refscite{Liu:2019oav,Liu:2020wbn} found agreement for $H\to \gamma\gamma$ at LL order with~\refscite{Kotsky:1997rq,Akhoury:2001mz}, but disagreed with the NLL prediction of~\refcite{Akhoury:2001mz}.
Factorization at subleading power in SCET has been studied extensively in recent years, see e.g.\ \refscite{Beneke:2002ph,Beneke:2002ni,Kolodrubetz:2016uim,Moult:2016fqy,Feige:2017zci,Moult:2018jjd,Ebert:2018gsn,Beneke:2018gvs,Beneke:2019mua,Moult:2019uhz}, and the treatment of quark-mass effects in SCET was discussed, e.g., in \refscite{Leibovich:2003jd,Rothstein:2003wh,Chay:2005ck,Gritschacher:2013pha,Pietrulewicz:2017gxc}.
A generic feature of subleading-power factorization theorems is the appearance of endpoint divergences in the convolutions between the different functions, which need to be removed by a suitable rearrangement of the factorization theorem before the large logarithms can be resummed~\cite{Liu:2019oav,Liu:2020wbn,Liu:2020tzd}.
The renormalization and a general treatment of endpoint divergences are still an active topic of research~\cite{Beneke:2020ibj,Bodwin:2021cpx,Beneke:2022obx,Bell:2022ott,Feldmann:2022ixt,Cornella:2022ubo,Hurth:2023paz} and were recently studied for the soft-quark functions in $gg \to H$ and $\gamma\gamma \to H$ in \refcite{Beneke:2024cpq}.

\begin{figure}
	\centering
	\begin{subfigure}{0.25\textwidth}
		\includegraphics[width=\textwidth]{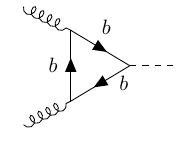}
		\subcaption{}  \label{fig:ggHLO0}
	\end{subfigure}%
	\hspace{3cm}
	\begin{subfigure}{0.25\textwidth}
		\includegraphics[width=\textwidth]{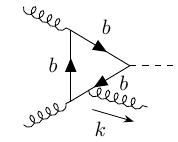}
		\subcaption{} \label{fig:ggHLO1}
	\end{subfigure}%
	\caption{Illustration of Higgs production via gluon fusion through a bottom-quark loop. (a) Leading-order contribution (LO$_0$). (b) Leading-order contribution with an extra emission $k$ (\loone). This diagram serves as an example for all possible gluon attachments to the leading-order diagram.}
\end{figure}

In this paper, we study the factorization of bottom-mass effects in the Higgs $q_T$ spectrum at small $q_T$ for the first time.
In contrast to the cases of the $H \to \gamma \gamma$ decay or the $gg \to H$ form factor investigated previously at the amplitude level \cite{Liu:2019oav,Liu:2020wbn,Liu:2022ajh}, the nontrivial $b$-mass effects are also resolved by real emissions and have to be considered within the context of $q_T$ factorization and resummation~\cite{Boussarie:2023izj}.
The $q_T$ measurement thus adds a third scale to the problem, and we consider the various hierarchies between $m_b$, $q_T$, and $m_H$, leading to different factorization setups.
To allow the Higgs boson to pick up a transverse momentum, we need to add an emission $k$ that the Higgs boson can recoil against, as illustrated in \fig{ggHLO1}, where $\vec{k}_T = - \vec{q}_T$.
We refer to the leading-order $gg\to H$ contribution without emission, \fig{ggHLO0}, as LO$_0$, and to the leading-order contribution with one additional real emission, \fig{ggHLO1}, as \loone.

This paper is organized as follows.
In \sec{theorysetup}, we review the SCET description of the $gg\to H$ amplitude mediated by a bottom-quark loop and derive the bare factorization theorems for the $y_b$-$y_t$-interference term for different hierarchies between $q_T$ and $m_b$.
In \secs{collinearemission}{softemission}, we compute the \loone{} amplitudes with a collinear and a soft emission, respectively, and show that the endpoint divergences cancel separately in the sum of the collinear and of the soft emission diagrams.
In \sec{FO}, we compare our results to the full QCD helicity amplitudes of \refcite{Baur:1989cm}.
In \sec{logstructure}, we discuss the implications of our results
for the structure of the $(q_T, m_b)$-dependent factorization theorems.
We conclude in \sec{ybytsummary}.

\section{Theory setup}
\label{sec:theorysetup}
In this section, we collect the effective-theory ingredients used throughout the paper. In \sec{operatorbasis}, we review the SCET description of the $gg\to H$ amplitude mediated by a bottom-quark loop developed in \refscite{Liu:2019oav,Liu:2020wbn,Liu:2022ajh} and fix our notation. In \sec{factorization_theorems}, we discuss how the transverse-momentum measurement modifies this picture in the different kinematic regimes and derive the bare factorization theorems for the $y_b$-$y_t$-interference contribution where the $q_T$ and $m_b$ dependence factorizes.
\subsection{Operator basis}
\label{sec:operatorbasis}

\subsubsection{Effective field-theory modes}
\label{sec:modes}

We introduce two light-like reference vectors $n^\mu$ and $\nb^\mu$ along the beam axis, with $n^2=\nb^2=0$ and $n\cdot\nb=2$, and decompose any four-vector as
\begin{align}
	p^\mu = n\cdot p\,\frac{\nb^\mu}{2} + \nb\cdot p\,\frac{n^\mu}{2} + p_\perp^\mu\,,
	\qquad
	p = (p^+, p^-, p_\perp) \equiv (n\cdot p,\, \nb\cdot p,\, p_\perp)\,.
\end{align}
The transverse components are projected out by
\begin{align}
	g_\perp^{\mu\nu} = g^{\mu\nu} - n^\nu  \frac{\nb^\mu}{2} - \nb^\nu \frac{n^\mu}{2}\,,
	\qquad
	\gamma_\perp^\mu = g_\perp^{\mu\nu}\gamma_\nu\,,
\end{align}
such that $p_\perp^\mu = g_\perp^{\mu\nu} p_\nu$ and $p^2 = p^+ p^- + p_\perp^2$. Since $p_\perp$ is spacelike, $p_\perp^2 = -\vec p_T^{\,2} \equiv -p_T^2 < 0$, where $\vec p_T$ denotes the two-dimensional Euclidean transverse momentum.

At leading order, the Higgs boson is produced from two incoming gluons with momenta $p_1^\mu = m_H\, n^\mu/2$ and $p_2^\mu = m_H\, \nb^\mu/2$, such that $q=p_1+p_2$ and $q^2=m_H^2$.
In the Standard Model, this process is loop induced.
For the top quark, $m_t \gg m_H$, the loop can be integrated out and the dynamics are described by the Lagrangian of Higgs effective theory~\cite{Wilczek:1977zn,Shifman:1979eb,Inami:1982xt} with Wilson coefficient $C_{ggt}$ which is known to ${\cal O}(\alpha_s^5)$
in QCD \cite{Djouadi:1991tka,DAWSON1991283,Chetyrkin:1997iv,  vanRitbergen:1997va, Czakon:2004bu, Chetyrkin:2005ia, Schroder:2005hy, Gerlach:2018hen}.
For the bottom quark, the hierarchy is reversed, $m_b \ll m_H$, and the loop gives rise to large logarithms of $m_H/m_b$~\cite{Akhoury:2001mz,Penin:2014msa,Liu:2017vkm,Liu:2018czl}.
Since the Yukawa interaction flips the chirality of the bottom quark, a second chirality flip, i.e.\ a mass insertion, is needed to obtain a nonvanishing amplitude.
At fixed Yukawa coupling, the bottom-loop amplitude is therefore suppressed by one power of $m_b/m_H$, and its description in SCET requires subleading-power operators~\cite{Liu:2019oav,Liu:2020wbn,Liu:2022ajh}.

The relevant operators can be identified from the amplitude for the leading-order gluon fusion diagram shown in \fig{ggHLO0} by considering the possible scalings of the loop momenta.
They can scale as
\begin{align}
	\text{hard:} \quad &p \sim m_H(1, 1, 1)\,,\nn \\
	\text{$n$-collinear:} \quad &p\sim m_H(\lambda ^2, 1, \lambda)\,,\nn \\
	\text{$\nb$-collinear:}\quad &p \sim m_H(1,\lambda ^2,  \lambda)\,,\nn \\
	\text{$n$-hard-collinear:} \quad &p\sim m_H(\lambda, 1, \lambda^{\frac{1}{2}})\,,\nn \\
	\text{$\nb$-hard-collinear:}\quad &p \sim m_H(1,\lambda,  \lambda^{\frac{1}{2}})\,,\nn \\
	\text{soft:} \quad &p \sim m_H(\lambda, \lambda, \lambda)\,, \nn \\
	\text{Glauber:} \quad &p \sim m_H(\lambda^a, \lambda^b, \lambda)\,, \ (a+b\ge 3)\,,
\end{align}
where our power-counting parameter is given by $\lambda=m_b/m_H$, such that collinear and soft quarks have virtuality $p^2\sim m_b^2$.
The incoming gluons are $n$- and $\nb$-collinear and scale as
\begin{align}
	p_1 \sim m_H(\lambda ^2, 1, \lambda), \qquad p_2 \sim m_H(1, \lambda ^2,\lambda).
\end{align}
The Higgs boson is always produced in a hard interaction which we denote by $\otimes$ and carries a hard momentum,
\begin{align}
	q \sim m_H (1,1,\lambda).
\end{align}
The last entry indicates that we take the Higgs transverse momentum $q_T\sim \lambda\,m_H$.

\begin{figure}
	\begin{align*}
		\raisebox{-0.5\height}{\includegraphics{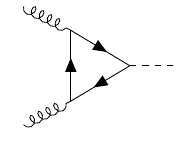}}\to
		\raisebox{-0.45\height}{\includegraphics{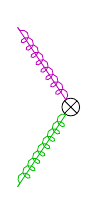}}
		+
		\quad
		\raisebox{-0.45\height}{\includegraphics{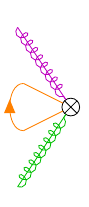}}
		\quad
		+
		\quad
		\raisebox{-0.45\height}{\includegraphics{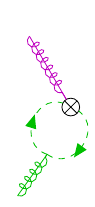}}
		\quad
		+
		\raisebox{-0.45\height}{\includegraphics{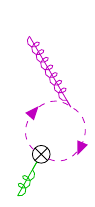}}
		+
		\raisebox{-0.45\height}{\includegraphics{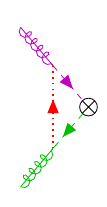}}
	\end{align*}
	\caption{Decomposition of full QCD diagram in terms of hard, soft, $n$- and $\nb$-collinear, and Glauber loop momenta. Springs with a line denote $n_i$-collinear gluons and dashed fermion lines $n_i$-collinear quarks, shown in green for $n$ and magenta for $\nb$. Solid fermion lines denote soft quarks in orange and dotted red lines Glauber quarks. The hard interaction is denoted by $\otimes$. The Glauber contribution (last diagram) coincides with the Glauber zero-bin of the soft contribution and is therefore already contained in it, see \app{ybytsoftLO}.}
	\label{fig:modesLO0}
\end{figure}
With these constraints, there are five possible scenarios which are diagrammatically shown in \fig{modesLO0}:
One possibility is that all loop momenta have hard scaling and can be integrated out.
This corresponds to the first term in the sum in \fig{modesLO0}.
It gives a contribution  $C(m_H)$ to the Wilson coefficient of the same two-gluon operator that arises from integrating out the top quark.
 The full $b$-mediated gluon form factor $C_{ggb}(m_b,m_H)$ is only obtained once the remaining soft and collinear contributions discussed below are added.
We can also have two hard-collinear quark momenta producing the Higgs boson.
Then the remaining quark line connecting the two gluons has soft scaling which is shown by the second term in the sum.
Integrating out the two hard-collinear propagators yields the soft operator $\cO_{bbgg}$, whose nonlocal prefactor $1/(\ell^+\ell^-)$, with $\ell$ the soft loop momentum, is the product of these two propagators.
Beyond tree level, the hard-collinear modes with virtuality $p^2\sim m_b m_H$ are integrated out at the hard-collinear scale $\mu_{hc}\sim\sqrt{m_b m_H}$, and the jet functions $\cJ(\ell^+)$ and $\cJ(\ell^-)$ arise as the corresponding matching coefficient~\cite{Liu:2019oav,Liu:2020wbn,Liu:2022ajh},
which is known to two loops~\cite{Liu:2021mac}.
At tree level, the hard-collinear propagators reduce to the prefactor $1/(\ell^+\ell^-)$ in the operator and $\cJ(\ell^\pm)=1$.
The third and fourth terms correspond to a scenario where only one of the quark momenta involved in the production of the Higgs boson exhibits hard scaling.
The remaining quark propagators then have $n$-collinear scaling (third term in the sum in \fig{modesLO0}) or $\nb$-collinear scaling (fourth term in the sum in \fig{modesLO0}).
Integrating out the hard propagator yields the collinear operator $\cO_{bbg}$ with an $n$-collinear quark-antiquark pair and an $\nb$-collinear gluon, and its $\nb$-collinear counterpart obtained by $n\leftrightarrow\nb$.
Finally, the quark line connecting the two gluons can also have Glauber scaling, with the two quarks producing the Higgs boson being $n$- and $\nb$-collinear, which is shown by the last term in \fig{modesLO0}.
In this case, no hard propagator is integrated out at this order, and the Higgs boson is produced by the scalar \emph{leading-power} current $\bar\chi_{\nb}\chi_n$ (and $n\leftrightarrow\nb$).

\subsubsection{Hard scattering operators}
\label{sec:hard_operators}

\begin{figure}
	\centering
	\begin{subfigure}{0.17\textwidth}
		\includegraphics[width=\textwidth]{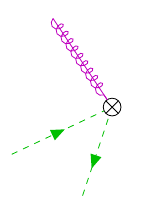}
		\subcaption{}
	\end{subfigure}%
	\hspace{3cm}
	\begin{subfigure}{0.17\textwidth}
		\includegraphics[width=\textwidth]{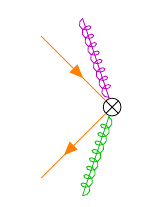}
		\subcaption{}
	\end{subfigure}%
	\caption{Tree-level diagrams associated with the collinear operator $\cO_{bbg}$ (a) and the soft operator $\mathcal{O}_{bbgg}$ (b).}
	\label{fig:operators}
\end{figure}

The relevant SCET operators at tree level are shown in \fig{operators}.
In terms of the gauge-invariant collinear quark and gluon fields $\chi_n = W_n^\dagger \xi_n$ and $g\cB^\mu_{n\perp} = \big[ W_n^\dagger\, \img D^\mu_{n\perp} W_n\big]$, where the collinear quarks are massive, the  $n$-collinear operator is given by \cite{Chang:2017atu,Liu:2019oav}
\begin{align}\label{eq:collinear_Op}
	\cO_{bbg}(\w_1, \w_2) = \bar{\chi}_{n , \omega_1}  \frac{\slashed{\nb}}{2} (g \slashed{\cB}_{\nb,\perp}^a)_{\bar{\omega}} ({\cal S}_{\bn}^{ba}{\cal S}_n^{bc}T^c) \chi_{n, \omega_2}
\,,\end{align}
where $\omega_{1,2}$ denote the large light-cone momenta of the collinear quarks and $\bar{\omega}$ denotes the gluon's large light-cone momentum, e.g.\ $\chi_{n,\omega} = \delta(\omega - \nb\cdot\mathcal{P})\,\chi_n$ with $\mathcal{P}$ the label-momentum operator.
Here ${\cal S}_{n,\bn}$ are soft Wilson lines in the adjoint representation, while we will use $S_{n,\bn}$ for soft Wilson lines in the fundamental representation below.
With the sign convention for the label momenta used in the Feynman rule in \eq{LOcFR}, momentum conservation fixes $-(\omega_1+\omega_2)= q^- \equiv \omega_n = m_H$ and $\bar{\omega} = q^+ = m_H$.
The collinear Wilson coefficient is given by
\begin{align}\label{eq:Cbbg_tree}
	 C_{bbg} =  - \frac{\omega_n}{\omega_1}-\frac{\omega_n}{\omega_2}+ \ord{\as}\,,
\end{align}
which at tree level follows from expanding the hard propagator, see \eq{LOcFR} in \app{feynman_rules} for reference.
Since $\omega_{1,2} < 0$ in this convention, $C_{bbg}$ is positive.
The one-loop QCD corrections to \eq{Cbbg_tree} are also known~\cite{Liu:2022ajh}.
The soft operator is given by \cite{Liu:2019oav, Liu:2022ajh}
\begin{align}\label{eq:soft_Op}
	\mathcal{O}_{bbgg}(\ell^+,\ell^-)=&\frac{1}{\ell^+\ell^-}
   \Bigl[\bar{\psi}S_n\,\delta\bigl(\ell^+ \!+ n\!\cdot\! \img\overleftarrow{\partial_s}\bigr)\Bigr]
  \Bigl[(g \slashed{\cB}_{n \perp})
\frac{\slashed{n}\slashed{\bar{n}}}{4}S_n^\dagger S_{\bar{n}} (g \slashed{\cB}_{\nb \perp})
\Bigr]
  \Bigl[\delta\bigl(\ell^- \!- \bar{n}\!\cdot\! \img\overrightarrow{\partial_s}\bigr)\,S^{\dagger}_{\bar{n}}\psi\Bigr]
 \nn \\
	& \qquad+(n \leftrightarrow \nb)\,,
\end{align}
where we have suppressed the $\img 0$ terms in the $1/(\ell^+ \ell^-)$ prefactor for now;
we will return to them around \eq{replacment_i0_prescr}
when discussing the contributions of Glauber exchanges.
Since all momentum components of the soft fields are of the same size, the soft momenta $\ell^\pm$ are measured by derivatives rather than label-momentum operators.
Note that here and in the following the total label momentum of collinear operators is set to $q^\pm$ in the end.
The arrows indicate the direction in which the derivatives act, and they act only on the soft fields within the respective square brackets, i.e.\ on $\bar\psi S_n$ and $S_{\nb}^\dagger\psi$, respectively, with $\ell$ the momentum of the soft quark flowing from $\psi$ to $\bar\psi$.
In addition to the hard-collinear jet functions discussed earlier,
the matching onto this operator also receives contributions from the hard scale.
At that scale, the matching onto hard-collinear modes (with external
massless quark states) proceeds exactly as at leading power,
and the associated Wilson coefficient
is given by the massless quark scalar form factor,
which is also present e.g.\ in the leading-power $q_T$ factorization
for the $b\bar{b} \to H$ annihilation process in \eq{higgsqTxsec}
and known to four loops~\cite{Gehrmann:2014vha, Chakraborty:2022yan}.
Using the notation and normalization from \refcite{Ebert:2017uel},
this matching coefficient is given by $C_{q\bar{q}}^S(m_H) = 1 + \ord{\as}$.

Summarizing the discussion so far,
we have identified three operator contributions
to the $g_n g_{\nb} \to H X$ transition amplitude
for a general final state $X$ consisting
of zero or more collinear and/or soft light partons.
With all prefactors, Wilson coefficients,
and momentum integrals made explicit, they read
\begin{align} \label{eq:summary_hard_operators}
\cM^{gg \to HX}_{h}
&= \frac{1}{m_H} \, C(m_H) \,
\bra{X} \cO_{gg}^\text{LP} \ket{g_n g_{\bn}}
\,, \nn \\
\cM^{gg \to HX}_{c}
&= \frac{1}{m_H} \int \! \df \xi \, C_{bbg}(m_H, \xi) \,
\bra{X} \cO_{bbg}\bigl(-(1-\xi)\,\omega_n, -\xi\,\omega_n\bigr) \ket{g_n g_{\bn}}
\,, \nn \\
\cM^{gg \to HX}_{s}
&= \frac{1}{m_H} C_{q\bar{q}}^S(m_H) \, \int \! \df \ell^+ \, \df \ell^- \,
\cJ(\ell^+) \, \cJ(\ell^-) \,
\bra{X} \cO_{bbgg}(\ell^+, \ell^-) \ket{g_n g_{\bn}}
\,,\end{align}
where $\cO_{gg}^\text{LP}$ is the standard leading-power SCET operator
mediating $g_n g_{\bn} \to H$.
Here we have introduced the shorthand $\xi \equiv -\w_2/\w_n$, see also \sec{collinearemission},
and have omitted an obvious fourth term $\cM^{gg \to HX}_{\bar{c}}$
where the power suppression
is contributed by the $\nb$-collinear sector instead.
Note that we set the (bare) Yukawa coupling to 1 in all three operators, following \eq{higgsqTxsec}.
The individual contributions of the collinear and soft operators are divergent at the endpoints: $C_{bbg}$ behaves as $1/\omega_i$ for $\omega_i \to 0$, and the $\ell^\pm$ integrals of the soft matrix element diverge in the corresponding $\ell^\pm\to 0$ limits. These endpoint divergences cancel between the soft and collinear contributions~\cite{Liu:2019oav,Liu:2020wbn}, and the remaining $\eps$ poles
are canceled by a hard coefficient counterterm.
The four contributions to the LO$_0$ ($X = \{ \}$) amplitude are collected in \app{LO0-calc}. Their sum reproduces the expansion of the full QCD LO$_0$ amplitude~\cite{Spira:1995rr}, including the double logarithm $\log^2(m_H^2/m_b^2)$. The same cancellation mechanism,
but now with an extra emission and transverse momentum scale,
is central to the real-emission calculation in \secs{collinearemission}{softemission}.

\subsubsection{Glauber Lagrangian insertions}
\label{sec:glauber_operators}

An important new feature of our analysis is that we will also need to consider subleading power Glauber exchange interactions involving exchange of offshell quarks with Glauber momentum scaling. At the level required for our analysis,
the relevant Lagrangian involves operators coupling two and three rapidity sectors
can be obtained from~\refcite{Moult:2017xpp} by including the Fermion mass with $\Sl\cP_\perp\to \Sl\cP_\perp -m$, giving
\begin{align}\label{eq:Glauber_Lagrangian_quark}
\cL_{Gq}^{\text{II}(1/2)}
&=  e^{-ix\cdot \cP}  \Big(
 \bar \cO_n  \frac{1}{\,\,\slash\!\!\!\!{\cP_\perp} - m} \cO_s^n
 + \bar \cO_s^n  \frac{1}{\,\,\slash\!\!\!\!{\cP_\perp} - m} \cO_n \Big)
  + (n \leftrightarrow \nb)
 \nn \,, \\
\cL_{Gq}^{\text{II}(1)}
&= e^{-ix\cdot \cP}
 \bar \cO_{\bar n} \frac{1}{\,\,\slash\!\!\!\!{\cP_\perp} - m} \cO_s
  \frac{1}{\,\,\slash\!\!\!\!{\cP_\perp} - m} \cO_{n}
  + (n \leftrightarrow \nb)
  \,,
\end{align}
where the collinear and soft composite operators are
\begin{align} \label{eq:Glaub_ops_quark}
  \cO_{\bar n} &=\Sl{\cB}_{\bar n\perp} \chi_{\bar n}\,,
  \qquad \cO_{n}=\Sl{\cB}_{n\perp} \chi_n \,,
  \qquad \cO^{n'}_s=-4\pi \alpha_s \Sl{\cB}_{S\perp}^{n'} \psi^{n'}_S
   \,,
 \nn\\
	\cO_s &= -2\pi \alpha_s \left[ S_{\bar n}^\dagger S_{n}
(\Sl{\cP}_\perp-m) + (\Sl{\cP}_\perp-m) S_{\bar n}^\dagger S_{n} - S_{\bar n}^\dagger S_n g\Sl{\cB}^{n}_{S\perp}-  g\Sl{\cB}^{\bar n}_{S\perp} S_{\bar n}^\dagger S_n \right]
  \,,
\end{align}
and $g{\cB}_{S\perp}^{n'} = \big[ S_{n'}^\dagger\, \img D^\mu_{S\perp} S_{n'}\big]$.
While these ``fermionic'' Glauber Lagrangians
are subleading in the power counting,
they can appear at our working $\ord{\la}$
as a single insertion in time-ordered products with a suitable hard operator.
For the three-sector Lagrangian $\cL_{Gq}^{\text{II}(1)} \sim \la$,
power counting dictates that this must be
the leading-power quark scalar current
$\bar{\chi}_{\nb} S_{\nb}^\dagger S_{n} \chi_n \sim \la^0$,
as in the example in \fig{modesLO0},
which results in a contribution of the following form,
\begin{align} \label{eq:glauber_contrib_three_sectors}
\cM^{gg \to HX}_{G}
&= \frac{1}{m_H} C_{q\bar{q}}^S(m_H) \,
\bra{X} T \Bigl\{
	\bigl[ \bar{\chi}_{\bn}S_{\nb}^\dagger S_{n}  \chi_{n} \bigr](0) \,
   \Bigl[ \img \int\!\! \df^4 x \, \cL_{Gq}^{\text{II}(1)}
	\Bigr]
	+ (n \leftrightarrow \bn)
\Bigr\} \ket{g_n g_{\bn}}
\,.\end{align}
Note that in this expression,
the leading-power current ensures that fermion number
can be conserved in each individual sector of the effective theory.
Interestingly, this means that the leading-power $b \bar{b} \to H$ current
remains dynamical even in the SCET$_\mathrm{II}$ theory at the low scale $\mu \sim \la Q$;
this is noteworthy because in the previously considered cases in \eq{summary_hard_operators},
the $b \bar{b} \to H$ current is integrated out at
either the hard or hard-collinear scale.

The second possible contribution, which we denote by $\cM_{G'}$,
arises from the fermionic Glauber Lagrangian coupling only two sectors
to each other. These operators emit (or absorb) a soft fermion,
so $T$ ordered products with the leading-power current
would violate fermion conservation in the soft sector.
Instead, fermion number conservation
and the expected overall $\ord{\la}$ scaling
are recovered by a time-ordered product
with an $\ord{\la^{1/2}}$ hard-scattering operator
that has the following form,
\begin{align}
\frac{1}{\ell^-}
\bar{\chi}_{n}
S_n^\dagger
\Bigl[
\frac{\slashed{\bar{n}}}{2}S_{\bar{n}} (g \slashed{\cB}_{\nb \perp})
\Bigr]
  \Bigl[\delta\bigl(\ell^- \!- \bar{n}\!\cdot\! \img\overrightarrow{\partial_s}\bigr)\,S^{\dagger}_{\bar{n}}\psi\Bigr]
\,.\end{align}
It is generated by first matching onto the scalar quark current at the hard scale,
and then performing a trivial matching (with an external massless quark state)
of $\nb$-hard-collinear onto $\nb$-collinear modes,
while the $n$-collinear modes are integrated out
and matched onto a soft quark field as before.
(An example diagram for this contribution for the case of a soft emission
is shown in \fig{LO1softGSS}.)
Accordingly, the total contribution to the amplitude
is given by
\begin{align} \label{eq:glauber_contrib_two_sectors}
\cM_{G'}^{gg \to HX}
&= \frac{1}{m_H} C_{q\bar{q}}^S(m_H) \,
\bra{X} T \Bigl\{
	\int \! \df \ell^- \,
	\frac{\cJ(\ell^-)}{\ell^-}
\bar{\chi}_{n}
S_n^\dagger
\Bigl[
\frac{\slashed{\bar{n}}}{2}S_{\bar{n}} (g \slashed{\cB}_{\nb \perp})
\Bigr]
  \Bigl[\delta\bigl(\ell^- \!- \bar{n}\!\cdot\! \img\overrightarrow{\partial_s}\bigr)\,S^{\dagger}_{\bar{n}}\psi\Bigr]
\nn \\ & \qquad \qquad \qquad \qquad \qquad \times
	\Bigl[ \img \int \df^4 x \, \cL_{Gq}^{\text{II}(1/2)}
	\Bigr]
	+ (n \leftrightarrow \bn)
\Bigr\} \ket{g_n g_{\bn}}
\end{align}
and involves $C_{q\bar{q}}^S$ as well as a single hard-collinear jet function.

One important purpose served
by the contributions in \eqs{glauber_contrib_three_sectors}{glauber_contrib_two_sectors}
in the EFT calculation of the amplitude
is that they effectively determine the correct $\img 0$ prescriptions
in the eikonal factors $1/\ell^+$ and $1/\ell^-$
multiplying the soft operator in \eq{soft_Op}.
Concretely, computing contributions from that operator
leads to loops over soft momenta
that are defined including a Glauber zero-bin subtraction
to avoid double counting.
We find that in all cases we considered,
the dependence on the $\img 0$ prescription exactly cancels between
the naive soft loop integral and its Glauber zero bin,
as expected because the $\img 0$ prescription
precisely matters in the limit where $\ell^+$
and/or $\ell^-$ become small and the loop momentum enters a Glauber region.
The correct causal structure and the crossing properties of the amplitude
are then recovered by adding the genuine Glauber operator contribution
which contains dynamical collinear propagators at the scale $\mu \sim m_H\la$.
We note that for this mechanism to be active,
it is crucial that the contribution in \eq{glauber_contrib_two_sectors} is present,
which involves ``one half'' of the usual soft operator
and precisely is able to set the $\img 0$ prescription in this way
in cases where only one of $\ell^+$ or $\ell^-$ is large.
Another important requirement for this mechanism to work
is that all soft and Glauber contributions indeed involve
the leading-power $C_{q\bar{q}}^S$ matching coefficient at the hard scale.

In practice, we find that these observations can be exploited
to simplify the calculation by including the ``correct'' $\img 0$
prescription in the soft operator from the start,
where one replaces
\begin{align} \label{eq:replacment_i0_prescr}
\frac{1}{\ell^+ \ell^-}
\mapsto
\frac{1}{(\ell^+ + \img 0) (\ell^- - \img 0)}
\,.\end{align}
To be specific,
this is the correct replacement for the first part of the operator
that we have written out
in \eq{soft_Op}, i.e., the case where the quark field carries
a definite momentum $\ell^-$ along the $n^\mu$ direction
and the antiquark field carries $\ell^+$ along $\bn^\mu$.
For the case of opposite fermion flow implemented by $n \leftrightarrow \bn$,
where the antiquark carries definite momentum along $n^\mu$,
the prescription is $1/(\ell^+ - \img 0) (\ell^- + \img 0)$ instead.
With this choice, the Glauber contributions and Glauber zero bins
for some (but not all!) of the amplitudes we consider
are simply opposite and equal,
such that the naive soft integral gives the correct total result.
Specifically, we find
that this is the case for the well-studied LO$_0$ amplitude,
see \app{ybytsoftLO} for details,
such that a dedicated Glauber contribution indeed need not be included.
It also holds in the same way for the case of a single soft-gluon emission,
see \sec{softGlauber}.
In that case we find that the Glauber contribution
to the amplitude is nontrivial and involves a gluon emission
from the soft operator in \eq{glauber_contrib_three_sectors}, akin to a ``fermionic'' Lipatov vertex.
Nevertheless, with the above choice of $\img 0$ prescription
the Glauber and Glauber zero-bin contributions can be made to cancel.

By contrast, none of this is the case when a collinear gluon is emitted,
see \sec{glaubercoll}.
There, we find that the Glauber contribution
in \eq{glauber_contrib_three_sectors}
on its own is nontrivial,
not cancelled by the Glauber zero bin of the corresponding
naive collinar loop integrals,
and in fact required to recover the full-theory amplitude.
This means that Glauber contributions
for our process indeed have to be considered explicitly,
and cannot just be thought of as being absorbed into eikonal factors or Wilson lines.%
\footnote{Of course, one could argue that the $\img 0$ prescription
of the eikonal factors in the soft operator
could also be fixed through the matching at the hard-collinear scale,
i.e., by choosing to retain it as part of the matching coefficient.
However, such an ``inherited'' causal structure
would again cancel against the respective Glauber zero bins,
leaving the results unchanged.
At the same time,
the presence of nontrivial Glauber contributions (starting with one collinear
emission) provides strong evidence that Glauber modes
need to be part of the EFT.
In our analysis, the same operator contributes to these nontrivial terms and to terms that can be cancelled against soft loop Glauber zero bin subtractions.
}

\subsection{Kinematic regimes and factorization theorems}
\label{sec:factorization_theorems}

\begin{figure}
	\centering
	\hfill
	\begin{subfigure}{0.22\textwidth}
		\centering
		\includegraphics[height=4.5cm]{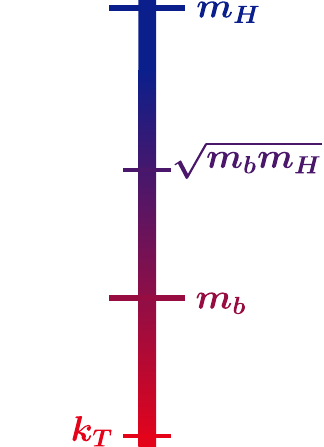}
		\subcaption{}
		\label{fig:kTllmllmH}
	\end{subfigure}%
	\hfill
	\begin{subfigure}{0.22\textwidth}
		\centering
		\includegraphics[height=4.5cm]{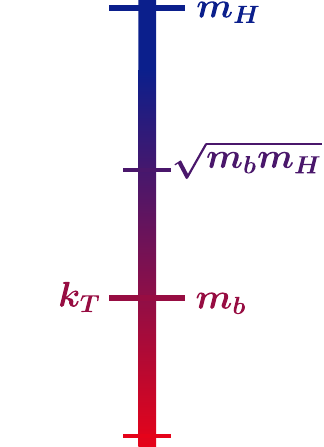}
		\subcaption{}
		\label{fig:kTsimmllmH}
	\end{subfigure}%
	\hfill
	\begin{subfigure}{0.22\textwidth}
		\centering
		\includegraphics[height=4.5cm]{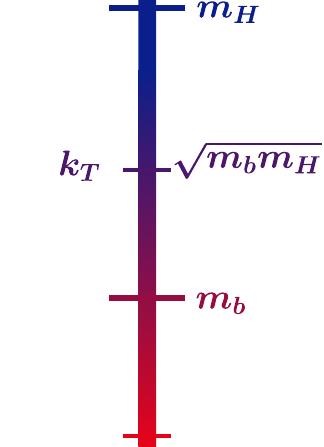}
		\subcaption{}
	\end{subfigure}%
	\hfill
	\begin{subfigure}{0.22\textwidth}
		\centering
		\includegraphics[height=4.5cm]{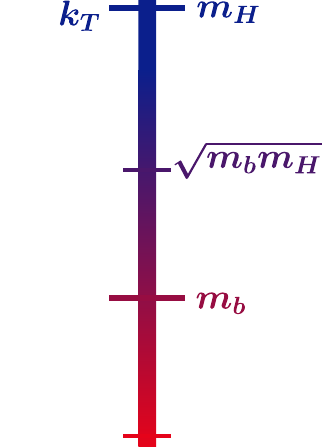}
		\subcaption{}
	\end{subfigure}%
	\caption{Illustration of different kinematic regimes with a hard scale $m_H$, an intermediate scale $\sqrt{m_b m_H}$ and a low scale $m_b$. The emission $k_T$ can have different scalings.}
	\label{fig:kTregimes}
\end{figure}

We now turn to the $y_b$-$y_t$-interference contribution to the Higgs $q_T$ spectrum. In contrast to the $2\to1$ amplitude discussed in \sec{operatorbasis}, the Higgs boson can only acquire a transverse momentum by recoiling against real radiation, so that at least one gluon crosses the cut.
At the order we work, this is a single emission with $\vec k_T=-\vec q_T$.
In addition to the masses  $m_b \ll m_H$, this emission $k_T$ adds a third scale to the problem.
As  the emission can have different scalings, we have to consider different kinematic regimes.
These regimes are illustrated in \fig{kTregimes}.
The hard scale $m_H$ and the low scale $m_b$ are always fixed. The hard-collinear scale is given by the intermediate scale $\sqrt{m_b m_H}$.
Where we give factorization theorems below, they are bare with respect to the endpoint divergences.
The renormalization of the leading-power hard, beam and soft functions is known~\cite{Chiu:2011qc,Chiu:2012ir,Echevarria:2011epo}, as is that of the bottom-quark form factor $C_{ggb}$~\cite{Liu:2022ajh}.
However, for $q_T\sim m_b$, the individual subleading-power terms contain endpoint divergences, which only cancel in their sum, so that they cannot be renormalized separately and are written in terms of unrenormalized functions.
\subsubsection{Small $q_T$: $q_T \ll m_b \ll m_H$}
We start by considering the kinematic region where the transverse momentum is smaller than any other scale $q_T \ll m_b \ll m_H$.
This scenario is sketched in \fig{kTllmllmH}.
The bottom-quark mass, $m_b =4.18 \GeV$~\cite{ParticleDataGroup:2024cfk} ($\MSbar$ scheme), is small compared to typical values of $q_T$.
This kinematic region therefore only covers the very beginning of the spectrum, close to the nonperturbative regime, and is phenomenologically not particularly relevant.
As $m_b$ is much larger than $q_T$, the bottom quark can be integrated out and we can describe this kinematic region using standard soft and beam functions for $n_f=4$ light flavors.
The factorization theorem is given by
\begin{align}\label{eq:factorization_small_qT}
	\frac{\df \sigma_{y_t y_b}}{\df q_T \, \df Y} = 2 \Re [C^*_{ggt}(m_H) \, C_{ggb}(m,m_H)]\,B_g(q_T, x_a)\otimes B_g(q_T, x_b)\otimes S_{gg}(q_T)\,,
\end{align}
where $Y$ is the rapidity of the Higgs boson and  $x_{a,b}= (m_H/E_{\mathrm{cm}}) e^{\pm Y}$ are  longitudinal momentum fractions.
Further, $\otimes$ denotes the convolutions in the transverse momentum.
Here and in the following, we use the shorthand $m \equiv m_b$ for the bottom-quark mass.
In this factorization theorem the $m$ and the $q_T$ dependence nicely factorize.
Here all emissions are described by leading-power beam and soft functions and the $m$-dependence is governed by the hard function.
The hard function has two contributions: $C_{ggt}$ is the Wilson coefficient that appears when integrating out the top quark and $C_{ggb}$ is the Wilson coefficient from integrating out the bottom quark which is given in terms of the form factor for the Born process from \refcite{Liu:2022ajh}. Note that $C_{ggb}$ contains the complete bottom-quark loop, i.e.\ the sum of the hard, collinear and soft contributions of \sec{operatorbasis}.
\subsubsection{Intermediate $q_T$: $q_T \sim m_b \ll m_H$}
\begin{figure}
	\begin{align*}
		\raisebox{-0.5\height}{\includegraphics[scale=0.8]{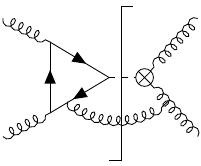}}\to
		& \; H_{gg}(m_H)\times\left[\,\raisebox{-0.5\height}{\includegraphics[scale=0.8]{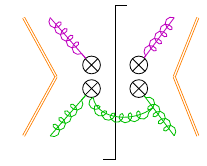}}
		+\raisebox{-0.5\height}{\includegraphics[scale=0.8]{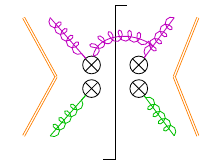}}
		+\raisebox{-0.5\height}{\includegraphics[scale=0.8]{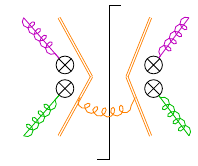}}\right]\\
		&+\raisebox{-0.5\height}{\includegraphics[scale=0.8]{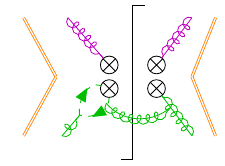}}
		+\raisebox{-0.5\height}{\includegraphics[scale=0.8]{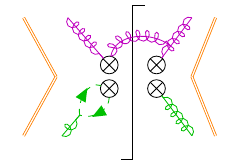}}
		+\raisebox{-0.5\height}{\includegraphics[scale=0.8]{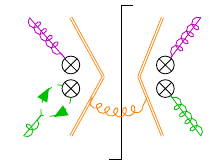}}\\
		&+\raisebox{-0.5\height}{\includegraphics[scale=0.8]{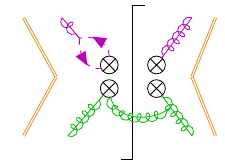}}
		+\raisebox{-0.5\height}{\includegraphics[scale=0.8]{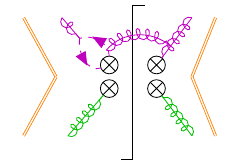}}
		+\raisebox{-0.5\height}{\includegraphics[scale=0.8]{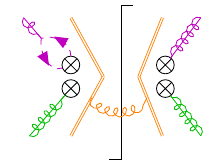}}\\
		&+\raisebox{-0.5\height}{\includegraphics[scale=0.8]{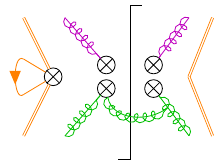}}
		+\raisebox{-0.5\height}{\includegraphics[scale=0.8]{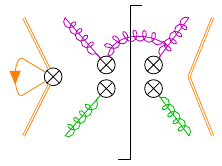}}
		+\raisebox{-0.5\height}{\includegraphics[scale=0.8]{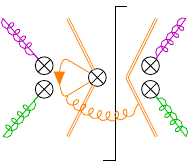}}
	\end{align*}
	\caption{Decomposition of the interference diagram in terms of hard, beam and soft functions.
		We only show example diagrams in each case. In each line, the mass dependence sits in the same place (hard function, $n$-collinear beam function, $\nb$-collinear beam function and soft function, respectively), and the emitted gluon is $n$-collinear (green), $\nb$-collinear (magenta) or soft (orange).
		The springs with a solid line represent collinear gluons. Collinear gluons attached to $\otimes$ represent a beam function.
		The solid double line represents a soft Wilson line where only soft gluons can attach. }
	\label{fig:modesLO1}
\end{figure}

Next, we consider the regime where the transverse momentum is of the size of the bottom-quark mass corresponding to \fig{kTsimmllmH}.
Phenomenologically, the regime is the most important one as the gluon fusion cross section peaks at roughly $q_T \approx 10 \GeV \approx 2 m$,
and hence it will be the main focus of our work.
The structure of the interference is illustrated in \fig{modesLO1}.
The first diagram shows a full QCD diagram which is then decomposed in terms of hard, collinear and soft contributions.
The right-hand side of the cut contains the top-quark loop that was integrated out.
In the effective theory, this contribution is given in terms of leading-power collinear and soft amplitudes.
The left-hand side of the cut shows the bottom-quark contributions.
We can write them in terms of next-to-leading-power (NLP) hard, collinear and soft amplitudes.
Once both sides are interfered, we will find new NLP beam and soft functions. The interference is thus always the product of a leading-power amplitude, proportional to $y_t$, and a next-to-leading-power amplitude, proportional to $y_b m$.
In the first line of the decomposition, the mass dependence is included in the Wilson coefficient $H_{gg}(m_H)$, i.e.\ the hard contribution. In the remaining lines the mass dependence is included in subleading-power beam functions, with a massive $n$-collinear (second line) or $\nb$-collinear (third line) loop, and in the subleading-power soft function (fourth line). Each line shows an $n$-collinear, an $\nb$-collinear and a soft emission. Note that the diagrams in \fig{modesLO1} symbolically stand for a full class of diagrams.
Although we always analyze the power corrections on the LHS of the cut, the corrections on the RHS are always included.
The schematic factorization theorem reads
\begin{align}\label{eq:factorization_intermediate_qT}
	\frac{\df \sigma_{y_t y_b}}{\df q_T\, \df Y} =&H_{gg}(m_H)\,B_g(q_T, x_a)\otimes B_g(q_T, x_b)\otimes S_{gg}(q_T)  \\
	&+ \int \df \xi H_{bbg}(\xi)\,\left[B_{n,\bar{\chi}\chi}(\xi,q_T,m,x_a)\otimes B_g(q_T, x_b)\otimes S_{gg}(q_T)\right.\nn \\
	&\left. \qquad
	+ B_g(q_T, x_a)\otimes B_{\bar{n},\bar{\chi}\chi}(\xi,q_T,m,x_b)\otimes S_{gg}(q_T)\right]\nn \\
	&+ \int   \df\ell^+\df\ell^- H_{bbgg}\mathcal{J}(\ell^+)\mathcal{J}(\ell^-)\,B_g(q_T, x_a)\otimes B_g(q_T, x_b)\otimes S_{\bar{\psi}\psi}(\ell^+, \ell^-,q_T,m)\nn \\
	&
  + \ldots
  \,. \nn
\end{align}
Here, $H_{gg}(m_H)$ is built from $C_{ggt}$ and the hard contribution $C(m_H)$, while $H_{bbg}(\xi)$ and $H_{bbgg}$ are built from $C_{ggt}$ and the Wilson coefficients $C_{bbg}$ and $C_{q\bar q}^S$ of \sec{operatorbasis}.
Note that $H_{gg}(m_H)$ differs from $2 \Re [C^*_{ggt} C_{ggb}]$ in \eq{factorization_small_qT}, since it only contains the hard part of the bottom-quark loop.
The ellipses in \eq{factorization_intermediate_qT} denote possible contributions from the Glauber quark exchange Lagragian in \eq{Glauber_Lagrangian_quark}. For the one-loop single-emission analysis carried out here, we will show that these Glauber quark contributions either can be absorbed into the soft function for soft emission, or,
for collinear emissions, give a nonvanishing amplitude that contributes only to its imaginary part and therefore drops out of the interference with the tree-level leading-power amplitude on the other side of the cut.
Since this cancellation relies only on the relative phase between the two sides of the cut,
the Glauber quark contributions could be nonvanishing at higher orders, where loop corrections give complex amplitudes.
These Glauber one-loop amplitudes are needed to reproduce known full theory amplitude results in the collinear limit.
The next-to-leading-power beam and soft functions are defined as
\begin{align}\label{eq:NLP_beam_soft}
	B_{n,\bar{\chi}\chi}(\xi, k_T, m,\frac{\omega}{P_N^-})&=\SumInt_X\bra{N} \mathcal{B}_\perp^{a, \mu}\delta^2(k_\perp-P_{X,\perp})\ket{X}\bra{X} \bar{\chi}_{n,\omega \xi}\frac{\slashed{\nb}}{2}T^{a}\gamma_{\perp\mu}\chi_{n,\omega (1-\xi)}\ket{N}
+ \mathrm{h.c.} \,, \nn \\
	S_{\bar{\psi}\psi}(\ell^+,\ell^-,k_T,m)&=\SumInt_X \frac{1}{N_c^2-1}\bra{0}\mathcal{O}_s^{(0)\, ab,\mu\nu}\delta^2(k_\perp-P_{X,\perp})\ket{X}\nn\bra{X}\mathcal{O}^{ba}_{s,\bar{\psi}\psi,\nu\mu}(\ell^+,\ell^-)\ket{0}\nn \\
	&\quad+ \mathrm{h.c.} \,
\end{align}
Here, $\omega$ is the total label momentum of the $n$-collinear fields entering the hard vertex, which by momentum conservation equals $q^- = m_H e^{Y}$ up to power corrections.
With $P_N^- = E_{\rm cm}$ this gives $\omega/P_N^- = x_a$, and analogously $\bar\omega/P_{\bar N}^+ = x_b$ for the $\nb$-collinear beam function.
In both functions, the first matrix element contains the leading-power operator from the top-quark side of the cut, whereas the second one contains the NLP operators obtained from \eqs{collinear_Op}{soft_Op}.
Here $\cO_s^{(0)\,ab,\mu\nu}=g_\perp^{\mu\nu}\,\big(S_n^\dagger S_{\nb}\big)^{ab}$ denotes the leading-power soft operator.
The NLP soft operator $\cO^{ab,\mu\nu}_{s,\bar\psi\psi}(\ell^+,\ell^-)$ is obtained from $\cO_{bbgg}$ in \eq{soft_Op} by removing the two collinear gluon fields, which are part of the beam functions, i.e.\ by replacing $(g\slashed{\cB}_{n\perp})\to T^a\gamma_\perp^\mu$ and $(g\slashed{\cB}_{\nb\perp})\to T^b\gamma_\perp^\nu$ (and correspondingly in the $n\leftrightarrow\nb$ term). Likewise, the operator $\bar\chi_{n}\frac{\slashed{\nb}}{2}T^a\gamma_{\perp\mu}\chi_{n}$ in the beam function is obtained from $\cO_{bbg}$ in \eq{collinear_Op} by removing the $\nb$-collinear gluon field.

We now discuss the three contributions in \eq{factorization_intermediate_qT} in detail.
The first term in the sum describes the scenario where all propagators in the loop are hard and the mass dependence is fully encoded in the hard function $H_{gg}(m_H)$.
Here $C(m_H)$ carries the subleading-power bottom quark
mass dependence whereas the emissions are described by leading-power soft and beam functions.
Since the emission is soft or collinear, it cannot resolve the hard loop, and $C(m_H)$ is given by the leading-order hard contribution given in \app{LO0-calc}.
With a single emission, exactly one of the beam and soft functions contributes at NLO, while the other two contribute at leading order.
The emission can be $n$- or $\nb$-collinear with the respective other beam function and the soft function contributing at leading order which is described by the first two diagrams in the first line of \fig{modesLO1}.
The emission can also have soft scaling, which is shown in the last diagram of this line. In this case, the two beam functions contribute at leading order.

Next, we take a closer look at the second term in \eq{factorization_intermediate_qT},
where the mass dependence is captured by either the $n$- or the $\nb$-collinear beam function.
In contrast to the first term, it suffers from endpoint divergences in the $\xi$ integral. Similar divergences arise in the $\ell^\pm$ integrals of the soft term, and they only cancel in the sum of both, see \secs{collinearemission}{softemission}.
The corresponding single-emission diagrams are shown in the second and third lines of \fig{modesLO1}.
If the emission is collinear to the massive quark loop, it comes from the subleading-power beam function, which also governs the mass dependence.
Otherwise, the emission-free (LO$_0$) matrix element of the subleading-power beam function multiplies a leading-power emission from the other beam function or the soft function.
Diagrammatically, the new beam function reads
\begin{align}\label{eq:NLP_beam_LO}
	B_{n,\bar\chi\chi}(\xi,k_T,m,z) = \raisebox{-0.45\height}{\includegraphics[height=2.2cm]{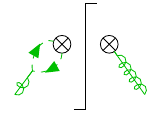}} + \raisebox{-0.45\height}{\includegraphics[height=2.2cm]{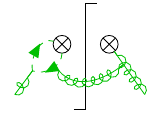}} + \dots\,,
\end{align}
where the first diagram is the lowest-order contribution and the second one an example of an $\ord{\alpha_s}$ real-emission correction. The left-hand side of the cut shows the massive $n$-collinear quark loop from the operator $\cO_{bbg}$ and the right-hand side the leading-power gluon beam operator. The $\nb$-collinear beam function is obtained by $n\leftrightarrow\nb$.

Finally, we discuss the last term in \eq{factorization_intermediate_qT},
where the mass dependence is included in the soft function.
The corresponding diagrams are shown in the last line of \fig{modesLO1}.
Similar to the collinear case, the emission either originates from the subleading-power soft function, or the lowest-order massive soft function multiplies a leading-power emission from one of the beam functions.
Diagrammatically, the new soft function reads
\begin{align}\label{eq:NLP_soft_LO}
	S_{\bar\psi\psi}(\ell^+,\ell^-,k_T,m) = \raisebox{-0.45\height}{\includegraphics[height=2.6cm]{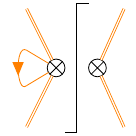}} + \raisebox{-0.45\height}{\includegraphics[height=2.6cm]{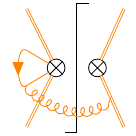}} + \dots\,,
\end{align}
where again the first diagram is the lowest-order contribution and the second one an example of an $\ord{\alpha_s}$ real-emission correction. The left-hand side of the cut shows the soft quark loop from $\cO_{bbgg}$ with its soft Wilson lines, and the right-hand side the leading-power soft operator $\cO_s^{(0)}$.
The two jet functions $\mathcal{J}(\ell^+)\mathcal{J}(\ell^-)$ in \eq{factorization_intermediate_qT} encode the effect of virtual and hard-collinear contributions that have been integrated out.
 At our working order the jet functions are simply given by $\mathcal{J}(\ell^\pm)=1$.

We note that for both $B_{n,\bar{\chi} \chi}$ and $S_{\bar{\psi}\psi}$ we will always focus on the case where no heavy flavor crosses the cut, adopting a theorist's definition of what we mean by ``gluon fusion''.
 (While vetoing this is experimentally unfeasible for $b$ hadrons,
 it is theoretically well defined
 and results in a finite partonic cross section since the bottom quark is massive.)
 There are, in principle, diagrams where the quarks emitted by the subleading operator
 enter the final state, and are created by a leading-power emission
 and a subsequent $g \to b \bar{b}$ splitting in the conjugate $y_t$ amplitude.
 These are already contained in the above matrix-element definitions,
 but we leave a dedicated study of these contributions to future work.

The most important feature of this regime is the cancellation of the endpoint divergences.
As at LO$_0$, the $\xi$ convolution and the $\ell^\pm$ convolutions in \eq{factorization_intermediate_qT} are individually endpoint divergent, and only their sum is finite.
Since all terms in \eq{factorization_intermediate_qT} are interfered with the same leading-power amplitude proportional to $y_t$, this cancellation already takes place at the amplitude level, i.e.\ between the collinear and soft contributions to the bottom-quark amplitude at fixed emission momentum $k$.
In contrast to LO$_0$, however, the coefficients of the endpoint divergences can now be nontrivial functions of $m/k_T$.
We therefore perform the calculations in the following \secs{collinearemission}{softemission} at the amplitude level and show explicitly how the endpoint divergences cancel.
\subsubsection{Large $q_T$: $m_b \ll q_T$}
Both remaining regimes, $m \ll q_T \ll m_H$ and $m \ll m_H \lesssim q_T$,
share the feature that the bottom-quark mass is the smallest scale in the
problem. One might therefore expect that a massless treatment is sufficient:
in the first case the factorization theorem would reduce to the massless one
since $m \ll q_T$, and in the second the logarithms of $q_T/m_H$ are no longer
large, so that the fixed-order result of \refcite{Ellis:1987xu} would suffice.
Neither expectation holds. The logarithms of $m/m_H$ remain large in both
regimes and require resummation, and subleading-power operators must be kept in
the high-scale matching in order to pick up the correct mass suppression at the
low scale where the bottom-quark loop is closed.

The two regimes differ in how the emission is generated. For
$m \ll m_H \lesssim q_T$, a transverse momentum of order $m_H$ requires the
gluon crossing the cut to have hard scaling, so that no amplitude-level
factorization is possible; instead, the factorization takes place at
cross-section level and assumes a PDF-like form. For $m \ll q_T \ll m_H$, a
promising way to determine the structure of this part of the cross section is
to approach it from the two neighboring regimes. A detailed discussion of both
regimes and their factorization theorems is beyond the scope of this paper.

\section{Collinear emission diagrams}
\label{sec:collinearemission}
Since we are computing a real-emission process, the diagrams must be organized according to two independent momentum scalings: that of the quarks in the loop and that of the emitted gluon $k$.
As \fig{modesLO1} shows, these two scalings are not tied to each other: the emission can be soft while the loop momenta are collinear, and vice versa.
A substantial part of the calculation consists in verifying the cancellation of the endpoint divergences, a known feature of subleading-power calculations.
These divergences must cancel point by point in $k$, which already
suggests the appropriate classification: the sum of all diagrams with a
collinear emission and the sum of all diagrams with a soft emission must be free of endpoint divergences on their own.
We therefore treat the $n$-collinear emission, $k \sim (\lambda^2, 1, \lambda)$, in this section and the soft emission in \sec{softemission}.
The contributions from an $\nb$-collinear emission, $k\sim(1,\lambda^2,\lambda)$, follow from those presented here by symmetry, i.e.\ by interchanging $n\leftrightarrow\nb$.
We organize the collinear-emission diagrams according to where the gluon is emitted: from the incoming collinear gluon via the triple-gluon vertex ($d_{ggg}$), from the effective vertex ($d_{VE}$), or from the quark loop ($d_{LE}$), using the same color coding as in \fig{modesLO0}.
We find that the endpoint divergences cancel within these diagram classes.
In particular, the endpoint divergences of the diagrams in which the gluon is emitted from the subleading-power beam function cancel against those of the LO$_0$ collinear and soft contributions multiplied by leading-power emissions.
The $\eps$ poles, in contrast, cancel only once the soft and collinear contributions are combined with the contribution from the hard loop.
The hard and Glauber contributions are discussed after the three diagram classes.

As in \sec{operatorbasis}, we can choose the incoming gluon momenta to be exactly aligned with the light-cone directions:
\begin{align}
	p_1^\mu =p_1^- \frac{n^\mu}{2}, \qquad p_2^\mu =p_2^+ \frac{\nb^\mu}{2},
\end{align}
such that $p_1 =(0, p_1^-,0)$ and $p_2 =(p_2^+,0,0)$.
Throughout this calculation we assume transverse gluon polarization for the incoming gluons and drop all terms proportional to $n^\mu$, $\nb^\mu$, $n^\nu$ and $\nb^\nu$, where $\mu$ and $\nu$ are the Lorentz indices of the two incoming gluons and $\rho$ is the one of the emitted gluon.
We have carried out the calculation both in Feynman gauge and in axial gauge with reference vector $\nb$, i.e.\ with $\nb\cdot\eps(k)=0$ for the emitted gluon.
Since the sum of all diagrams satisfies the Ward identity (see \app{wardcoll}), the two gauges give the same physical amplitudes, and using axial gauge amounts to dropping all terms proportional to $\nb^\rho$. In particular, the emission from the Wilson lines in the effective vertex, i.e.\ the vertex-emission diagrams $d_{VE}$, then does not contribute.
In the following we present the results in Feynman gauge, keeping the $\nb^\rho$ terms.
All external gluons are on shell, $p_1^2=p_2^2=k^2=0$.
For convenience, we change variables from  $(\omega_{1},\omega_{2})$ $\to (\omega_n, \xi)$ where $\omega_n = -(\omega_{1}+\omega_2) = q^-$
 denotes the overall minus momentum of the process which coincides with the minus momentum of the Higgs boson.
  With the emission, we have
\begin{align}
	\omega_n = p_1^--k^-,
\end{align}
and $\xi= -\omega_2/\omega_n$, such that $\xi$ and $1-\xi$ are the fractions of $\omega_n$ carried by the two collinear quark lines.
In this notation, and including the Jacobian $\df\omega_1\df\omega_2=\omega_n\,\df\omega_n\df\xi$ as well as the overall sign in the Feynman rule \eq{LOcFR}, the Wilson coefficient is given by
\begin{align}\label{eq:Wcoeff}
	C_{bbg}(\xi)=\frac{1}{\xi}+ \frac{1}{1 -\xi}.
\end{align}
For simplicity, we will omit the $\delta$-function constraining the overall $+$ and $-$ momenta in our calculation as they can be trivially reinstated.

For the evaluation of the real emission diagrams, we proceed  as follows: For the diagrams involving a collinear loop, we start by using the $\delta$-function to perform the $\ell^-$ integral. Next, we do the $\ell^+$ integral by contours. This constrains the $\xi$ integral since the contour integral only has support in specific regions for $\xi$. After doing the contour integral we can use standard formul\ae~to evaluate the $\ell_\perp$ integral.
In the last step, we evaluate the $\xi$ integral which suffers from endpoint divergences and, thus, needs to be regulated.
For the $1/\xi$ and $1/(1-\xi)$ terms of $C_{bbg}(\xi)$, we use the factors $|\xi\omega_n/\nu|^{-\eta}$ and $|(1-\xi)\omega_n/\nu|^{-\eta}$, respectively, which we refer to as the absolute-value regulator. Here, $\nu$ is an auxiliary scale introduced by the regulator. Although these are endpoint divergences and not rapidity divergences, we regulate them in the same way as rapidity divergences are regulated in \refscite{Chiu:2011qc,Chiu:2012ir,Echevarria:2011epo}, i.e.,
we are inserting an explicit regulator mostly as a bookkeeping device.
For the diagrams with a soft or $\nb$-collinear loop, the loop is not affected by the collinear emission, and we can reuse the leading-order results of \app{LO0-calc}, which are obtained with the same regulator.

\subsection{Emission from the incoming gluon}
\label{sec:gggsection}
\begin{figure}
	\begin{subfigure}{0.31\textwidth}
		\centering
		\includegraphics[width=0.8\textwidth]{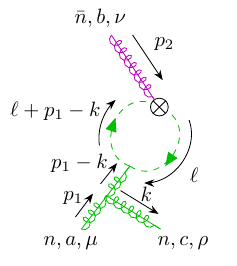}
		\caption{d$_{ggg}^{n}$}
		\label{fig:d2diasn}
	\end{subfigure}
	\hfill
	\begin{subfigure}{0.31\textwidth}
		\centering
		\includegraphics[width=0.8\textwidth]{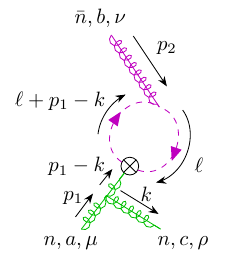}
		\caption{d$_{ggg}^{\nb}$}
		\label{fig:d2diasnb}
	\end{subfigure}
	\begin{subfigure}{0.31\textwidth}
		\centering
		\includegraphics[width=0.8\textwidth]{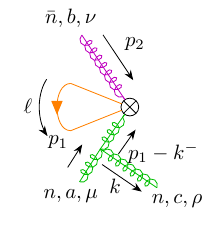}
		\caption{d$_{ggg}^{s}$}
		\label{fig:d2diass}
	\end{subfigure}
	\caption{d$_{ggg}$-type diagrams: all diagrams where the collinear emission comes from the interaction of three collinear gluons.}
	\label{fig:d2dias}
\end{figure}
We start with the diagrams where the collinear gluon is emitted from the incoming gluon via the three-gluon vertex.
Using the Feynman rules in \app{feynman_rules} as well as the massive SCET Feynman rules of \refcite{Leibovich:2003jd}, the contribution depicted in \fig{d2diasn} is given by
\begin{align}
	\cM_{d_{ggg}^{n}} =&-\frac{ g^3}{2}  T^d T^b f^{dac} \int \! \df \xi \int \!\frac{\df^{d}\ell}{(2\pi)^d}
		C_{bbg}\big(\xi\big)
		\delta(\omega_n \xi -\ell ^-- p_1^- +k^-\big)
		\nn\\
		&	\times \frac{g^{\alpha\beta}}{\big(p_1-k\big)^{2}}\;
		\Big[g^{\beta\mu}\big(k-2p_1\big)^{\rho}
		+g^{\beta\rho}\big(p_1-2k\big)^{\mu}
		+g^{\mu\rho}\big(k+p_1\big)^{\beta}\Big]\nn \\
	&\times\Tr\Bigg\{
	\frac{\slashed n}{2}\frac{\img\,\ell \cdot\bar n}{\big(\ell^2-m^2+\img 0\big)}\;
	\frac{\slashed{\bar n}}{2}\,\gamma^{\nu}
	\frac{\slashed n}{2}\frac{\img\bar n\cdot(\ell-k+p_1)}
	{\big[(\ell-k+p_1)^2-m^2+\img 0\big]} \nn\\
	&\qquad\quad\Bigg[
	n^{\alpha}\,\slashed{\bar n}
	+\frac{\gamma^{\alpha}_\perp\big(m+\slashed{\ell}_\perp\big)\slashed{\bar n}}{\ell\cdot\bar n}
	+\frac{\big(\slashed{\ell}_\perp-\slashed{k}_\perp+\slashed{p_1}_\perp-m\big)\gamma^{\alpha}_\perp\,\slashed{\bar n}}{\bar n\cdot(\ell-k+p_1)} \nn\\
	&\hspace{3.5cm}-\,\frac{\big(\slashed{\ell}_\perp-\slashed{k}_\perp+\slashed{p_1}_\perp-m\big)\big(m+\slashed{\ell}_\perp\big)\slashed{\bar n}\;\bar n^{\alpha}}{\big(\ell\cdot\bar n\big)\big[\bar n\cdot(\ell-k+p_1)\big]}
	\Bigg]\Bigg\}
.
\end{align}
After taking the trace using the \texttt{FeynCalc} package~\cite{Mertig:1990an, Shtabovenko:2016sxi, Shtabovenko:2020gxv, Shtabovenko:2023idz}, we integrate over $\ell^-$ by resolving the $\delta$ function and the expression simplifies to
\begin{align}
	\cM_{d_{ggg}^{n}}  =& -\frac{g^3 m(1-z)f^{abc}}{8\pi z\,k_\perp^2} (4 \pi)^{-\eps} e^{\gamma_E \eps}\mu^{2\eps}\int \! \df \xi \int \!\frac{\df \ell^+\df^{d-2}\ell_\perp}{(2\pi)^{d-1}} \frac{C_{bbg}(\xi)}
	{\,\Big[\ell_\perp^2+\omega_n(\xi-1)\,\ell^{+}-m^2+\img 0\Big]} \nn\\[4pt]
	&\times\frac{1}
	{\Big[\big(1+(\xi-1)z\big)k_\perp^2-(z-1)\big(\ell_\perp^2-2k_\perp\cdot\ell_\perp+\omega_n\xi\,\ell^{+}-m^2+\img 0\big)\Big]} \nn\\[8pt]
	&\times\Bigg\{-4\omega_n(z-1)\,g^{\mu\rho}_\perp\,{k}^{\nu}_\perp
	+4z(z-1)\,{k}^{\mu}_\perp\Big(\omega_n\,g^{\nu\rho}_\perp+{k}^{\nu}_\perp\,\bar n^{\rho}\Big) \nn\\
	&\qquad+g^{\mu\nu}_\perp\Big[-2\omega_n z(z-1)\,{k}^{\rho}_\perp+\omega_n^2(z^2-1)\,n^{\rho}-z^2 k_\perp^2\,\bar n^{\rho}\Big]\Bigg\},
\end{align}
Here, we used that the incoming gluons $p_{1,2}$ and the emission $k$ are on-shell, and we write the external momenta as
\begin{align}
	p_1^-= \frac{\omega_{n}}{z}, \qquad k^-= \frac{\omega_{n}(1-z)}{z}, \qquad k^+=- \frac{k_\perp^2}{k^-},
\end{align}
where $z \equiv \omega_n/p_1^- \in [0,1]$.
In the next step, we do the $\ell^+$ integral by contours. This constrains the $\xi$ integral since the contour integral only has support for $0<\xi<1$.
After doing the contour integral we can use standard formul\ae~to solve the $\ell_\perp$ integral. Now, we are left with the endpoint-divergent $\xi$ integral which we regulate using the absolute-value regulator
\begin{align}
		\cM_{d_{ggg}^{n}}  =&\frac{\img g^3f^{abc} m}{32\,\pi^2\,k_\perp^2\,\omega_n\,z} e^{\eps\gamma_E}(\mu^2)^{\eps}\,\Gamma(\eps)\nn \\
	&	\times \int_0^1 \! \df \xi
		\Bigg[\frac{1}{\xi} \Big \vert \frac{\xi\omega_n}{\nu}\Big \vert^{-\eta} +\frac{1}{1-\xi}\Big \vert \frac{(1-\xi)\omega_n}{\nu}\Big \vert^{-\eta} \Bigg]
		\Bigg[\frac{1}{\frac{\xi(\xi-1)k_\perp^2}{1-z}+\left(m^2-\img 0\right)}\Bigg]^{\eps}\nn \\
		&\hspace{2cm}\times
		\Bigg\{
		g^{\mu\nu}_\perp\Big[2\omega_n z(z-1)\,{k}^{\rho}_\perp-\omega_n^2(z^2-1)\,n^{\rho}+z^2 k_\perp^2\,\bar n^{\rho}\Big]
		\nn \\
		&\hspace{3cm}
		 +4\omega_n(z-1)\,g^{\mu\rho}_\perp\,{k}^{\nu}_\perp -4z(z-1)\,{k}^{\mu}_\perp\Big(\omega_n\,g^{\nu\rho}_\perp+{k}^{\nu}_\perp\,\bar n^{\rho}\Big)\Bigg\}\nn \\
		= &\frac{\img g^3\, m f^{abc}}
		 {16\,\pi^2\,k_T^2\,\omega_n\,z\,\eta}  \left|\dfrac{\omega_n}{\nu}\right|^{-\eta}  e^{\eps\gamma_E} \Big(\frac{\mu^2}{m^2}\Big)^{\eps} \Gamma(\eps)
		 \Bigg\{
		 4z(1-z)\,{k}^{\mu}_\perp\Big(\omega_n\,g^{\nu\rho}_\perp+{k}^{\nu}_\perp\,\bar n^{\rho}\Big) \nn\\
		 &\qquad+4\omega_n(z-1)\,g^{\mu\rho}_\perp\,{k}^{\nu}_\perp+g^{\mu\nu}_\perp\Big[2\omega_n z(z-1)\,{k}^{\rho}_\perp-\omega_n^2(z^2-1)\,n^{\rho}+z^2 k_\perp^2\,\bar n^{\rho}\Big]\Bigg\}\nn \\
		 &\times
		 F_1\!\left(-\eta;\eps,\eps;1-\eta;\;
		 \frac{2k_T}{k_T+\sqrt{k_T^2-4m^2(z-1)}},\;
		 \frac{-2k_T}{-k_T+\sqrt{k_T^2-4m^2(z-1)}}\right).
\end{align}
In the above expression $F_1 (...)$ denotes the $F_1$ Appell function, $\eta$ regulates the endpoint divergences at $\xi=0$ and $\xi=1$ and $\nu$ is the auxiliary scale introduced by the regulator.
 The resulting expression for $\cM_{d_{ggg}^{n}}$ features poles in both regulators $\eta$ and $\eps$.
 The $\eta$ poles cancel against the contributions with soft and $\nb$-collinear loops, while the cancellation of the $\eps$ poles also requires the hard contribution.

For the remaining two diagrams of this class (\fig{d2diasnb} and \fig{d2diass}), we can simply use the leading-order results and contract them with the triple-gluon vertex of QCD.
The leading-order results in \app{LO0-calc} were derived for on-shell gluons with physical polarization.
We thus need to make a slight modification and replace $\slashed{n}\to\slashed{n}+{\slashed{p}_{1,\perp}}/{p_1^-}$ as well as shift the momenta according to the \loone~kinematics which yields
\begin{align}
	\cM_{\mathrm{d}_{ggg}^{\nb}}=
	&\frac{\img\,g^3 f^{abc}m}{16\pi^2\,k_T^2\,\omega_n\,z\,\eta}\left|\dfrac{\omega_n}{\nu}\right|^{-\eta} e^{\eps\gamma_E}
	\left(\dfrac{\mu^2}{m^2}\right)^{\eps}\Gamma(\eps)
	\nn\\
	&\times\Bigg\{g^{\mu\nu}_\perp\Big[2\omega_n(z-1)z\,{k}^{\rho}_\perp-k_T^2 z^2\,\bar n^{\rho}-\omega_n^2(z^2-1)\,n^{\rho}\Big] \nn\\
	&\qquad-4(z-1)z\,{k}^{\mu}_\perp\Big(\omega_n\,g^{\nu\rho}_\perp+\bar n^{\rho}\,{k}^{\nu}_\perp\Big)
	+4\omega_n(z-1)\,{k}^{\nu}_\perp\,g^{\mu\rho}_\perp\Bigg\}\\
\cM_{\mathrm{d}_{ggg}^{s}}=&\frac{\img\,g^3 f^{abc}  m}{\pi^{5/2}\,k_T^2\,\omega_n^2}2^{-\eta-4} \, e^{\eps\gamma_E} \left(\dfrac{\mu^2}{m^2}\right)^{\eps}\nn\\
	&\times\Bigg\{\pi^{3/2}\Gamma(\eps)\,\nu^{\eta}\,\csc(\pi\eta)\Big[({-}\img 0)^{-\eta}-(\img 0)^{-\eta}\Big]
	-\frac{2}{\eta}\,\Gamma\!\left(\dfrac12-\dfrac\eta2\right)\Gamma\!\left(\eps+\dfrac\eta2\right)\left(\dfrac{m}{\nu}\right)^{-\eta}\Bigg\} \nn\\[8pt]
	&\times\Bigg\{g^{\mu\nu}_\perp\Big[2\omega_n^2(z-1)\,{k}^{\rho}_\perp-\omega_n k_T^2 z\,\bar n^{\rho}-\frac{\omega_n^3(z^2-1)}{z}\,n^{\rho}\Big] \nn\\
	&\qquad-4\omega_n(z-1)\,{k}^{\mu}_\perp\Big(\omega_n\,g^{\nu\rho}_\perp+\bar n^{\rho}\,{k}^{\nu}_\perp\Big)
	+\frac{4\omega_n^2(z-1)}{z}\,{k}^{\nu}_\perp\,g^{\mu\rho}_\perp\Bigg\}
\end{align}
We can now combine all three contributions
\begin{align}
	\cM_{d_{ggg}}=\cM_{\mathrm{d}_{ggg}^{n}}+\cM_{\mathrm{d}_{ggg}^{\nb}}+\cM_{\mathrm{d}_{ggg}^{s}}=\Ord{ \eta^0}\,,
\end{align}
and find that this subset is indeed free of endpoint divergences.
\subsection{Vertex emission diagrams}
\label{sec:VEsection}
\begin{figure}
	\centering
	\hfill
	\begin{subfigure}{0.24\textwidth}
		\centering
		\includegraphics[width=\textwidth]{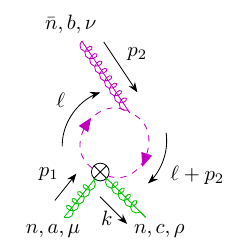}
		\caption{$d_{VE}^{\nb}$}
		\label{fig:LO1d1diasnb}
	\end{subfigure}
	\hfill
	\begin{subfigure}{0.24\textwidth}
		\centering
		\includegraphics[width=\textwidth]{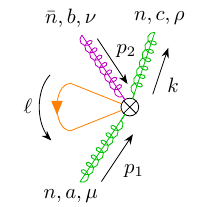}
		\caption{$d_{VE}^{s}$}
		\label{fig:LO1d1diass}
	\end{subfigure}
	\hfill
	\begin{subfigure}{0.24\textwidth}
		\centering
		\includegraphics[width=0.9\textwidth]{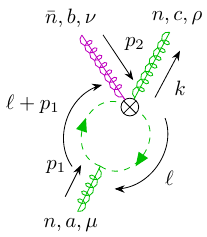}
		\caption{$d_{VE}^n$}
		\label{fig:LO1d1diasn}
	\end{subfigure}
	\hfill
	\begin{subfigure}{0.24\textwidth}
		\centering
		\includegraphics[width=\textwidth]{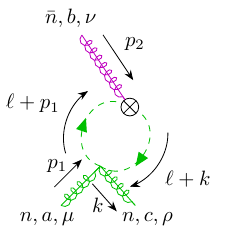}
		\caption{$d_{VE}^{\text{qqgg}}$}
		\label{fig:LO1d1diasFFGG}
	\end{subfigure}
	\hfill
	\caption{$d_{VE}$ diagram class: all diagrams where the emission is attached to an effective vertex.}
	\label{fig:LO1d1dias}
\end{figure}

In this subsection, we sketch the calculation of contributions where the gluon is emitted from an effective vertex. These diagrams are shown in \fig{LO1d1dias}.
\subsubsection{Contribution from $\nb$-collinear and soft loops}
The $\nb$-collinear and soft loop contributions in this class of diagrams are analogous to the leading-order calculation.
The contribution from the $d_{VE}^{\nb}$ and the $d_{VE}^{s}$ can be obtained by contracting the leading-order Feynman rule with the gluon field strength at $\ord g$ which is given in the appendix of \refcite{Berger:2010xi}.
For completeness, we give the resulting Feynman rules in \eqs{d1acFR}{d1sFR}.
This leads to the following results
\begin{align}
	\cM_{d_{VE}^{\nb}}=&- \frac{\img g^3f^{abc}}{8\pi ^{2}} \frac{\Gamma(\eps)}{\eta}\frac{z}{\omega_n(1-z)} m^{1-2\eps}e^{\eps \gamma_E}\mu^{2\eps} \left|\frac{\nu}{\omega_n}\right|^\eta g_\perp^{\mu \nu}\nb^\rho \\
	\cM_{\mathrm{d}_{VE}^s}=&-\frac{\img g^3f^{abc}}{8 \pi ^{\frac{5}{2}}} \frac{z}{\omega_n(1-z)} 2^{-\eta } m^{1-2\eps}e^{\eps \gamma_E}\mu^{2\eps} g_\perp^{\mu \nu}\nb^\rho  \\
	&\times\Bigg\{\pi^{3/2}\Gamma(\eps)\,\nu^{\eta}\,\csc(\pi\eta)\Big[({-}\img 0)^{-\eta}-(\img 0)^{-\eta}\Big]
	-\frac{2}{\eta}\,\Gamma\!\left(\dfrac12-\dfrac\eta2\right)\Gamma\!\left(\eps+\dfrac\eta2\right)\left(\dfrac{m}{\nu}\right)^{-\eta}\Bigg\} \nn
\end{align}
\subsubsection{\boldmath{$d_{VE}^{qqgg}$} contribution }
For the $d_{VE}^{qqgg}$ diagram  (shown in \fig{LO1d1diasFFGG}), the additional emission happens via the massive SCET vertex connecting two collinear quarks and two collinear gluons, which arises from the massive collinear quark Lagrangian~\cite{Leibovich:2003jd}.
Although this vertex does not stem from the operator $\cO_{bbg}$, we include the diagram in the vertex-emission class since its endpoint divergences are required for the cancellation within this class.

The evaluation of the $\ell^-$, $\ell^+$ and $\ell_\perp$ integrals is completely analogous to the calculation presented in \sec{gggsection}.
\begin{align}
	\cM_{d_{VE}^{qqgg}} =- \frac{\img g^3 m}{32\pi ^{2}} \frac{z}{\omega_n} & e^{\eps \gamma_E}\mu^{2\eps}\Gamma(\eps)\int_0^1 \df \xi \Big(\frac{1}{\xi} \Big \vert \frac{\xi\omega_n}{\nu}\Big \vert^{-\eta} +\frac{1}{1-\xi}\Big \vert \frac{(1-\xi)\omega_n}{\nu}\Big \vert^{-\eta} \Big)
	\\
	&\times \left(m^2+\frac{k_\perp^2\xi(\xi-1)}{1-z}\right)^{-\eps} \frac{f^{abc}(2 \xi^2 z-2 \xi z+1)+\img d^{abc}(2 \xi -1)}{(\xi z -1)[(\xi -1)z +1]} \,. \nn
\end{align}
For the first time in this calculation, we  have an integrand with nontrivial $\xi$ dependence.
This adds extra complication since the integral cannot be evaluated in closed form for general $\eps$.
Importantly, the endpoint divergences are isolated from the rest of the integral which is regular for  $\xi \to 1$ and $\xi \to 0$.
Thus, we can isolate the endpoint divergences using
\begin{align} \label{eq:extractpoles}
	\cI=&\int_0^1 \df \xi \left(\frac{1}{\xi}+\frac{1}{1-\xi}\right)	f(\xi)
	\nn \\
	=&\int_0^1 \df \xi \Bigg\{\frac{1}{\xi}\Big[ \Big \vert \frac{\xi\omega_n}{\nu}\Big \vert^{-\eta} f(0)\underbrace{-f(0)+f(\xi)}_{\text{regular for } \xi \to 0}\Big] +\frac{1}{1-\xi}\Big[\Big \vert \frac{(1-\xi)\omega_n}{\nu}\Big \vert^{-\eta} f(1)\underbrace{-f(1)+f(\xi)}_{\text{regular for } \xi \to 1}\Big]
	\Bigg\},
\end{align}
where the subtractions $f(\xi)-f(0)$ and $f(\xi)-f(1)$ cancel in the respective divergent limit and contribute a finite result.
Extracting the poles as in \eq{extractpoles} allows us to keep the full $\eps$ dependence in the endpoint divergence terms.
For the finite contributions, it is sufficient to expand in $\eps$ before integrating. Our final result is given by
\begin{align}
	\cM_{d_{VE}^{qqgg}}
	=& \frac{\img g^3f^{abc}}{16 \pi^{2}} \frac{z}{\omega_n(1-z)} m \, e^{\eps \gamma_E}\mu^{2\eps}  g_\perp^{\mu \nu}\nb^\rho
	\Biggl[
	\frac{\log(1-z)}{\eps}
	-\frac{\Gamma(\eps)}{\eta}
	\left(\frac{\mu^2}{m^2}\right)^\eps \left|\frac{\nu}{\omega_n} \right|^\eta
	\Biggr]
	+\cM_{d_{VE}^{qqgg}} ^{\mathrm{fin.}},
\end{align}
where the first term in the sum is the $\ord{1/\eps}$ contribution of the $\eta$-finite result.
The remaining term $\cM_{d_{VE}^{qqgg}} ^{\mathrm{fin.}}$ is free of $\eps$ and  $\eta$ poles.
We give an explicit form of $\cM_{d_{VE}^{qqgg}} ^{\mathrm{fin.}}$  in the auxiliary material.
\subsubsection{\boldmath{$d_{VE}^{n}$} contribution }
Similar to the soft and $\nb$-collinear contributions, the $d_{VE}^{n}$ diagram (shown in \fig{LO1d1diasnb}) differs from the  leading-order diagram by a modified effective vertex. However, in this case it is not sufficient to simply modify the prefactor of our leading-order result:
In the full theory, the gluon can be emitted from either one of the quarks or the gluon. This leads to the modified Feynman rule~\eq{d1cFR} featuring two different sets of $\delta$ functions and different color structure, which leads to two distinct terms after resolving the $\delta$ functions
\begin{align}
	\cM_{d_{VE}^{n}}=& \int \df \xi \int\frac{\df \ell^+ \df^{d-2}\ell_\perp}{(2\pi)^{d-1}}\frac{g^3 m}{8 \pi} \frac{z}{1-z}\mu^{2\eps} (4\pi)^{-\eps} e^{\eps\gamma_{E}}C_{bbg}(\xi) \nb^\rho g_\perp^{\mu\nu}\nn\\
	& \left[\frac{f^{abc}- \img d^{abc}}{\left[\omega_n(\xi-1) \ell^+ +\ell_\perp^2-m^2 +\img 0\right] \left[\omega_n((\xi-1)z+1)\ell^+ +z (\ell_\perp^2-m^2 +\img 0)\right]}
	\right. \nn \\
	&\left.
	+\frac{f^{abc}+ \img d^{abc}}{\left[\omega_n\xi \ell^+ +\ell_\perp^2-m^2 +\img 0\right] \left[\omega_n(\xi z-1)\ell^+ +z (\ell_\perp^2-m^2 +\img 0)\right]}\right].
\end{align}
We will call the different terms the $\cM_{d_{VE}^{n}}^+$ and $\cM_{d_{VE}^{n}}^-$ contributions corresponding to the relative sign between $f^{abc}$ and $\img d^{abc}$.
In the next step, we do the $\ell^+$ integral by contours where we have to consider both contributions separately.

\begin{figure}
	\centering
	\begin{subfigure}{0.45\textwidth}
		\begin{tikzpicture}
			\draw [->] (-3.5,0) -- (3.5,0) node [anchor=south]  {$\xi$};
			\draw [|-|] (-2,0.8) -- (2,0.8) ;
			\draw (2.,-0.35) node[anchor=south] {$|$};
			\draw (2.,-0.35) node[anchor=north] {$1$};
			\draw (-0.,-0.35) node[anchor=south] {$|$};
			\draw (-0.,-0.3) node[anchor=north] {${0}$};
			\draw (-2,-0.35) node[anchor=south] {$|$};
			\draw (-2,-0.3) node[anchor=north] {$1-\frac{1}{z}$};
			\draw[red] (2.8,0.4) node[anchor=south] {$\times$};
			\draw[blue] (2.8,0.1) node[anchor=south] {$\times$};
			\draw[red] (-3.0,-0.1) node[anchor=north] {$\times$};
			\draw[blue] (-3.0,-0.4) node[anchor=north] {$\times$};
			\draw[red] (-1.,-0.1) node[anchor=north] {$\times$};
			\draw[blue] (-1.,0.3) node[anchor=south] {$\times$};
			\draw[red] (1.,-0.1) node[anchor=north] {$\times$};
			\draw[blue] (1.,0.3) node[anchor=south] {$\times$};
			\draw[red] (-4.,1.6) node[anchor=west] {$\omega_n(\xi-1)$};
			\draw(-2.4,1.5) node[anchor=west] {,};
			\draw[blue] (-2.,1.6) node[anchor=west] {$\omega_n[(\xi-1)+\frac{1}{z}]$};
		\end{tikzpicture}
		\subcaption{}
	\end{subfigure}
	\hfill
	\begin{subfigure}{0.45\textwidth}
		\begin{tikzpicture}
			\draw [->] (-3.5,0) -- (3.5,0) node [anchor=south]  {$\xi$};
			\draw [|-|] (-2,0.8) -- (2,0.8) ;
			\draw (2.,-0.35) node[anchor=south] {$|$};
			\draw (2.,-0.35) node[anchor=north] {$\frac{1}{z}$};
			\draw (-0.,-0.35) node[anchor=south] {$|$};
			\draw (-0.,-0.3) node[anchor=north] {${1}$};
			\draw (-2,-0.35) node[anchor=south] {$|$};
			\draw (-2,-0.3) node[anchor=north] {$0$};
			\draw[blue] (2.8,0.4) node[anchor=south] {$\times$};
			\draw[red] (2.8,0.1) node[anchor=south] {$\times$};
			\draw[blue] (-3.0,-0.1) node[anchor=north] {$\times$};
			\draw[red] (-3.0,-0.4) node[anchor=north] {$\times$};
			\draw[blue] (-1.,-0.1) node[anchor=north] {$\times$};
			\draw[red] (-1.,0.3) node[anchor=south] {$\times$};
			\draw[blue] (1.,-0.1) node[anchor=north] {$\times$};
			\draw[red] (1.,0.3) node[anchor=south] {$\times$};
			\draw[red] (-4.,1.6) node[anchor=west] {$\omega_n\xi$};
			\draw(-3.3,1.5) node[anchor=west] {,};
			\draw[blue] (-3,1.6) node[anchor=west] {$\omega_n(\xi-\frac{1}{z})$};
		\end{tikzpicture}
		\subcaption{}
	\end{subfigure}
	\caption{Pole analysis for the collinear $d_{VE}^{n}$ contribution. (a) shows the poles of $\cM_{d_{VE}^{n}}^-$ and (b) shows the poles of the $\cM_{d_{VE}^{n}}^+$ contribution. The black interval marks the region where the contour integral is non-zero.}
	\label{fig:poleanalysisd1}
\end{figure}
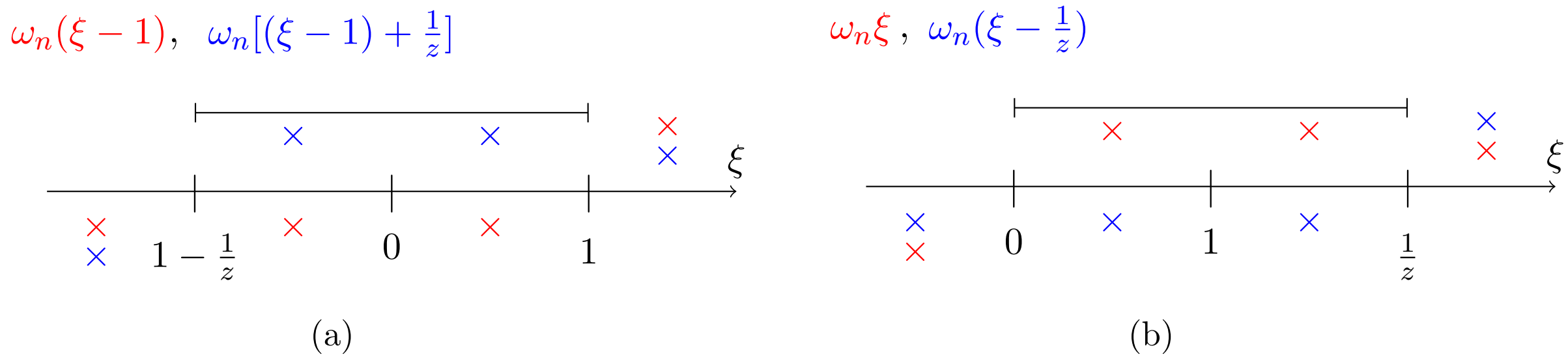
In \fig{poleanalysisd1}, the coefficients of the $\img 0$ are shown as a function of $\xi$.
The sign of these coefficients determines whether the corresponding pole in $\ell^+$ is in the upper or lower half-plane when doing the $\ell^+$ integral.
The contour integral only has support for specific values of $\xi$ which imposes a constraint on the $\xi$ integral.
It is convenient to define three different regions of $\xi$,
\begin{align}
	\cR^-= \left[1-\frac{1}{z},0\right], \qquad \cR^0=\left[0,1\right], \qquad \cR^+=\left[1,\frac{1}{z}\right],
\end{align}
with
\begin{align}
	\xi \in \cR^- \quad  \mathrm{for }\quad \cM_{d_{VE}^{n}}^-, \quad  \mathrm{and}\quad 	\xi \in \cR^+ \quad  \mathrm{for }\quad \cM_{d_{VE}^{n}}^+.
\end{align}
Including the central region, $\cM_{d_{VE}^{n}}^\pm$ thus has support for $\xi\in\cR^0\cup\cR^\pm$.
After doing the $\ell^+$ integral it is straightforward to evaluate the $\ell_\perp$ integral and only the endpoint-divergent $\xi$ integral remains.
Note that both contributions, $\cM_{d_{VE}^{n}}^+$ and $\cM_{d_{VE}^{n}}^-$, have support in the central region $\cR^{0}$ where the endpoint divergences appear in both integral limits.
Additionally, each term also gets a contribution from one of the outer regions, where the integrand diverges only at one endpoint, $\xi=0$ in $\cR^-$ and $\xi=1$ in $\cR^+$.

As before, we regulate the endpoint divergences with the absolute-value regulator and use \eq{extractpoles} to isolate the endpoint divergences.
After integrating, we can recombine the different contributions and arrive at our result for this diagram
\begin{align}
	\cM_{d_{VE}^{n}}=& -\frac{\img g^3f^{abc}}{16 \pi^{2} } \frac{\Gamma(\eps)}{\eta}\frac{z}{\omega_n(1-z)} m^{1-2\eps}e^{\eps \gamma_E}\mu^{2\eps}  g_\perp^{\mu \nu}\nb^\rho \nn \\
	&\quad\times
	\left[2\left|\frac{\nu}{\omega_n}\right|^\eta - \left(\frac{1}{z}-1\right)^{-\eta}\left|\frac{\nu}{\omega_n}\right|^\eta + \eta \log z  \right],
\end{align}
where the $d^{abc}$ terms have canceled between $\cM_{d_{VE}^{n}}^+$ and $\cM_{d_{VE}^{n}}^-$.
\subsubsection{Combining all \boldmath{$d_{VE}$} diagrams}
We have now calculated all diagrams in \fig{LO1d1dias} and can finally combine them
\begin{align}
	\cM_{d_{VE}}= &\cM_{d_{VE}^{\nb}} +\cM_{d_{VE}^{s}}+\cM_{d_{VE}^{qqgg}} +\cM_{d_{VE}^{n}}\nn  = \cO(\eta^0),
\end{align}
and, indeed, all endpoint divergences cancel in this diagram class.
\subsection{Loop emission diagrams}
\label{sec:LEsection}
\begin{figure}[h]
	\begin{subfigure}{0.45\textwidth}
		\centering
		\includegraphics[width=0.7\textwidth]{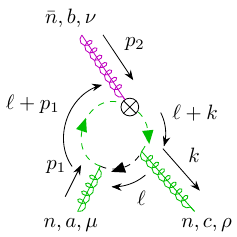}
		\caption{$d_{LE}^{1}$}
	\end{subfigure}
	\hfill
	\begin{subfigure}{0.45\textwidth}
		\centering
		\includegraphics[width=0.8\textwidth]{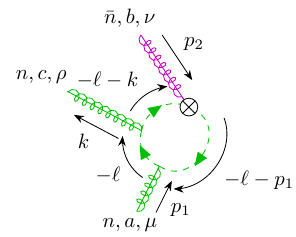}
		\caption{$d_{LE}^{2}$}
	\end{subfigure}
	\caption{$d_{LE}$-type diagrams: the gluon is emitted from the loop.}
	\label{fig:nontrivialLO1coll}
\end{figure}

This subclass of diagrams is the most complex one, since the emission from the loop introduces a third propagator and changes the momentum routing in the $\delta$-functions fixing the $\ell^-$ and $\xi$ integrals.
Using the choice of momenta in \fig{nontrivialLO1coll}, both diagrams have the same denominators which simplifies the calculation. As before, we start by doing the $\ell^-$ integral by resolving the $\delta$-functions.

\begin{figure}
	\centering
	\begin{subfigure}{0.45\textwidth}
		\begin{tikzpicture}
			\draw [->] (-3.5,0) -- (3.5,0) node [anchor=south]  {$\xi$};
			\draw [|-|] (-0.75,1.5) -- (2,1.5) ;
			\draw (2.,-0.35) node[anchor=south] {$|$};
			\draw (2.,-0.35) node[anchor=north] {$\frac{1}{z}$};
			\draw (-0.75,-0.35) node[anchor=south] {$|$};
			\draw (-0.75,-0.3) node[anchor=north] {${0}$};
			\draw (+0.75,-0.35) node[anchor=south] {$|$};
			\draw (+0.75,-0.3) node[anchor=north] {${1}$};
			\draw (-2,-0.35) node[anchor=south] {$|$};
			\draw (-2,-0.3) node[anchor=north] {$1-\frac{1}{z}$};
			\draw[Green] (2.8,-0.7) node[anchor=north] {$\times$};
			\draw[blue] (2.8,-0.4) node[anchor=north] {$\times$};
			\draw[red] (2.8,-0.1) node[anchor=north] {$\times$};
			\draw[red] (1.3,-0.1) node[anchor=north] {$\times$};
			\draw[Green] (1.3,-0.4) node[anchor=north] {$\times$};
			\draw[blue] (1.3,0.3) node[anchor=south] {$\times$};
			\draw[red] (0.,-0.1) node[anchor=north] {$\times$};
			\draw[Green] (0,0.6) node[anchor=south] {$\times$};
			\draw[blue] (0.,0.3) node[anchor=south] {$\times$};
			\draw[Green] (-1.3,0.9) node[anchor=south] {$\times$};
			\draw[blue] (-1.3,0.6) node[anchor=south] {$\times$};
			\draw[red] (-1.3,0.3) node[anchor=south] {$\times$};
			\draw[Green] (-3.0,0.9) node[anchor=south] {$\times$};
			\draw[blue] (-3.0,0.6) node[anchor=south] {$\times$};
			\draw[red] (-3.0,0.3) node[anchor=south] {$\times$};
			\draw[red] (-4.,1.8) node[anchor=west] {$-\omega_n\xi$};
			\draw(-3.,1.7) node[anchor=west] {,};
			\draw[blue] (-2.8,1.8) node[anchor=west] {$\omega_n(\frac{1}{z}-\xi)$};
			\draw(-1.2,1.7) node[anchor=west] {,};
			\draw[Green] (-1,1.8) node[anchor=west] {$\omega_n(1-\xi)$};
		\end{tikzpicture}
		\subcaption{}
	\end{subfigure}
	\hfill
	\begin{subfigure}{0.45\textwidth}
		\begin{tikzpicture}
			\draw [->] (-3.5,0) -- (3.5,0) node [anchor=south]  {$\xi$};
			\draw [|-|] (-2,0.8) -- (0.75,0.8) ;
			\draw (2.,-0.35) node[anchor=south] {$|$};
			\draw (2.,-0.35) node[anchor=north] {$\frac{1}{z}$};
			\draw (-0.75,-0.35) node[anchor=south] {$|$};
			\draw (-0.75,-0.3) node[anchor=north] {${0}$};
			\draw (+0.75,-0.35) node[anchor=south] {$|$};
			\draw (+0.75,-0.3) node[anchor=north] {${1}$};
			\draw (-2,-0.35) node[anchor=south] {$|$};
			\draw (-2,-0.3) node[anchor=north] {$1-\frac{1}{z}$};
			\draw[red] (2.8,0.4) node[anchor=south] {$\times$};
			\draw[Green] (2.8,0.7) node[anchor=south] {$\times$};
			\draw[blue] (2.8,0.1) node[anchor=south] {$\times$};
			\draw[red] (1.3,0.4) node[anchor=south] {$\times$};
			\draw[Green] (1.3,0.7) node[anchor=south] {$\times$};
			\draw[blue] (1.3,0.1) node[anchor=south] {$\times$};
			\draw[red] (0.,-0.1) node[anchor=north] {$\times$};
			\draw[Green] (0,0.7) node[anchor=south] {$\times$};
			\draw[blue] (0.,0.4) node[anchor=south] {$\times$};
			\draw[red] (-1.3,-0.1) node[anchor=north] {$\times$};
			\draw[Green] (-1.3,-0.4) node[anchor=north] {$\times$};
			\draw[blue] (-1.3,0.3) node[anchor=south] {$\times$};
			\draw[red] (-3.0,-0.1) node[anchor=north] {$\times$};
			\draw[Green] (-3.0,-0.7) node[anchor=north] {$\times$};
			\draw[blue] (-3.0,-0.4) node[anchor=north] {$\times$};
			\draw[red] (-4.,1.8) node[anchor=west] {$\omega_n(\xi-1)$};
			\draw(-2.4,1.7) node[anchor=west] {,};
			\draw[blue] (-2.,1.8) node[anchor=west] {$\omega_n[(\xi-1)+\frac{1}{z}]$};
			\draw(0.5,1.7) node[anchor=west] {,};
			\draw[Green] (0.8,1.8) node[anchor=west] {$\omega_n\xi$};
		\end{tikzpicture}
		\subcaption{}
	\end{subfigure}
	\caption{Pole analysis for the $d_{LE}$ contributions. (a) shows the poles in the complex $\ell^+$-plane of $\cM_{d_{LE}^{1}}$ and (b) shows the poles of $\cM_{d_{LE}^{2}}$. The black interval marks the region where the contour integral is non-zero.}
	\label{fig:poleanalysisd3}
\end{figure}
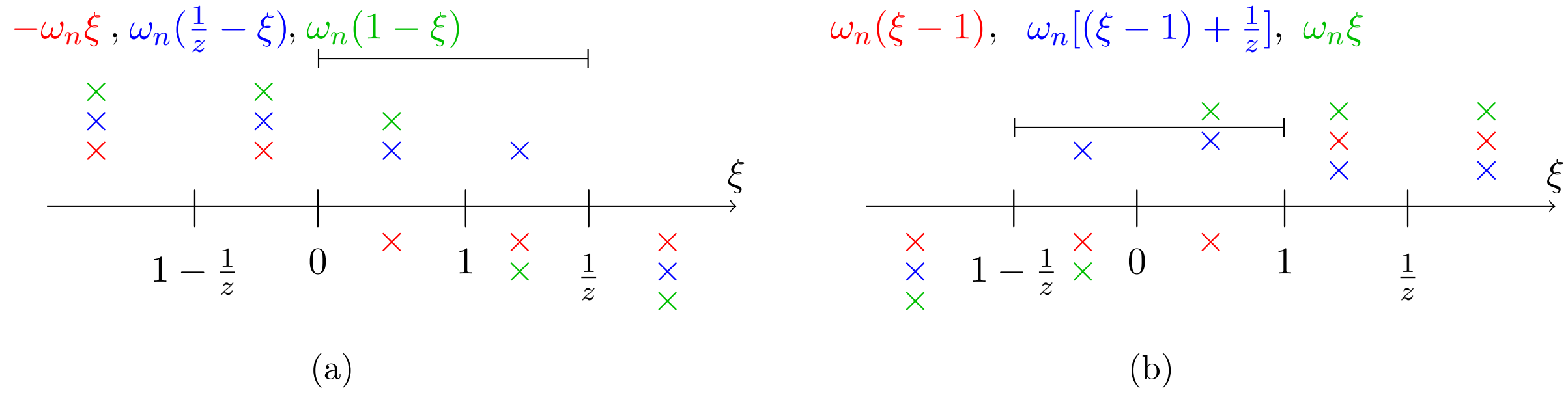

Next, we do the $\ell^+$ integral by contours. Note that as we have three propagators in the loop, there will also be three poles in the complex plane.
The coefficients of the $\img0$ in the denominators determine whether the pole will be in the upper or lower half-plane.
These coefficients are shown in \fig{poleanalysisd3} as a function of $\xi$.
As for $d_{VE}^{n}$, we find non-zero contributions for $\xi\in\cR^0\cup\cR^\pm$.
However, the third pole switches sign when going from the central region $\cR^0$ to one of the outer regions $\cR^\pm$.
Therefore all contributions must be considered separately and can only be combined after integrating over $\xi$.
In the next step, we evaluate the $\ell_\perp$ integral where we need Feynman parameters to combine the denominators.
Now, the only remaining integral is the divergent $\xi$ integral. Once again, we utilize the formula in \eq{extractpoles} to isolate the endpoint divergences. This approach remains valid even when the integral diverges only in one of the limits, as is the case for the $\cR^\pm$ regions.

For this class of diagrams, the function $f(\xi)$ is rather complex, preventing a straightforward integration of its finite part over $\xi$. However, since we only require the full $\eps$-dependence of the endpoint-divergent contribution, we can expand the $\eta$-finite piece in $\eps$ before integrating. This allows us to extract the $1/\eps$ term, which is free of endpoint divergences. The $\eps$- and $\eta$-finite contributions are given in the auxiliary files.

For the central region $\cR^0$, we find
\begingroup
\allowdisplaybreaks
\begin{align}
	\cM_{d_{LE}^{1}}^0=& \frac{\img g^3m\,z\,(f^{abc}+ \img d^{abc})}{32\pi ^{2}}
	\Bigg\{-\frac{\log (1-z)}{\eps\,\omega_n(1-z)}\,g_\perp^{\mu\nu}\nb^\rho +{\left|\frac{\nu}{\omega_n}\right|^\eta} e^{\eps \gamma_E} \mu^{2 \eps}\frac{\Gamma(\eps)}{\eta} \nn \\
	&\hspace{1.3cm} \times\Bigg[  \frac{m^{-2 \eps}}{\omega_n(1-z)}g_\perp^{\mu\nu}\nb^\rho +
	 4^{1+\eps}\,\eps\,
	\Big(k_\perp^\mu  g_\perp^{\nu \rho}- \frac{k_\perp^\nu  g_\perp^{\mu \rho}}{z }+\frac{k_\perp^\mu k_\perp^\nu \nb^{\rho}}{\omega_n}
	\Big)
	\nn \\
	&\hspace{2cm}\times (4 m^2- k_\perp^2)^{-1-\eps}{}_2F_1\Big(\frac{1}{2},\eps+1, \frac{3}{2},\frac{k_\perp^2}{k_\perp^2-4m^2}\Big)\Bigg]\Bigg\}
	+\cM_{d_{LE}^{1}}^{0, \mathrm{fin.}}.
\end{align}
\endgroup
The result for $\cM_{d_{LE}^{2}}^0$ follows from $\cM_{d_{LE}^{1}}^0$ by replacing $d^{abc}\to -d^{abc}$, including the finite part.
In the sum the $d^{abc}$ terms will cancel, leaving a pure $f^{abc}$ part as final result as expected.
For the outer regions, $\cR^+$ and $\cR^-$, we find
\begin{align}
	\cM_{d_{LE}^{1}}^+=& -\frac{\img g^3m\,z\,(f^{abc}+ \img d^{abc})}{32\pi ^{2}}
	\Bigg\{-\frac{\log z}{\eps\,\omega_n(1-z)}\,g_\perp^{\mu\nu}\nb^\rho +\Big(\frac{1}{z}-1\Big)^{-\eta}{\left|\frac{\nu}{\omega_n}\right|^\eta} e^{\eps \gamma_E} \mu^{2 \eps}\nn \\
	&\hspace{0.5cm} \times\frac{\Gamma(\eps)}{\eta} \Bigg[  \frac{m^{-2 \eps}}{\omega_n(1-z)}g_\perp^{\mu\nu}\nb^\rho +
	4^{1+\eps}\,\eps\,
	\Big(k_\perp^\mu  g_\perp^{\nu \rho}- \frac{k_\perp^\nu  g_\perp^{\mu \rho}}{z }+\frac{k_\perp^\mu k_\perp^\nu \nb^{\rho}}{\omega_n}
	\Big)
	 \nn \\
	&\hspace{2cm}\times (4 m^2- k_\perp^2)^{-1-\eps}{}_2F_1\Big(\frac{1}{2},\eps+1, \frac{3}{2},\frac{k_\perp^2}{k_\perp^2-4m^2}\Big)\Bigg]\Bigg\}
	+\cM_{d_{LE}^{1}}^{+, \mathrm{fin.}}.
\end{align}
The contribution $\cM_{d_{LE}^{2}}^-$ follows again from $\cM_{d_{LE}^{1}}^+$ by $d^{abc}\to -d^{abc}$.
We can now combine all contributions and find that this class of diagrams is free of endpoint divergences
\begin{align}
	\cM_{d_{LE}}=\cM_{d_{LE}^{1}}^0+\cM_{d_{LE}^{2}}^0+\cM_{d_{LE}^{1}}^++\cM_{d_{LE}^{2}}^-= \Ord{\eta^0}.
\end{align}

\subsection{Hard contributions}
\label{sec:hardcoll}
\begin{figure}[h]
	\centering
	\begin{minipage}{0.45\textwidth}
		\begin{align*}
			C(m_H) \times \raisebox{-0.45\height}{\includegraphics{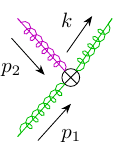}}
		\end{align*}
		\subcaption{{$d_{VE}^{h}$}: emission from the vertex.}
	\end{minipage}
	\begin{minipage}{0.45\textwidth}
		\begin{align*}
			C(m_H) \times \raisebox{-0.45\height}{\includegraphics{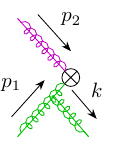}}
		\end{align*}
		\subcaption{{$d_{ggg}^{h}$}: emission via the triple-gluon vertex.}
	\end{minipage}
	\caption{Hard contributions with collinear gluon emission.}
	\label{fig:LO1hard}
\end{figure}
There are two hard contributions that can emit a collinear gluon shown in \fig{LO1hard}.
The hard contributions can be obtained by using the leading-order result in \app{ybytLO0-hard}.
To obtain our result for $d_{VE}^{h}$, we can simply multiply the Wilson coefficient $C(m_H)$ with the gluon field strength at $\ord g$~\cite{Berger:2010xi}  which we already used for the calculation of the vertex emission diagram $d_{VE}^{n}$.
This  yields
\begin{align}
	\cM_{d_{VE}^{h}}=& \img 2^{4 \eps -3} \pi^{\eps-\frac{1}{2}}  \frac{ g^3  f^{abc} g_\perp^{\mu\nu}\nbar^\rho mz}{(z-1)\omega_n}\frac{\csc (\pi \eps)}{\eps \Gamma(\frac{1}{2}-\eps)}\Big(\frac{\mu^2 e^{\gamma_E}}{4\pi}\Big)^{\eps}(-m_H^2 - \img 0)^{ -\eps} \nn \\
	&\hspace{3.5cm}\times
	\left[\frac{2\eps^2 }{2 \eps ^2- 3 \eps+1}-1\right].
\end{align}
For the remaining diagram $d_{ggg}^{h}$, we contract the leading-order result with the triple-gluon vertex of QCD.
As in \sec{gggsection}, we need to take into account that the LO result was derived for on-shell gluons and shift the momenta according to the \loone~kinematics. We find
\begin{align}
	\cM_{d_{ggg}^{h}}=&{\img\, 2^{4 \eps -4}} \pi^{\eps-\frac{1}{2}}  \frac{ g^3  f^{abc}  m}{z k_T^2\omega_n}\frac{\csc (\pi \eps)}{\eps \Gamma(\frac{1}{2}-\eps)}\Big(\frac{\mu^2 e^{\gamma_E}}{4\pi}\Big)^{\eps}(-m_H^2 - \img 0)^{ -\eps}
	\left[\frac{2\eps^2 }{{2 \eps ^2- 3 \eps+1}}-1\right]\nn \\
	&\times
	\left[2g_\perp^{\mu \nu}k_\perp^\rho \omega_n z(z-1)+g_\perp^{\mu \nu}\nb^\rho k_\perp^2z^2
	+g_\perp^{\mu \nu}n^\rho\omega_n^2(1-z^2) +4g_\perp^{\nu\rho}k_\perp^\mu \omega_n z(1-z)
	\right.\nn \\
	&+ \left. 4 g_\perp^{ \mu\rho}k_\perp^\nu\omega_n(z-1)+4 k_\perp^\mu k_\perp^\nu \nb^\rho z(1-z)\right].
\end{align}

\subsection{Glauber contributions}
\label{sec:glaubercoll}
\begin{figure}
	\centering
	\begin{minipage}{0.45\textwidth}
		\centering
		\includegraphics[width=0.9\textwidth]{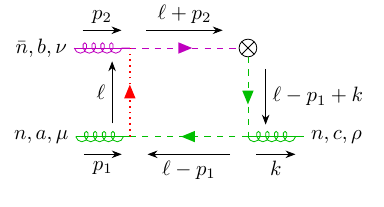}
		\subcaption{$d_G^{1a}$}
	\end{minipage}
	\hspace{5ex}
	\begin{minipage}{0.45\textwidth}
		\centering
		\includegraphics[width=0.9\textwidth]{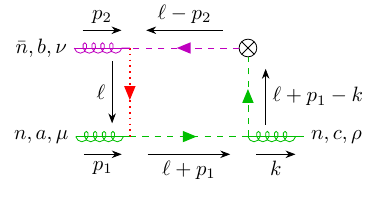}
		\subcaption{$d_G^{1b}$}
	\end{minipage}\\
	\begin{minipage}{0.45\textwidth}
		\centering
		\includegraphics[width=0.9\textwidth]{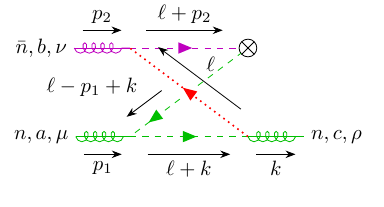}
		\subcaption{$d_G^{2a}$}
	\end{minipage}
	\hspace{5ex}
	\begin{minipage}{0.45\textwidth}
		\centering
		\includegraphics[width=0.9\textwidth]{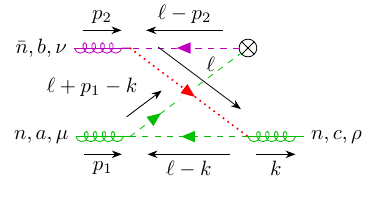}
		\subcaption{$d_G^{2b}$}
	\end{minipage}
	\caption{Glauber contributions with collinear emission. Colors as in \fig{modesLO0}; Glauber quarks are shown in red. The Higgs boson is produced at the hard vertex $\otimes$ from an $n$-collinear and an $\nb$-collinear quark, as at LO$_0$.}
	\label{fig:Galuberdias}
\end{figure}

Finally, a collinear gluon can also be emitted in diagrams where one of the quark propagators in the loop has Glauber scaling, $\ell \sim (\lambda^2,\lambda^2,\lambda)$, see \fig{Galuberdias}.
At LO$_0$, this region coincides with the Glauber zero-bin of the soft contribution and does not have to be added separately.
With a collinear emission this is no longer the case, and the Glauber diagrams give a finite correction to the amplitude.
For the vertices of the Glauber quark we use the Lagrangian in \eq{Glauber_Lagrangian_quark} whose Feynman rules are given in \refcite{Moult:2017xpp}, and the Higgs boson is produced at the hard vertex $\otimes$, as at LO$_0$.
We begin by evaluating the traces and expanding the Glauber momentum in $\lambda$.
As before, we perform the $\ell^+$ integral by contours. For $d_G^{1a,b}$, all poles lie on the same side of the real axis, so that the integral vanishes,
\begin{align}
	\cM_G^{1a} =0, \qquad \cM_G^{1b} =0\,.
\end{align}

For $d_G^{2a,b}$, the two $n$-collinear propagators have poles on opposite sides of the real axis, and we use the results of \refcite{Rothstein:2016bsq} for the remaining $\ell^+$ and $\ell^-$ integrals. We find
\begin{align}
	\cM^{2a}_G =&-\frac{\img g^3 m \left(d^{abc}+\img f^{abc}\right) }{8 \omega_n (1-z)}\!\int \!\frac{\df^{d-2} \ell_\perp}{(2 \pi)^{d-2}}
	\frac{1}{ \left(\ell_\perp^2-m^2+ \img 0\right)
		\left(\ell_\perp^2+k_\perp^2+2 k_\perp \ell_\perp-m^2 + \img 0\right)} \nn \\
	& \hspace{1.1cm} \times
		\Big\{(1-z)\Big[
		\omega_n \big( g_\perp^{\mu\rho} k_\perp^\nu - g_\perp^{\mu\nu} k_\perp^\rho
		- 2 g_\perp^{\mu\nu} \ell_\perp^\rho \big)
		- 2z \big( k_\perp^\mu + \ell_\perp^\mu \big) k_\perp^\nu \nb^\rho
		\nn \\
		&\hspace{1.7cm}
		+ \omega_n \big( (1-2z)\, k_\perp^\mu + 2(1-z)\, \ell_\perp^\mu \big) g_\perp^{\nu\rho}
		\Big]
		- z\, k_\perp \!\cdot\! \left( k_\perp + 2 \ell_\perp \right) g_\perp^{\mu\nu} \nb^\rho\Big \}\,.
\end{align}
For the remaining integral, we introduce a Feynman parameter to combine the denominators.
$\cM^{2b}_G$ is computed analogously.
Their sum is
\begin{align}
	\cM_G=\cM^{2a}_G+\cM^{2b}_G \;=&\frac{g^3 m f^{abc}  }{4 \pi \omega_{n}}
	4^{\eps} e^{\eps \gamma_E}  \mu ^{2\eps}
	\Gamma (\eps+1)
		{}_2F_1\left(\frac{1}{2},\eps+1;\frac{3}{2};\frac{k_T^2}{k_T^2+4 m^2}\right)  \nn \\
	&\times \,
	\left(k_T^2+4 m^2\right)^{-\eps-1}
 \left[z k^{\mu }_\perp \left( \omega_ng^{\nu \rho
	}_\perp + k_\perp^\nu \nb^\rho\right)-\omega_{n} k^{\nu }_\perp g^{\mu \rho }_{\perp}\right]\,.
\end{align}
This expression is free of $\eps$ poles. Expanding in $\eps$ yields
\begin{align}
	\cM_G=\frac{g^3 m f^{abc}}{4 \pi\,\omega_n k_T^2 \sqrt{1+\tfrac{4 m^2}{k_T^2}}} \tanh ^{-1}\Bigg(\frac{1}{\sqrt{1+\tfrac{4 m^2}{k_T^2}}}\Bigg)   \left[z k^{\mu }_\perp \left( \omega_ng^{\nu \rho
	}_\perp + k_\perp^\nu \nb^\rho\right)-\omega_{n} k^{\nu }_\perp g^{\mu \rho }_{\perp}\right],
\end{align}
which contributes only to the imaginary part of the full collinear emission amplitude.

We also computed the zero-bin subtractions of all collinear-emission diagrams and find that each of them is either scaleless, power suppressed, or has a vanishing contour integral.
\subsection{Combination and cross checks}
\label{sec:endpointdiv}
After the rather technical calculation of the collinear emission diagrams in the previous sections, we now take a step back to reflect on the conclusions we can draw from the cancellation of the endpoint divergences.
For the form factor, this cancellation is relatively straightforward since the problem involves only two scales, $m$ and $m_H$.
However, when considering the $q_T$ spectrum, an additional scale is introduced, which affects the coefficient functions of the endpoint divergences. In general, these coefficients can be nontrivial functions of $m/k_T$.
Here, we investigate how the additional $q_T$ dependence influences the structure of the endpoint divergences. To do so, we express our diagrams in the factorized form already introduced in our discussion of factorization theorems in \sec{factorization_theorems}.
The collinear-emission diagrams with soft and collinear loops can be categorized into two distinct classes.
In the first class, the gluon is emitted from a subleading-power collinear amplitude:
\begin{align} \label{eq:coll-LO1}
	\left(\raisebox{-0.45\height}{\includegraphics[height=2.5cm]{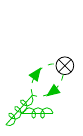}}+
	\raisebox{-0.45\height}{\includegraphics[height=2.5cm]{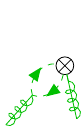}} +
	\raisebox{-0.45\height}{\includegraphics[height=2.5cm]{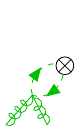}} +
	\raisebox{-0.45\height}{\includegraphics[height=2.5cm]{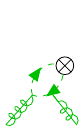}} +
	\raisebox{-0.45\height}{\includegraphics[height=2.5cm]{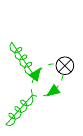}} \right)
	\times
	\hspace{-2ex}
	\raisebox{-0.45\height}{\includegraphics[height=2.5cm]{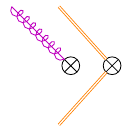}} = \cO\left(\frac{1}{\eta}\right).
\end{align}
For this class of diagrams alone, the endpoint divergences remain uncanceled.
In the second class of collinear emission diagrams, the gluon is emitted from a leading-power collinear vertex or Wilson line. Like the previous class, these diagrams also exhibit endpoint divergences:
\begin{align} \label{eq:coll-LO0}
	\left(	\raisebox{-0.45\height}{\includegraphics[height=2.5cm]{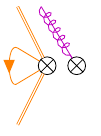}}
	+
	\raisebox{-0.45\height}{\includegraphics[height=2.5cm]{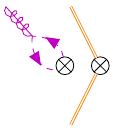}}\right)
	\times
	\left(\raisebox{-0.45\height}{\includegraphics[height=2.5cm]{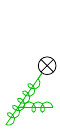}}
	+
	\raisebox{-0.45\height}{\includegraphics[height=2.5cm]{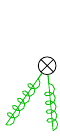}}\right) = \cO\left(\frac{1}{\eta}\right)
\end{align}
Only the sum of both classes is free of such divergences.
 This is consistent with the cancellation within the diagram classes $d_{ggg}$, $d_{VE}$ and $d_{LE}$ found above, since each of these classes contains diagrams of both types.
Remarkably, the endpoint divergences from the nontrivial subleading-power diagrams cancel against those from leading-power emissions multiplying the LO$_0$ next-to-leading-power contributions.
This is a highly nontrivial result:
although the additional scale allows the coefficients of the endpoint divergences to depend on $m/k_T$, they are completely fixed by the LO$_0$ contributions times a leading-power emission.
The endpoint-divergent part of the amplitude therefore factorizes into a mass-dependent and a $k_T$-dependent factor, while the finite remainder retains a nontrivial dependence on $m/k_T$, see \sec{logstructure}.
For a physical polarization of the emitted gluon with $\nb\cdot\eps(k)=0$, the vertex-emission class does not contribute and $\cM_{d_{LE}}$ is free of poles, so that the endpoint divergences then reside entirely in the $d_{ggg}$ class; the cancellation between $d_{VE}$ and $d_{LE}$ discussed above concerns the $g_\perp^{\mu\nu}\nb^\rho$ structure, which is fixed by the Ward identity.
Up to now, we have only combined contributions with soft and collinear scaling of the loop momenta. The endpoint divergences cancel between these contributions as expected, but the sum of soft and collinear contributions will still have $\eps$ poles and depend on the scale $\mu$.
Both will only cancel once we combine the soft, collinear and hard contributions.
Doing so, we find that the $\eps$ poles cancel separately in each of the three classes
\begin{align}
	\cM_{d_{ggg}}=&
	\raisebox{-0.45\height}{\includegraphics[height=3cm]{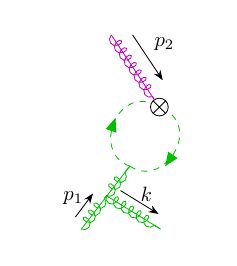}}
	+\raisebox{-0.45\height}{\includegraphics[height=3cm]{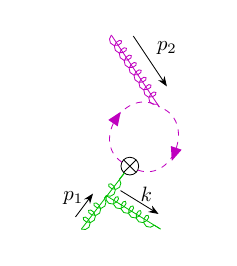}}
	+\raisebox{-0.45\height}{\includegraphics[height=3cm]{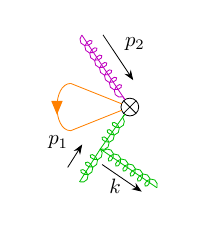}}
	+C(m_H) \times \raisebox{-0.45\height}{\includegraphics[height=2cm]{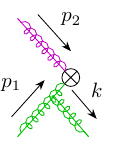}}
		\nn \\
	=& \cO(\eta^0,\eps^0)
	\label{eq:d2_cancellation}
	\\
	\cM_{d_{VE}}=&
	\raisebox{-0.45\height}{\includegraphics[height=3cm]{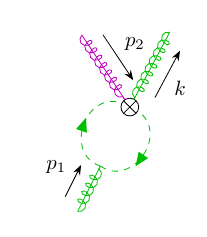}}
	+\raisebox{-0.45\height}{\includegraphics[height=3cm]{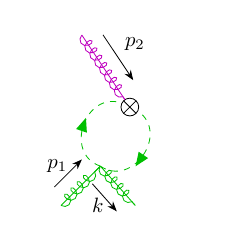}}
	+
	\raisebox{-0.45\height}{\includegraphics[height=3cm]{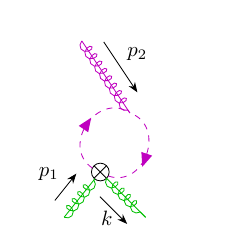}}\nn \\
	&
	+\raisebox{-0.45\height}{\includegraphics[height=3cm]{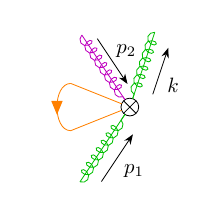}}
	+C(m_H) \times \raisebox{-0.45\height}{\includegraphics[height=2cm]{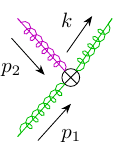}}
	= \cO(\eta^0,\eps^0), \\
	\cM_{d_{LE}}=&
	\raisebox{-0.45\height}{\includegraphics[height=3cm]{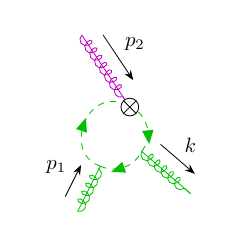}}
	+\raisebox{-0.45\height}{\includegraphics[height=3cm]{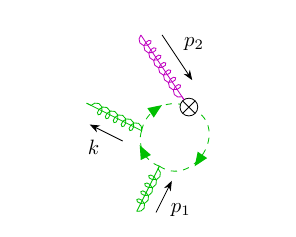}}
	=\cO(\eta^0,\eps^0).
	\label{eq:d3_cancellation}
\end{align}
Additionally, we find that the scale dependence cancels in the class $\cM_{d_{ggg}}$ and for the sum $\cM_{d_{VE}}+\cM_{d_{LE}}$.
This is expected and provides a strong cross-check for our results as the sum of all contributions has to agree with the full-theory result in~\refcite{Baur:1989cm} which is finite and has no scale dependence.

As a further check, we verified the Ward identity for the emitted gluon, $\eps_\mu(p_1)\,\eps_\nu(p_2)\,k_\rho\\\cM^{\mu\nu\rho}=0$, i.e.\ the amplitude vanishes when the polarization vector of the emitted gluon is replaced by its momentum, while the incoming gluons carry physical, transverse polarizations.
In our calculation, this is implemented by keeping only the transverse components in $\mu$ and $\nu$.
The Ward identity is fulfilled separately for the hard, soft and $\nb$-collinear contributions and for the Glauber contribution, and for the sum of all diagrams with an $n$-collinear loop, which contribute to the same operator.
For the latter, it is essential to include the finite parts, which we did not give explicitly in \secs{VEsection}{LEsection}.
The individual identities are given in \app{wardcoll}.

We have now computed all the contributions featuring a collinear emission.
The endpoint divergences cancel between soft and collinear contributions as required.
The remaining $\eps$ poles and the associated $\mu$ dependence cancel once the hard contributions are included.
We now turn to the diagrams in which the emitted gluon has soft
scaling, which we treat in \sec{softemission}, before comparing both
contributions to the full QCD result in \sec{FO}.
\section{Soft emission diagrams}
\label{sec:softemission}

\begin{figure}[htbp]
\centering
	\begin{subfigure}[b]{0.25\textwidth}
		\centering
		\raisebox{-2em}{
		\hspace{-2em}
		\includegraphics[width=0.75\textwidth]{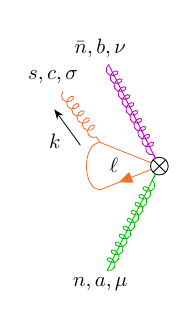}}
		\caption{Loop emission.}
		\label{fig:LO1softloop}
	\end{subfigure}
	\hspace{5em}
	\begin{subfigure}[b]{0.25\textwidth}
		\centering
		\raisebox{1em}{
		\includegraphics[width=0.8\textwidth]{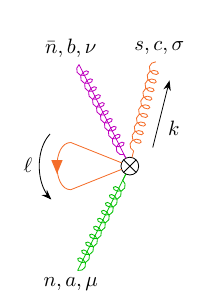}}
		\caption{Vertex emission. }
		\label{fig:LO1softvertex}
	\end{subfigure}
	\newline
	\begin{subfigure}{0.5\textwidth}
	\raisebox{1em}{
	\hspace{-8em}\raisebox{4em}{$\Bigg( $} \hspace{-1em}
		\raisebox{-1em}{\begin{subfigure}[b]{0.22\textwidth}
		        	\centering
        		\includegraphics[width=\textwidth]{figs/diagrams/LO0/hardLO0-womom-color.pdf}
		\end{subfigure}}
	\raisebox{4em}{$+$}
		\begin{subfigure}[t]{0.22\textwidth}
		        	\centering
        		\includegraphics[width=\textwidth]{figs/diagrams/LO0/collLO0-womom-color.pdf}
		\end{subfigure}
	\raisebox{4em}{$+$}
		\begin{subfigure}[t]{0.22\textwidth}
		        	\centering
        		\includegraphics[width=\textwidth]{figs/diagrams/LO0/antiLO0-womom-color.pdf}
		\end{subfigure}
	\raisebox{4em} {$\Bigg) \times \, \dfrac{1}{2} \, \Bigg ($} \hspace{-2.5em}
        	\begin{subfigure}[b]{0.4\textwidth}
        		\centering
        		\includegraphics[width=\textwidth]{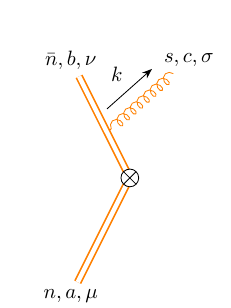}
        	\end{subfigure}
        	\hspace{-1em}\raisebox{4em}{$+$}
		\hspace{-2em}
        	\begin{subfigure}[b]{0.4\textwidth}
        		\centering
        		\includegraphics[width=\textwidth]{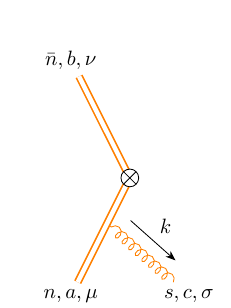}
        	\end{subfigure}
        	\hspace{-2em} \raisebox{4em}{$\Bigg)$} \hspace{-2em}}
	\caption{Wilson line emission.}
	\label{fig:LO1softwilson}
	\end{subfigure}
	\hspace{-2em}
\caption{All diagrams where the emission $k$ has soft scaling. All particles with $n$ ($\nb$)-collinear scaling are drawn in green (magenta), whereas the soft particles are drawn in orange.
In the two nontrivial diagrams, the soft gluon is emitted by the soft quark loop (a) or the vertex (b), respectively.
The diagrams where the soft gluon is emitted by Wilson lines are factorized as in (c). }
\label{fig:LO1diassoft}
\end{figure}

In this section, we calculate the sum of all diagrams with a soft emission as shown in \fig{LO1diassoft},
and show that it is free of endpoint and UV divergences.
To distinguish soft from collinear emissions, we denote the Lorentz index of the soft gluon by $\sigma$ instead of $\rho$.

A soft gluon emission can only change the diagrammatic structure of the LO$_{0}$ soft loop diagram in \fig{modesLO0},
whereas for both the hard vertex and collinear loop diagrams it couples only to the total color charge and the light-cone directions of the external legs.
Therefore, we have two nontrivial soft LO$_{1}$ diagrams where the emission is from the loop and from the vertex, respectively,
and a class of factorized diagrams where the emission is described by Wilson lines.
We first discuss the structure of each (class of) diagram, then prove the finiteness of the sum and give the final result.

\subsection{Soft loop emission}
\label{sec:softloop}

The diagram with soft gluon emission from the soft loop is evaluated as
\begin{align}\label{eq:softloop_diagram}
\mathcal{M}_{s, L}^{\mloone} = \hspace{-2em}
\raisebox{-0.45\height}{
   \includegraphics[scale=0.9]{figs/diagrams/LO1/NLP-emission/softLO1-softloop.pdf}
}
&= \frac{\img g^{3} m }{4} \Bigg\{  d^{abc} \cdot  \bigg[ 2 g_{\perp}^{\mu\nu}\Big( \cI_{L+}^{1\sigma} + \cI_{L+}^{2\sigma}\Big)
+ S_{-}^{\mu\nu\sigma} \Big(\cI_{L-}^{1} - \cI_{L-}^{2} \Big) \bigg]
\nn \\[-4em]
&\qquad
+  \img f^{abc} \cdot \bigg[ 2 g_{\perp}^{\mu\nu}\Big( \cI_{L+}^{1\sigma} - \cI_{L+}^{2\sigma}\Big)
+ S_{-}^{\mu\nu\sigma} \Big(\cI_{L-}^{1} + \cI_{L-}^{2} \Big) \bigg]
\Bigg\}
\,.\end{align}

We have defined the Lorentz structure
\begin{align}
S_{-}^{\mu\nu\sigma}
&\equiv
2 k^{\mu}  g_{\perp}^{\nu\sigma}
- 2 k^{\nu} g_{\perp}^{\mu\sigma}
\nn \\
& \quad
- k^{+} \Big( \nbar^{\sigma} (n^{\mu}\nbar^{\nu} - g^{\mu\nu} ) + \nbar^{[\mu}g^{\nu] \sigma}\Big)
- k^{-} \Big(n^{\sigma} ( g^{\mu\nu} - n^{\mu}\nbar^{\nu} ) + n^{[\mu}g^{\nu] \sigma} \Big)
\nn \\
&
\stackrel{\perp\perp}{\longrightarrow}
- 2 \Big( g_{\perp}^{\mu\sigma}  k_{\perp}^{\nu}   - g_{\perp}^{\nu\sigma}  k_{\perp}^{\mu}  \Big)
- k^{+} k^{-} g_{\perp}^{\mu\nu} \bigg ( \frac{n^{\sigma}}{k^{+}} - \frac{\nbar^{\sigma}}{k^{-}}   \bigg)
\label{eq:loop_lorentz_minus}
\,, \end{align}
where in the last line we keep only $\perp$ polarizations for the external gluons $\mu,\nu$.
With this, the two Dirac traces relevant for this diagram can be expressed as
\begin{align}
\Tr \big[ \gamma^{\mu} \Sl{n} \Sl{\nbar} \gamma^{\nu} (\Sl \ell - \Sl k + m) \gamma^{\sigma} (\Sl \ell + m)\big]
&= 4 m \big[ 2g_{\perp}^{\mu\nu} (2\ell^{\sigma} - k^{\sigma}) + S_{-}^{\mu\nu\sigma} \big]
 \,,\nn \\
\Tr \big[ \gamma^{\nu} \Sl{\nbar} \Sl{n} \gamma^{\mu} (\Sl \ell - \Sl k + m) \gamma^{\sigma} (\Sl \ell + m)\big]
& = 4 m \big[ 2g_{\perp}^{\mu\nu} (2\ell^{\sigma} - k^{\sigma}) - S_{-}^{\mu\nu\sigma} \big]
\,. \end{align}

The four (unregulated) loop integrals in \eq{softloop_diagram} are defined as
\begin{align}
\cI_{L-}^{1} = &\int \frac{\df^{d} \ell}{(2\pi)^{d}} \frac{1}{ [(\ell-k)^{2} - m^{2} + \img 0] [\ell^{2} - m^{2} + \img 0] } \,
\frac{1}{(\ell^{+} - \img 0 ) (\ell^{-}-k^{-} + \img 0)}
\,,
\label{eq:softloopint_m1}
\\
\cI_{L-}^{2} =&\int \frac{\df^{d} \ell}{(2\pi)^{d}} \frac{1}{ [(\ell-k)^{2} - m^{2} + \img 0] [\ell^{2} - m^{2} + \img 0] } \,
\frac{1}{ (\ell^{-} -\img 0) (\ell^{+} - k^{+} +\img 0)}
\,,
\label{eq:softloopint_m2}
\\
\cI_{L+}^{1\sigma} = &\int \frac{\df^{d} \ell}{(2\pi)^{d}} \frac{2\ell^{\sigma} - k^{\sigma}}{ [(\ell-k)^{2} - m^{2} + \img 0] [\ell^{2} - m^{2} + \img 0] } \,
\frac{1}{(\ell^{+} - \img 0 ) (\ell^{-}-k^{-} + \img 0)}
\,,
\label{eq:softloopint_p1}
\\
\cI_{L+}^{2\sigma} =&\int \frac{\df^{d} \ell}{(2\pi)^{d}} \frac{2\ell^{\sigma} - k^{\sigma}}{ [(\ell-k)^{2} - m^{2} + \img 0] [\ell^{2} - m^{2} + \img 0] } \,
\frac{1}{ (\ell^{-} -\img 0) (\ell^{+} - k^{+} +\img 0)}
\,.
\label{eq:softloopint_p2}
\end{align}

We can compute the first two scalar integrals analytically by performing the $\ell^{+}$ integral by contours,
combine the remaining denominators using Feynman parameters before performing the standard $\ell_{\perp}$ integral,
and finally perform the $\ell^{-}$ integral with a clever change of variable.
The final result is a finite integral over the Feynman parameter $x$ from the second step,
\begin{align}
\cI_{L-}^{1} = \cI_{L-}^{2} = \frac{- \img}{16 \pi^{2} k_{T}^{2}}
\int_{0}^{1} \df x \frac{\log \big[ (1-x) x / A   \big] + \img \pi}{A + x(1-x)}
\end{align}
where $A =m^{2}/ k_{T}^{2}$.
The Lorentz structure $S_{-}^{\mu\nu\sigma}$ which accompanies these finite integrals satisfies the Ward identity by itself.
We label this finite part of the loop emission amplitude as
\begin{align}
\cM_{s,L-}^{\mloone} =& \,\frac{\img g^{3} m}{4} \, (\img f^{abc} )\, S_{-}^{\mu\nu\sigma} \, 2 \cI_{L-}^{1}
\nn \\
=& \, \frac{\img g^{3} m}{4} \, (\img f^{abc} )\,  \frac{- \img S_{-}^{\mu\nu\sigma}}{8 \pi^{2} k_{T}^{2}}
\int_{0}^{1} \df x \frac{\log \big[ (1-x) x / A   \big] + \img \pi}{A + x(1-x)}
\label{eq:softloop_finite}
\,.\end{align}

The two remaining integrals $\cI_{L+}^{1\sigma}, \cI_{L+}^{2\sigma}$ with additional numerator momenta are divergent,
but satisfy $\cI_{L+}^{2\sigma} = - \cI_{L+}^{1\sigma}$, which can be checked by the change of variable $\ell' = k - \ell$.
We see that the $d^{abc}$ terms in \eq{softloop_diagram} indeed vanish, a manifestation of charge conjugation.
A closer look at these two integrals shows that the $\perp$ components of each integral vanish independently,
and the combination $\cI_{L+}^{1\sigma} - \cI_{L+}^{2\sigma}$ that shows up in the final amplitude produces the Lorentz structure
$\big( \frac{n^{\sigma}}{k^{+}} - \frac{\nbar^{\sigma}}{k^{-}} \big)$, satisfying the Ward identity%
\footnote{The integrals in particular have vanishing $k^{\sigma}$ component.}.

\subsection{Soft vertex emission}
\label{sec:softvertex}

The diagram with soft gluon emission from the vertex is written as
\begin{align} \label{eq:softvertex_diagram}
\mathcal{M}_{s, V}^{\mloone} = \raisebox{-0.45\height}{
   \includegraphics[scale=0.9]{figs/diagrams/LO1/NLP-emission/softLO1-softvert.pdf}
}
&= \frac{- \img g^{3} m }{4} \big( 2 g_{\perp}^{\mu\nu} \big)
\bigg \{ \frac{\nbar^{\sigma}}{k^{-}}
\Big[ d^{abc} \Big( - \cI_{V1} + \cI_{V2} \Big)
- \img f^{abc} \Big( \cI_{V1} + \cI_{V2} \Big)  \Big ]
\nn \\[-4em]
& \qquad \qquad\qquad \qquad
+\frac{n^{\sigma}}{k^{+}}
 \Big[ d^{abc} \Big( \cI_{V3} - \cI_{V4} \Big)
+ \img f^{abc} \Big( \cI_{V3} + \cI_{V4} \Big)  \Big ]\bigg \}
\,.\end{align}
The two Dirac structures in this diagrams reduce to the same result after tracing,
like the case in LO$_{0}$.

The four loop integrals involved in \eq{softvertex_diagram} are defined as
\begin{align}
\cI_{V1} &= \int \frac{\df^{d} \ell}{(2\pi)^{d}} \frac{1}{\ell^{2} - m^{2}+\img 0} \frac{1}{(\ell^{+}-\img 0) (\ell^{-}-k^{-} + \img0)}
\label{eq:softvertint_1}
\,,\\
\cI_{V2} &= \int \frac{\df^{d} \ell}{(2\pi)^{d}} \frac{1}{\ell^{2} - m^{2}+\img 0} \frac{1}{(\ell^{+}+ \img 0) (\ell^{-}+ k^{-} - \img0)}
\label{eq:softvertint_2}
\,,\\
\cI_{V3} &= \int \frac{\df^{d} \ell}{(2\pi)^{d}} \frac{1}{\ell^{2} - m^{2}+\img 0} \frac{1}{(\ell^{-} + \img0)[\ell^{+}+k^{+} - \img 0]}
\label{eq:softvertint_3}
\,,\\
\cI_{V4} &= \int \frac{\df^{d} \ell}{(2\pi)^{d}} \frac{1}{\ell^{2} - m^{2}+\img 0} \frac{1}{(\ell^{-} - \img0)[\ell^{+}-k^{+} + \img 0]}
\label{eq:softvertint_4}
\,. \end{align}
Shifting the loop momenta by $\ell' = -\ell$, it is simple to see that $\cI_{V1} = \cI_{V2}$ and $\cI_{V3} = \cI_{V4}$;
and with $\ell' = \ell- k$, we can check that $\cI_{V1} = \cI_{V3}$.
Therefore, we again see that charge conjugation and Ward identity are satisfied, and the only nontrivial loop integral is $\cI_{V1}$,
which again is divergent.

\subsection{Wilson line emission}
\label{sec:softwilson}

For both the hard vertex and collinear loop diagrams, a soft gluon emission cannot change the scalings at LO$_{0}$.
Therefore, we can compute the results in a refactorized fashion as in \fig{LO1softwilson}.
The factor of $\frac{1}{2}$ included in \fig{LO1softwilson} is the normalization arising from this refactorization,
which can be derived by comparing the refactorized diagrams and the effective vertex diagrams.

The Wilson line emission Feynman rule is
\begin{align}
\raisebox{-0.45\height}{
   \includegraphics[trim={0.4cm 0 0 0},clip,scale=0.9]{figs/diagrams/LO1/LP-emissions/softWilson_nbarc.pdf}
}
+
\raisebox{-0.45\height}{
   \includegraphics[trim={0.4cm 0 0 0},clip,scale=0.9]{figs/diagrams/LO1/LP-emissions/softWilson_nc.pdf}
}
= \, \img g f^{abc} \Big( \frac{n^{\sigma}}{k^{+}} - \frac{\nbar^{\sigma}}{k^{-}} \Big)
\,.\end{align}
We use the LO$_{0}$ result with the pure rapidity regulator~\cite{Ebert:2018gsn}, for which the LO$_{0}$ soft loop diagram vanishes (see \app{ybytsoftLO} for details).
Therefore, $\cM^{\mathrm{LO}_{0}}_{c+\bar{c}+h, \mathrm{p}\eta}$ is equal to the full LO$_{0}$ result in \eq{full_LO0}.

We conclude that the soft Wilson line emission contribution is
\begin{align}
\cM_{s, W}^{\mloone} = \frac{\img g^{3} m}{4} \, ( \img f^{abc}) \, \frac{- \img g_{\perp}^{\mu\nu}}{4\pi^{2}} \Big( \frac{n^{\sigma}}{k^{+}} - \frac{\nbar^{\sigma}}{k^{-}} \Big)
\times
\Bigg[ - \frac{1}{2}\bigg( \log \frac{m^{2}}{m_{H}^{2}} + \img \pi \bigg)^{2} +2 \Bigg]
\,.\end{align}

\subsection{Integrand level cancellation}
\label{sec:soft_intcancel}

To show the finiteness and compute the final result of the soft LO$_{1}$ diagrams,
instead of regulating and computing each divergent diagram individually,
we combine all the diagrams at loop integrand level.
In particular, we want to compute the combination
\begin{align} \label{eq:Msoft_cancellation}
&\,\, \cM_{s,L}^{\mloone} + \cM_{s,V}^{\mloone} + \cM_{s,W}^{\mloone}
\nn \\
=& \,\, \cM_{s,L}^{\mloone} + \cM_{s,V}^{\mloone} + \cM_{c+\bar c+h} ^{\mlozero} \times \frac{1}{2}\, \text{Wilson line}
\nn \\
=& \,\, \bigg[ \cM_{s,L}^{\mloone} + \cM_{s,V}^{\mloone} - \cM_{s}^{\mlozero} \times \frac{1}{2}\,  \text{Wilson line}  \bigg] + \cM_{s+c+\bar c+h}^{\mlozero} \times \frac{1}{2}\, \text{Wilson line}
\,.\end{align}
The task therefore is to show the combination in the square bracket is finite,
as the total LO$_{0}$ amplitude is already finite.

We omit the finite parts of $\cM_{s,L}^{\mloone}$ given in \eq{softloop_finite},
and consider only the remaining part which we label as $\cM_{s,L+}^{\mloone}$,
as it involves the divergent integral $\cI_{L+}^{1\sigma}$ given in \eq{softloopint_p1}.
As analyzed in \sec{softloop}, the Lorentz structure of $\cM_{s,L+}^{\mloone}$ is proportional to $\big( \frac{n^{\sigma}}{k^{+}} - \frac{\nbar^{\sigma}}{k^{-}} \big)$,
which is the same as the Lorentz structure of $\cM_{s,V}^{\mloone}$ and the Wilson line.
Therefore, we simply consider the scalar integral proportional to this structure.
Explicitly,
\begin{align}
\cM_{s, \mathrm{fin}}^{\mloone}
\equiv & \, \cM_{s,L+}^{\mloone} + \cM_{s,V}^{\mloone} - \cM_{s}^{\mlozero} \times \frac{1}{2}\,  \text{Wilson line}
\nn \\
=& \, \frac{\img g^{3} m }{4} \, \img f^{abc} \, 2 g_{\perp}^{\mu\nu} \Big( \frac{n^{\sigma}}{k^{+}} - \frac{\nbar^{\sigma}}{k^{-}} \Big)
\Big[ 2 \cI_{L+}^{1} - 2 \cI_{V1} + 2 \cI_{0} \Big]
\,,\end{align}
where we have defined the scalar version of $\cI_{L+}^{1\sigma}$ as
\begin{align}
\cI_{L+}^{1} = \frac{k^{+}}{2} \, \nbar^{\sigma} \cdot \cI_{L+}^{1\sigma}
\,,\end{align}
$\cI_{V1}$ is defined in \eq{softvertint_1},
and $\cI_{0}$ is the LO$_{0}$ loop integral (without prefactors) in \eq{softLO0intnoreg},
$\cM^{s}_{\mathrm{LO}_0}=-2\img g^2m\,\delta^{ab}g_\perp^{\mu\nu}\cI_0$.
In the above expression, we have also used the symmetry properties of the integrals discussed in \secs{softloop}{softvertex}.

Our strategy is to combine the loop integrals under the integrand, and use the freedom of shifting loop momenta to achieve common eikonal propagators.
Writing the integrands out, we get
\begin{align}
\cM_{s, \mathrm{fin}}^{\mloone} =&
\cA^{abc,\mu\nu\sigma}  \int \frac{\df^{d} \ell}{(2\pi)^{d}}
\frac{1}{(\ell^{+} + \img0)(\ell^{-}-\img0)} \times
\nn \\ &
\Bigg(
\frac{k^{+}(k^{-} - 2 \ell^{-})}{\big [\ell^{+} (\ell^{-}-k^{-}) + \ell_{\perp}^{2} - m^{2} + \img 0 \big ]
\big[\ell^{-} (\ell^{+}+k^{+}) + (\ell_{\perp} - k_{\perp})^{2} - m^{2} + \img 0 \big]}
\nn \\ & \quad
- \frac{2}{\ell^{-} (\ell^{+}+k^{+}) + \ell_{\perp}^{2} - m^{2} + \img 0}
+ \frac{2}{\ell^{-} \ell^{+} + \ell_{\perp}^{2} - m^{2} + \img 0}
\Bigg)
\,,\end{align}
where we have defined
$\cA^{abc,\mu\nu\sigma} = \frac{\img g^{3} m }{4} \, (\img f^{abc} )\, 2 g_{\perp}^{\mu\nu} \big( \frac{n^{\sigma}}{k^{+}} - \frac{\nbar^{\sigma}}{k^{-}} \big)$.
We compute the loop integral by first doing the $\ell^{+}$ integral by contours, then the $\perp$ integral in $d-2$ dimensions,
and finally the $\ell^{-}$ integral.
We do not need an explicit regulator for the endpoint divergences, but the cancellation of divergences is rather intricate in the $\ell^{-}$ integral:
the loop integral contributes both for $0<\ell^{-}<k^{-}$ and $\ell^{-}<0$,
and the divergence at $\ell^{-} \to 0$ cancels between these two pieces;
the vertex and LO$_{0}$ soft integrals contribute only for $\ell^{-}<0$,
and the divergence at $\ell^{-} \to -\infty$ cancels between the vertex plus LO$_{0}$ piece and the loop integral piece.
We find that the final result is indeed finite:
\begin{align}
\cM_{s, \mathrm{fin}}^{\mloone} =
\frac{\img g^{3}m}{4} (\img f^{abc} ) \, \frac{ -\img g_{\perp}^{\mu\nu}  }{ 8 \pi^{2}} \, \Big( \frac{n^{\sigma}}{k^{+}} - \frac{\nbar^{\sigma}}{k^{-}} \Big)
\times \int_{0}^{1} \df x \frac{\log \big[ (1-x) x / A   \big] + \img \pi}{A + x(1-x)}
\label{eq:softintegrand_finite}
\,.\end{align}
Note that the $x$ integral is the same as the one in \eq{softloop_finite} for $\cI_{L-}^{1}$.

\subsection{Glauber diagrams and zero-bin}
\label{sec:softGlauber}

In this subsection, we consider the potential Glauber diagrams with a soft gluon emission, and the zero-bin region of the soft emission diagrams in \fig{LO1diassoft}.
We will show that these two contributions cancel each other, and we do not have any nontrivial Glauber contribution.
\subsubsection{Glauber diagrams with soft emission}

\begin{figure}
\centering
	\begin{subfigure}[b]{0.48\textwidth}
		\centering
		\includegraphics[width=0.85\textwidth]{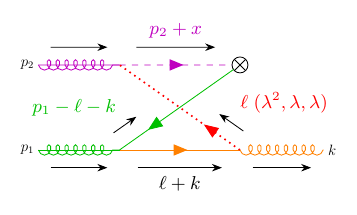}
		\caption{Sample Glauber diagram with GSS vertex.}
		\label{fig:LO1softGSS}
	\end{subfigure}
	\hfill
	\begin{subfigure}[b]{0.48\textwidth}
		\centering
		\includegraphics[width=0.85\textwidth]{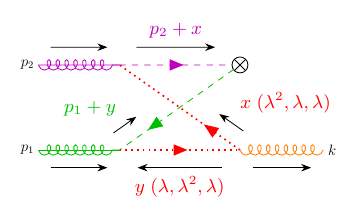}
		\caption{Sample Glauber diagram with GGS vertex.}
		\label{fig:LO1softGGS}
	\end{subfigure}
\caption{The Glauber diagrams that contribute with a soft gluon emission.
As before, $n (\bn)$-collinear gluons are colored green (magenta), and soft gluons are colored orange.
The vertex with an outgoing hard Higgs is labeled with $\otimes$.
The fermions in the crossed-box loop can have $n (\bn)$-collinear scaling,
which is dashed green (magenta),
$n (\bn)$-hard-collinear scaling, which is solid green (magenta),
soft scaling, which is solid orange,
or Glauber scaling, which is dotted red.
The arrows on the fermion lines indicate direction of fermion flow.
Diagram (a) has a Glauber-soft-soft vertex, and diagram (b) has a Glauber-Glauber-soft vertex.
We have to consider 1) reversing fermion flow and 2) exchanging $n \leftrightarrow \nbar$ for each diagram.
These two operations result in the same diagram for the GGS diagram.
Therefore, we have a total of four diagrams with scalings like diagram (a), and two with scalings like diagram (b).}
\label{fig:LO1diassoft_Glauber}
\end{figure}

For completeness, we also consider soft emission diagrams with Glauber scaling, $\ell^{+}\ell^{-} \ll \ell_{T}^{2}$, in the fermion loop, which are generated from the Lagrangian in \eq{Glauber_Lagrangian_quark}. The two classes of Glauber diagrams are drawn in \fig{LO1diassoft_Glauber},
one with the two-rapidity Lagrangian and a Glauber-soft-soft (GSS) vertex,
and one with the three-rapidity Lagrangian and Glauber-Glauber-soft (GGS) vertex.
There are a total of four GSS diagrams with scalings like \fig{LO1softGSS} (resulting from reversing fermion flow or exchanging $n\leftrightarrow \nbar$), and two GGS diagrams with scalings like \fig{LO1softGGS} (as the two operations are redundant for this diagram).

We find that the sum of the four GSS diagrams evaluates to 0, because the two subdiagrams with the same hard-collinear propagator but reversed fermion flow cancel each other.
On the other hand, the sum of the two GGS diagrams is nonzero, and evaluates to
\begin{align}
\cM_{s, GGS}^{\mloone} = -\frac{\img g^{3} m}{4} \big( \img f^{abc}\big) \Big( 2 g_{\perp}^{\mu\sigma } k_{\perp}^{\nu}
- 2 g_{\perp}^{\nu\sigma} k_{\perp}^{\mu} \Big) (2\cI_{G})
\,,\end{align}
where the Glauber integral $\cI_{G}$ is defined as
\begin{align}
\cI_{G} &= \int\frac{\df^{d}\ell_{G}}{(2\pi)^{d}}
\frac{1}
{\big[\ell_{\perp}^{2}-m^{2} + \img 0 \big]
\big[ (\ell_{\perp} + k_{\perp})^{2} - m^{2} +\img 0\big]}
\frac{1}{\big[ \ell^{+} - \Delta^{+} - \img 0\big]}
\frac{1}{\big[\ell^{-} - \Delta^{-} + \img 0\big]}
\,,\end{align}
which is a standard Glauber integral and the computation can be found in \refcite{Rothstein:2016bsq}.
It is also shown in \refcite{Rothstein:2016bsq} (with their regulator of choice) that the exact values of $\Delta^{+}$ and $\Delta^{-}$ in the eikonal denominators do not matter, only the signs of $\img 0$'s do; we will use this fact to show that the Glauber zero-bin of the soft loop cancels these Glauber diagrams.

\subsubsection{Glauber zero-bin of soft emission diagrams}

We also need to compute the Glauber zero-bin subtraction for the soft emission diagrams themselves.
Amongst the three (classes of) diagrams we have considered in \fig{LO1diassoft}, only the loop emission \fig{LO1softloop} can generate a Glauber zero-bin that depends on the transverse momentum $k_{T}$.
We compute
\begin{align}
\label{eq:G0bin_loop_final}
\mathcal{M}_{s, L}^{\mloone} \bigg|_\text{Glauber 0-bin} = - \frac{\img g^{3} m}{4} \big( \img f^{abc}\big)
\Big( 2 g_{\perp}^{\mu\sigma}  k_{\perp}^{\nu}
- 2 g_{\perp}^{\nu\sigma}  k_{\perp}^{\mu} \Big) ( 2\, \cI_{G})
\,,\end{align}
which is exactly the same as the Glauber diagram computed above.
Note however that the $\cI_{G}$ here does not have the same $\Delta^{+}$ and $\Delta^{-}$ as the Glauber diagrams,
but the $\img 0$ prescriptions are indeed the same, which guarantees the final integral to be the same.
This agreement should not be surprising, as diagrammatically, the zero-bin kinematics is exactly the same as the GGS Glauber diagram considered above.
Therefore, because the zero-bin needs to be subtracted and Glauber graphs added, we get a vanishing contribution when these are added.

For the vertex emission as well as collinear loop with Wilson line emission in \figs{LO1softvertex}{LO1softwilson},
changing the loop momenta to Glauber scaling does not change the diagram's dependence on the momentum $k$.
Therefore, the Glauber diagram and zero-bin subtraction work the same way as in the LO$_{0}$ case,
which does not give any net contribution.

\subsection{Final result for soft emission}
\label{sec:softfinal}

We are ready to present the final answer for the soft emission:
\begin{align}
\cM_{s}^{\mloone}
&= \cM_{s,L}^{\mloone} + \cM_{s,V}^{\mloone} + \cM_{s,W}^{\mloone}
\nn \\
&= \cM_{s,L-}^{\mloone} + \cM_{s, \mathrm{fin}}^{\mloone} + \cM^{\mlozero} \times \frac{1}{2}\,  \text{Wilson line}
\,,\end{align}
where $\cM_{s,L-}^{\mloone}$ is defined in \eq{softloop_finite},
$\cM_{s, \mathrm{fin}}^{\mloone}$ in \eq{softintegrand_finite}
and $\cM^{\mlozero}$ in \eq{full_LO0}.
The division of the full result into these three finite parts is discussed at the beginning of \sec{soft_intcancel}.
The Glauber contribution cancels against subtractions and is not included.
Adding all the terms explicitly, we get
\begin{align}
\cM_{s}^{\mloone} = &\frac{\img g^{3}m}{4} (\img f^{abc} ) \frac{-\img}{4\pi^{2}}
\Bigg\{ g_{\perp}^{\mu\nu}  \Big( \frac{n^{\sigma}}{k^{+}} - \frac{\nbar^{\sigma}}{k^{-}} \Big) \bigg[ - \bigg( \log \frac{m^{2}}{m_{H}^{2}} + \img \pi \bigg)^{2} +4 \bigg]
\nn \\
&\qquad\qquad\qquad\qquad
- \Big ( g_{\perp}^{\mu\sigma}  k_{\perp}^{\nu}
- g_{\perp}^{\nu\sigma}  k_{\perp}^{\mu} \Big)
\frac{1}{k_{T}^{2}}  \int_{0}^{1} \df x \frac{\log \big[ (1-x) x / A   \big] + \img \pi}{A + x(1-x)}
\Bigg\}
\,,\end{align}
where we have used the $\perp$-only components of $S_{-}^{\mu\nu\sigma}$ as written in \eq{loop_lorentz_minus}.

The integral can be expressed in terms of dilogarithms:
\begin{align}
\frac{1}{k_{T}}\int_{0}^{1} \df x \frac{\log \big[ (1-x) x / A   \big] + \img \pi}{A + x(1-x)}
= \frac{2}{\sqrt{X}} \bigg[ \mathrm{Li}_{2} \Big( \frac{\sqrt{X} - k_{T}}{\sqrt{X} + k_{T}} \Big)  -  \mathrm{Li}_{2} \Big( \frac{\sqrt{X} + k_{T}}{\sqrt{X} - k_{T}} \Big) \bigg]\,,
\end{align}
with $A = \frac{m^{2}}{k_{T}^{2}}$ and $\sqrt{X} = \sqrt{k_{T}^{2} + 4m^{2}}$.

\section{Comparison to fixed order}
\label{sec:FO}
In this section, we validate our collinear and soft emission results against the full QCD helicity amplitudes of \refcite{Baur:1989cm}, expanded in the corresponding limits.
Throughout, we account for the overall factor $g^3 f^{abc}/(64\pi^2 v)$ that is pulled out of the amplitudes in \refcite{Baur:1989cm}, and suppress writing the overall Yukawa couplings.
\subsection{Polarization vectors}
The full QCD result is expressed in terms of helicity amplitudes $\cM_{\lambda_1\lambda_2\lambda_k}$, where $\lambda_1$, $\lambda_2$, and $\lambda_k$ denote the helicities of the external gluons with momenta $p_1$, $p_2$, and $k$.
There are three independent amplitudes, $\cM_{+++}$, $\cM_{++-}$, and $\cM_{+--}$, which we obtain by contracting our results with the polarization vectors
\begin{align}\label{eq:polvec}
	\eps(p_1)^\mu =& \frac{1}{\sqrt{2}} (e_\perp^\mu + \img \lambda_1 y^\mu), \nn \\
	\eps(p_2)^\mu =& \frac{1}{\sqrt{2}} (e_\perp^\mu - \img \lambda_2 y^\mu), \nn \\
	\eps(k)^\mu =& -\frac{\lambda_k}{\sqrt{2}} \left(e_\perp^\mu - \frac{e_\perp \cdot k_\perp}{k \cdot \nb} \nb^\mu- \img \lambda_k y_\perp^\mu\right),
\end{align}
where $e_\perp^\mu$ is the unit vector in $k_\perp$ direction with $e_\perp^2 =-1$ and the unit vector $y^\mu$ ($y^2=-1$) defines an overall axis in the $n$-collinear sector, with
\begin{align}
	y \cdot n =0, \quad y \cdot \nb =0, \quad y \cdot k =0,
\end{align}
as well as
\begin{align}
	y_\perp \cdot e_\perp =0, \quad y_\perp \cdot k_\perp =0, \quad y_\perp \cdot y_\perp =-1.
\end{align}
The polarization vector of the emitted gluon satisfies $\nb\cdot\eps(k)=0$. This choice is allowed because our amplitudes satisfy the Ward identity (see \sec{endpointdiv} and \sec{softfinal}), and it removes all contributions proportional to $\nb^\rho$: for collinear emission, the vertex emission amplitudes $\cM_{d_{VE}}$ drop out, and for soft emission, the terms proportional to $\nb^\sigma/k^-$.

The phases of the polarization vectors are conventional and drop out of physical quantities.
For the emitted gluon, we include the factor $-\lambda_k$ in \eq{polvec}, such that $\eps_-(k) = -\eps_+(k)^*$, following the Condon--Shortley convention for helicity states~\cite{Jacob:1959at, Haber:1994pe}.
This reproduces the relative phase between $\cM_{+++}$ and $\cM_{++-}$ in \refcite{Baur:1989cm}, whereas $\eps_-(k) = \eps_+(k)^*$ would introduce a relative sign between them.

\subsection{Collinear emission}
Adding all collinear emission diagrams of \sec{collinearemission} and contracting the resulting amplitude $\cM^{\mathrm{coll.}}$ with the polarization vectors, we obtain
\begin{align}
	\cM^{\mathrm{coll.}}_{+++}&=-\frac{\img g^3 f^{abc}m}{\sqrt{2} \,8 \pi ^2 k_T }
	\left\{\frac{1}{2}
	\log ^2\left(\frac{\sqrt{1+\frac{4 m^2 (1-z)}{k_T^2}}+1}{\sqrt{1+\frac{4 m^2 (1-z)}{k_T^2}}-1}\right)
	\right. \nn \\
	&\left.
	+\frac{1}{z} \left[
	\log ^2\left(-\frac{\omega_n^2}{m^2-\img 0}\right)- \log ^2\left(\frac{\sqrt{1+\frac{4 m^2 (1-z)}{k_T^2}}+1}{\sqrt{1+\frac{4 m^2 (1-z)}{k_T^2}}-1}\right)
	-4
	\right] \right. \nn \\
	&\left.
	+\frac{k_T}{\sqrt{k_T^2+4 m^2}}\right. \nn \\
	&\left. \times \left[\text{Li}_2\left(\frac{\sqrt{\frac{4 m^2}{k_T^2}+1}+1}{\sqrt{\frac{4 m^2}{k_T^2}+1}-\sqrt{1+\frac{4 m^2 (1-z)}{k_T^2}}}\right)
	- \text{Li}_2\left(\frac{\sqrt{\frac{4
				m^2}{k_T^2}+1}-1}{\sqrt{\frac{4 m^2}{k_T^2}+1}-\sqrt{1+\frac{4 m^2 (1-z)}{k_T^2}}}\right)
	\right.\right. \nn \\
	&\left. \left.
	-\text{Li}_2\left(\frac{\sqrt{\frac{4 m^2}{k_T^2}+1}-1}{\sqrt{\frac{4
				m^2}{k_T^2}+1}+\sqrt{1+\frac{4 m^2 (1-z)}{k_T^2}}}\right)+
	\text{Li}_2\left(\frac{\sqrt{\frac{4 m^2}{k_T^2}+1}+1}{\sqrt{\frac{4 m^2}{k_T^2}+1}+\sqrt{1+\frac{4 m^2 (1-z)}{k_T^2}}}\right)\right]\right. \nn \\
	&\left.
	+\frac{ k_T}{\sqrt{k_T^2+4 m^2(1-z)^2}}\right. \nn \\
	&\left.
	\times \left[
	\text{Li}_2\left(\frac{\sqrt{\frac{4 m^2 (z-1)^2}{k_T^2}+1}-1}{\sqrt{\frac{4 m^2
				(z-1)^2}{k_T^2}+1}-\sqrt{1+\frac{4 m^2 (1-z)}{k_T^2}}}\right)
	-
	\text{Li}_2\left(\frac{\sqrt{\frac{4 m^2 (z-1)^2}{k_T^2}+1}+1}{\sqrt{\frac{4 m^2 (z-1)^2}{k_T^2}+1}-\sqrt{1+\frac{4 m^2 (1-z)}{k_T^2}}}\right)
	\right.	\right. \nn \\
	&\left. \left.
	+\text{Li}_2\left(\frac{\sqrt{\frac{4 m^2
				(z-1)^2}{k_T^2}+1}-1}{\sqrt{\frac{4 m^2
				(z-1)^2}{k_T^2}+1}+ \sqrt{1+\frac{4 m^2 (1-z)}{k_T^2}}}\right)- \text{Li}_2\left(\frac{\sqrt{\frac{4 m^2
				(z-1)^2}{k_T^2}+1}+1}{\sqrt{\frac{4 m^2
				(z-1)^2}{k_T^2}+1}+\sqrt{1+\frac{4 m^2 (1-z)}{k_T^2}}}\right)
	\right]\right\}\,,
	\label{eq:Mppp_coll}
\end{align}
\begin{align}
	\cM^{\mathrm{coll.}}_{++-}&=-\frac{\img g^3 f^{abc}m\, z }{\sqrt{2} \,16 \pi ^2 k_T }
	\left\{ 8
	+\log ^2\left(\frac{\sqrt{1+\frac{4 m^2 (1-z)}{k_T^2}}+1}{\sqrt{1+\frac{4 m^2 (1-z)}{k_T^2}}-1}\right)
	-2 \log ^2\left(-\frac{\omega_n^2}{m^2-\img 0}\right)
	\right. \nn \\
	&\left.
	+\frac{2 k_T}{\sqrt{k_T^2+4 m^2}}\right. \nn \\
	&\left. \times \left[\text{Li}_2\left(\frac{\sqrt{\frac{4 m^2}{k_T^2}+1}+1}{\sqrt{\frac{4 m^2}{k_T^2}+1}-\sqrt{1+\frac{4 m^2 (1-z)}{k_T^2}}}\right)
	- \text{Li}_2\left(\frac{\sqrt{\frac{4
				m^2}{k_T^2}+1}-1}{\sqrt{\frac{4 m^2}{k_T^2}+1}-\sqrt{1+\frac{4 m^2 (1-z)}{k_T^2}}}\right)
	\right.\right. \nn \\
	&\left. \left.
	-\text{Li}_2\left(\frac{\sqrt{\frac{4 m^2}{k_T^2}+1}-1}{\sqrt{\frac{4
				m^2}{k_T^2}+1}+\sqrt{1+\frac{4 m^2 (1-z)}{k_T^2}}}\right)+
	\text{Li}_2\left(\frac{\sqrt{\frac{4 m^2}{k_T^2}+1}+1}{\sqrt{\frac{4 m^2}{k_T^2}+1}+\sqrt{1+\frac{4 m^2 (1-z)}{k_T^2}}}\right)\right]\right. \nn \\
	&\left.
	+\frac{2 k_T}{\sqrt{k_T^2+4 m^2(1-z)^2}}\right. \nn \\
	&\left.
	\times \left[
	\text{Li}_2\left(\frac{\sqrt{\frac{4 m^2 (z-1)^2}{k_T^2}+1}+1}{\sqrt{\frac{4 m^2
				(z-1)^2}{k_T^2}+1}-\sqrt{1+\frac{4 m^2 (1-z)}{k_T^2}}}\right)
	-
	\text{Li}_2\left(\frac{\sqrt{\frac{4 m^2 (z-1)^2}{k_T^2}+1}-1}{\sqrt{\frac{4 m^2 (z-1)^2}{k_T^2}+1}-\sqrt{1+\frac{4 m^2 (1-z)}{k_T^2}}}\right)
	\right.	\right. \nn \\
	&\left. \left.
	+\text{Li}_2\left(\frac{\sqrt{\frac{4 m^2
				(z-1)^2}{k_T^2}+1}+1}{\sqrt{\frac{4 m^2
				(z-1)^2}{k_T^2}+1}+ \sqrt{1+\frac{4 m^2 (1-z)}{k_T^2}}}\right)- \text{Li}_2\left(\frac{\sqrt{\frac{4 m^2
				(z-1)^2}{k_T^2}+1}-1}{\sqrt{\frac{4 m^2
				(z-1)^2}{k_T^2}+1}+\sqrt{1+\frac{4 m^2 (1-z)}{k_T^2}}}\right)
	\right]\right\}\,,
		\label{eq:Mppm_coll}
\end{align}
while the third independent amplitude vanishes,
\begin{align}
 \cM^{\mathrm{coll.}}_{+--} =0.
\end{align}
To compare with \refcite{Baur:1989cm}, we expand the full QCD amplitudes in the limit where the emitted gluon is $n$-collinear, $k_T \sim m \sim \lambda\, m_H$. In terms of the Mandelstam variables used there, this corresponds to
\begin{align}
	s &= (p_1 + p_2 )^2= \frac{\omega_n^2 }{z} + \ord{\lambda^2} ,\\
	t &=(p_2-k)^2 = -\frac{1-z}{z}\omega_n^2+ \ord{\lambda^2}\,,\\
	u &= (p_1-k)^2 = \ord{\lambda^2}\,.
\end{align}
At leading power in $\lambda$, we find full analytic agreement for all three helicity amplitudes, provided the Glauber contribution of \sec{glaubercoll} is included.

\begin{figure}
	\centering
	\includegraphics[width=\WidthTwoSubfigs]{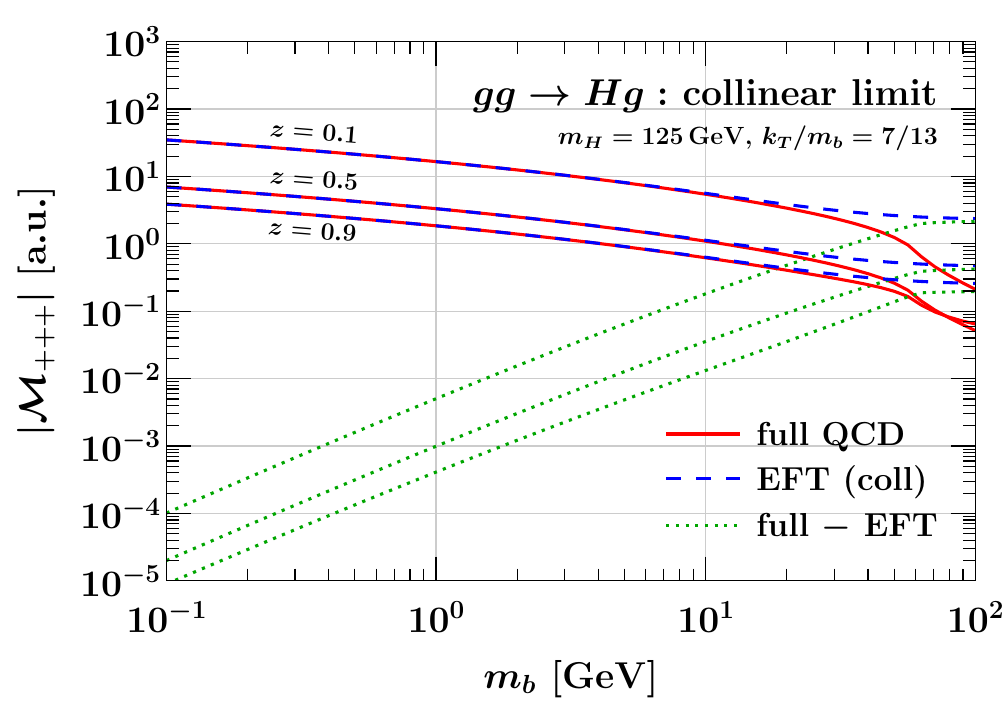}
	\hfill
	\includegraphics[width=\WidthTwoSubfigs]{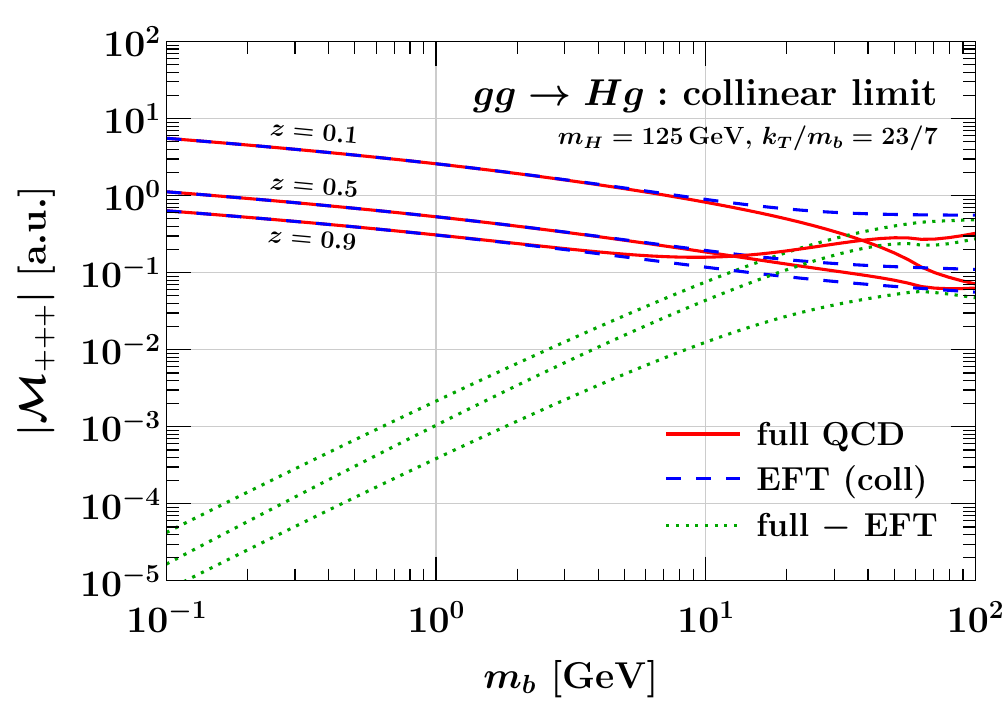}
	\\
	\includegraphics[width=\WidthTwoSubfigs]{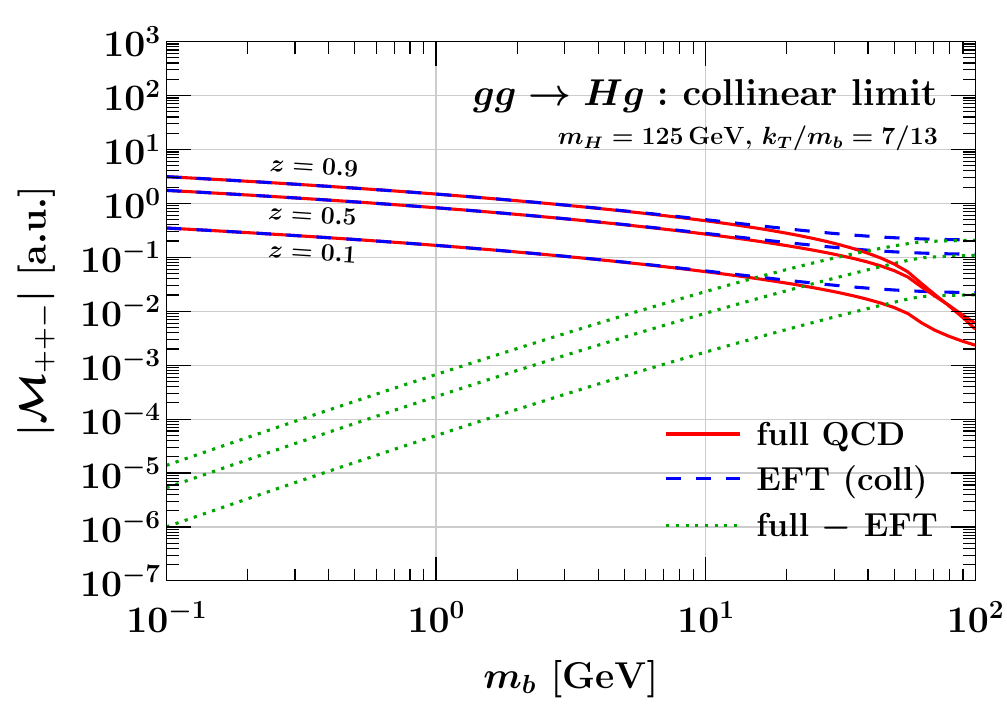}
	\hfill
	\includegraphics[width=\WidthTwoSubfigs]{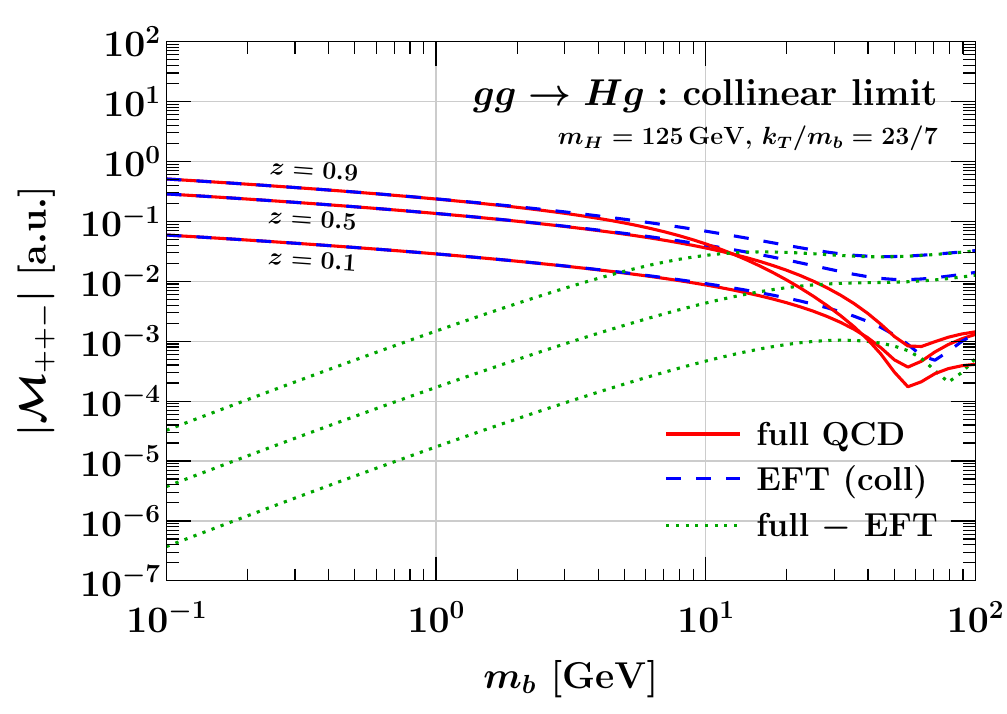}
	\\
	\includegraphics[width=\WidthTwoSubfigs]{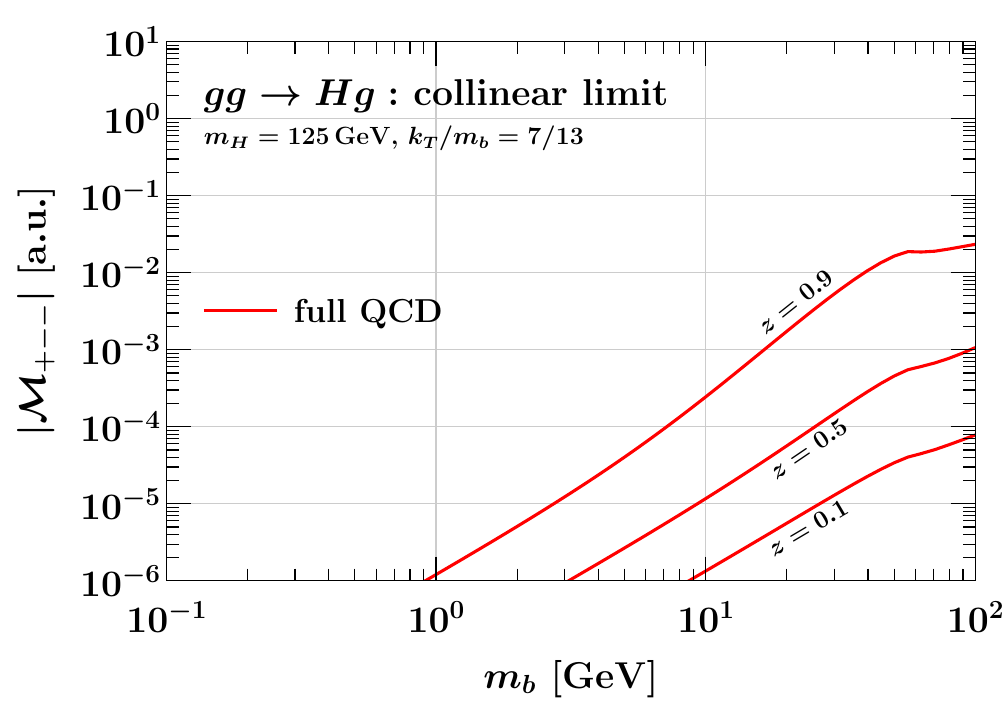}
	\hfill
	\includegraphics[width=\WidthTwoSubfigs]{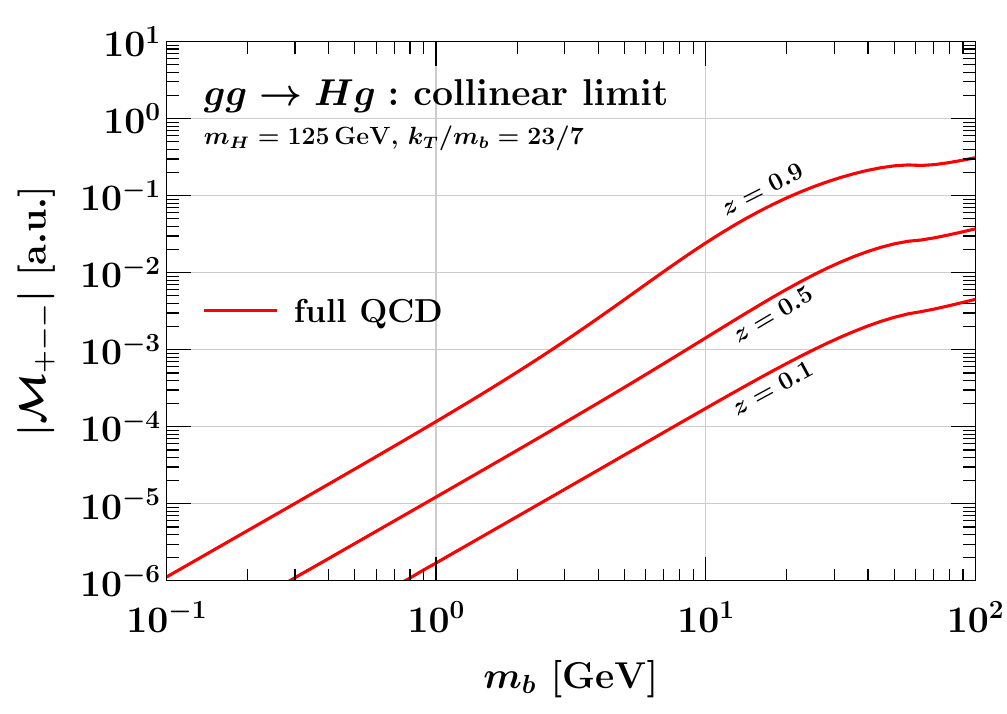}
	\caption{Comparison of the full QCD helicity amplitudes (solid red) to the SCET result (dashed blue) in the collinear emission limit as $m \to 0$,
		holding $k_T/m$ fixed at $7/13$ (left) or $23/7$ (right), for different values of the gluon light-cone momentum fraction $z$. The dotted green curves show the nonsingular remainder. From top to bottom: $\cM_{+++}$, $\cM_{++-}$, and $\cM_{+--}$.}
	\label{fig:comparison_fixed_order_coll}
\end{figure}

\Fig{comparison_fixed_order_coll} shows the numerical comparison of the SCET amplitudes (dashed blue) with the full QCD amplitudes (solid red) as $m\to 0$ at fixed $k_T/m$, for three values of the momentum fraction $z$ and two values of $k_T/m$ (left and right panels).
In all cases, the SCET amplitudes exactly reproduce the singular behavior of the QCD amplitudes, including the complete $z$ dependence and the large logarithms of $m_H/m$.
Their difference, the nonsingular remainder (dotted green), is suppressed by $\ord{\lambda}$ relative to the leading-power result.
Likewise, $\cM_{+--}$ is power suppressed (but nonzero) relative to $\cM_{+++}$ and $\cM_{++-}$,
consistent with $\cM^{\mathrm{coll.}}_{+--} = 0$.

\subsection{Soft emission}

\begin{figure}
\centering
\includegraphics[width=\WidthTwoSubfigs]{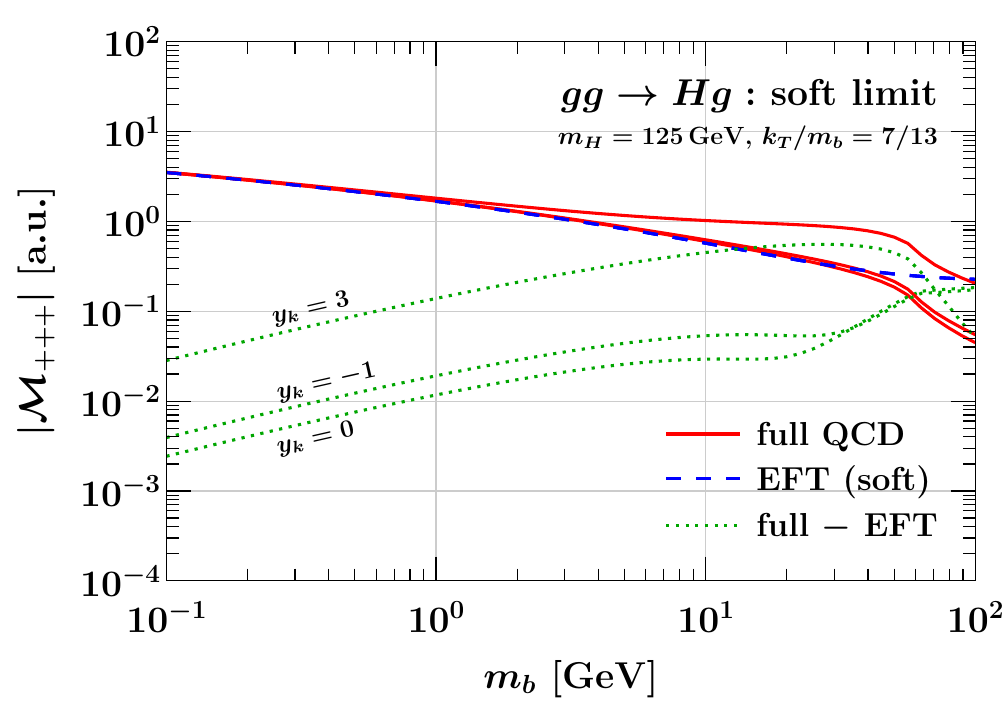}
\hfill
\includegraphics[width=\WidthTwoSubfigs]{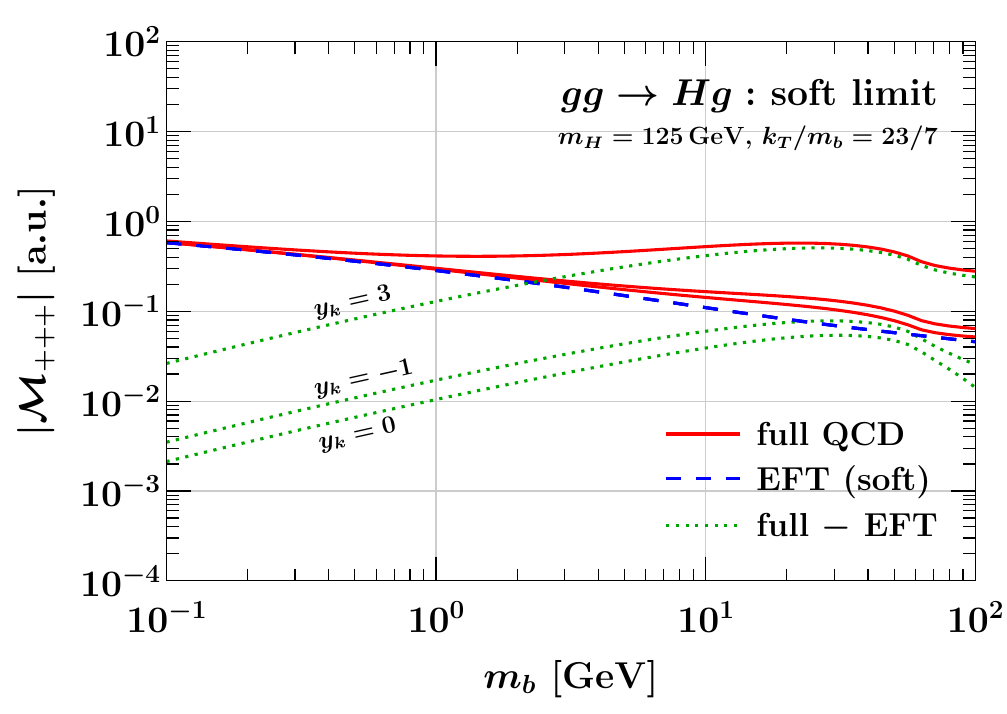}
\\
\includegraphics[width=\WidthTwoSubfigs]{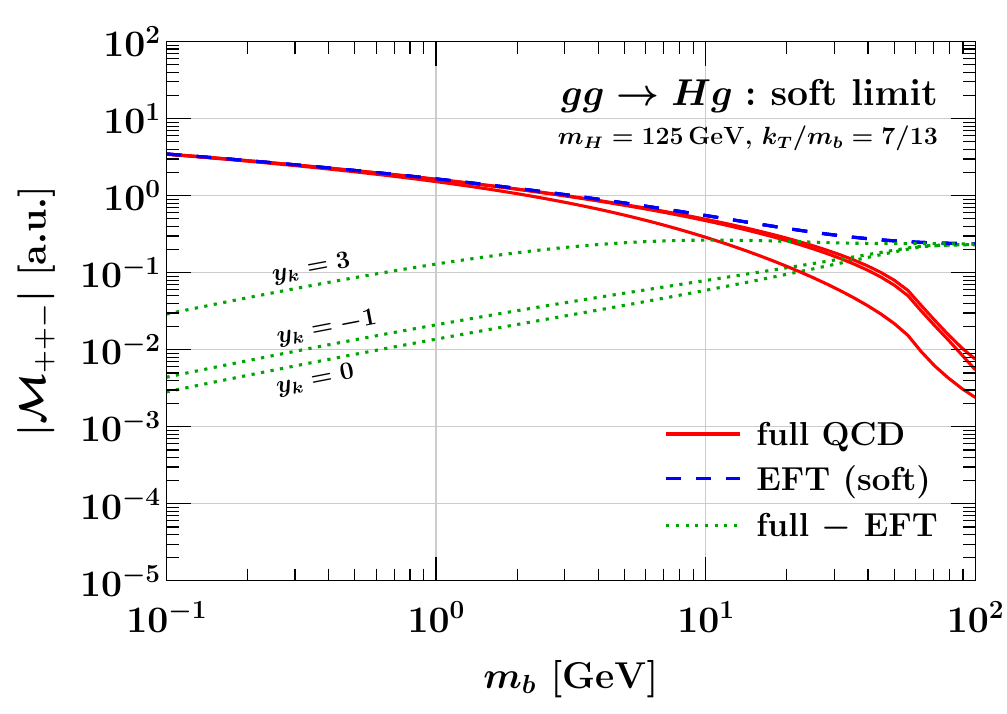}
\hfill
\includegraphics[width=\WidthTwoSubfigs]{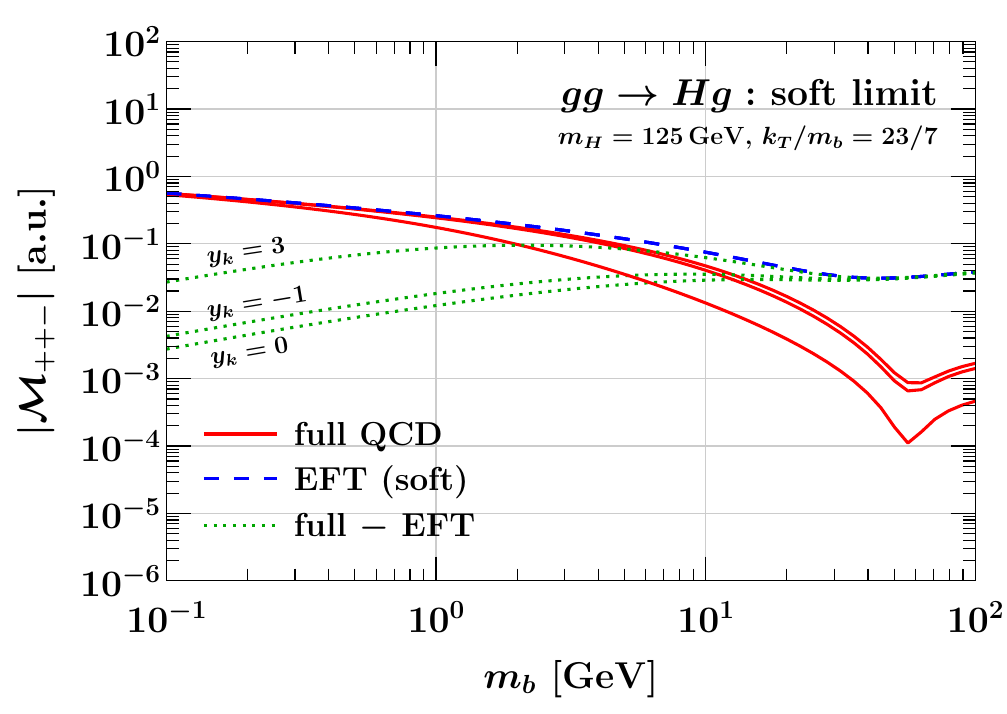}
\\
\includegraphics[width=\WidthTwoSubfigs]{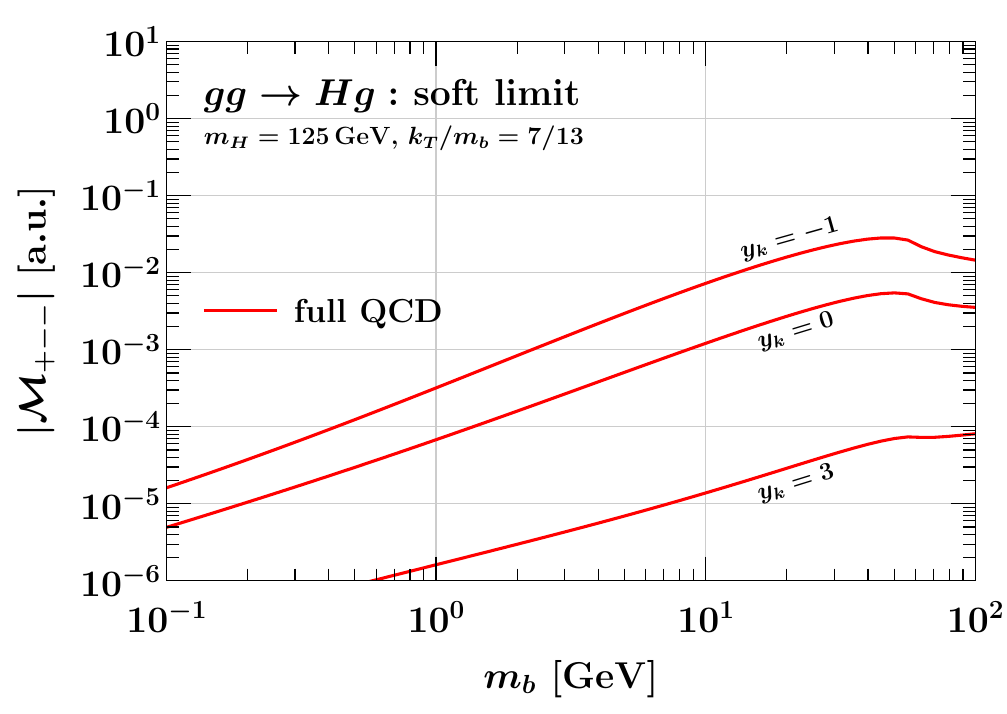}
\hfill
\includegraphics[width=\WidthTwoSubfigs]{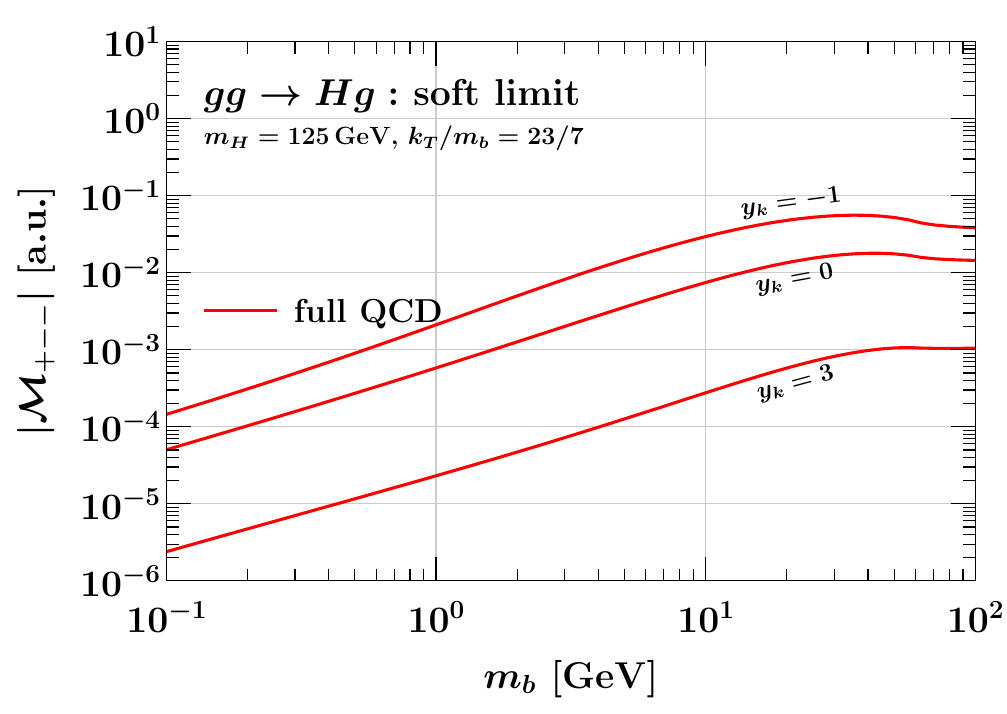}
\caption{Same as \fig{comparison_fixed_order_coll}, but in the soft emission limit and for different values of the gluon rapidity $y_k$.}
\label{fig:comparison_fixed_order_soft}
\end{figure}

For soft gluon emission, we obtain
\begin{align} \label{eq:helicityAmps_soft}
\cM^\mathrm{soft}_{+++}
&= -\frac{\img g^3 f^{abc}m}{\sqrt{2} \,8 \pi ^2 }\Bigg[ \frac{\log^{2}\Big( \frac{- m_{H}^{2}}{ m^{2} - \img 0}\Big) - 4}{k_{T}}
- \frac{\mathrm{Li}_{2} \Big( \frac{\sqrt{k_{T}^{2} + 4m^{2}} - k_{T}}{\sqrt{k_{T}^{2} + 4m^{2}} + k_{T}} \Big) - \mathrm{Li}_{2} \Big( \frac{\sqrt{k_{T}^{2} + 4m^{2}} + k_{T}}{\sqrt{k_{T}^{2} + 4m^{2}} - k_{T}} \Big)}{\sqrt{k_{T}^{2} + 4m^{2}}}
\Bigg]
\,,
\nn \\
\cM^\mathrm{soft}_{++-}
&= \frac{\img g^3 f^{abc}m}{\sqrt{2} \,8 \pi ^2 } \Bigg[ \frac{\log^{2}\Big( \frac{- m_{H}^{2}}{ m^{2} - \img 0}\Big) - 4}{k_{T}}
+ \frac{\mathrm{Li}_{2} \Big( \frac{\sqrt{k_{T}^{2} + 4m^{2}} - k_{T}}{\sqrt{k_{T}^{2} + 4m^{2}} + k_{T}} \Big) - \mathrm{Li}_{2} \Big( \frac{\sqrt{k_{T}^{2} + 4m^{2}} + k_{T}}{\sqrt{k_{T}^{2} + 4m^{2}} - k_{T}} \Big)}{\sqrt{k_{T}^{2} + 4m^{2}}}
\Bigg]
\,,
\end{align}
and, as in the collinear case,
\begin{equation}
\cM^\mathrm{soft}_{+--} = 0 \,.
\end{equation}

To expand the full QCD amplitudes, note that in the soft limit, the Mandelstam variables reduce to
\begin{align}
	s = m_H^2+ \ord{\lambda^2}, \qquad
	t = -m_H\,k_T\,e^{y_k}+ \ord{\lambda^2}\,, \qquad
	u = -m_H\,k_T\,e^{-y_k}+ \ord{\lambda^2}\,,
\end{align}
where $y_k$ is the rapidity of the emitted gluon.
At leading power, we again find full agreement with \eq{helicityAmps_soft}.
The rapidity dependence drops out in the soft limit because the soft gluon couples to the incoming partons only through eikonal factors, which are invariant under boosts along the beam axis.
Hence, the amplitude can depend on $t$ and $u$ only through the boost-invariant combination $t\,u = s\,k_T^2$, i.e., on the gluon momentum only through $k_T$.

\Fig{comparison_fixed_order_soft} shows the corresponding numerical comparison, now for three values of the gluon rapidity $y_k$. Since the SCET amplitude is independent of $y_k$, the rapidity dependence of the full QCD amplitudes is purely a power correction, and the singular behavior as $m\to 0$ is again exactly reproduced. As in the collinear case, the nonsingular remainders are suppressed by $\ord{\lambda}$.

Finally, in the limit $z\to 1$, the collinear helicity amplitudes in \eqs{Mppp_coll}{Mppm_coll} reduce to the soft ones in \eq{helicityAmps_soft}, as required by the overlap of the soft and collinear regions. Together with the agreement with full QCD in both limits, this provides a three-way consistency check of our results.

\section{Discussion of main results}
\label{sec:logstructure}

In this section, we briefly discuss the implications of the results presented in
\secs{collinearemission}{softemission} on the structure of the factorization
theorems from \sec{factorization_theorems}. In \sec{consistency}, we discuss the
relation to the $q_T \ll m$ limit. In \sec{rapidityadm}, we discuss the
structure of the rapidity anomalous dimension that can be inferred from our
calculations.

\subsection{Relation between \texorpdfstring{$q_T \ll m$}{qT << m} and \texorpdfstring{$q_T \sim m$}{qT sim m} hierarchies}
\label{sec:consistency}

We now discuss how the ingredients in the factorization theorems in \eqs{factorization_small_qT}{factorization_intermediate_qT} for the two hierarchies $q_T \ll m \ll m_H$ and
$q_T \sim m \ll m_H$ are related.
This is closely analogous to the relation and matching between hierarchies for secondary
quark-mass effects in the leading-power $q_T$ factorization as discussed in
detail in \refcite{Pietrulewicz:2017gxc}.
Similarly to that case, requiring consistency between the two limits allows
us to derive nontrivial relations between the individual ingredients.

Consistency between the two factorization theorems requires that they agree
in the stronger of the two limits, $q_T \ll m$, i.e., we must have
\begin{equation} \label{eq:sigma_consistency}
\df\sigma = \df\sigma_{{q_T} \ll m} \biggl[1 + \ORd{\frac{q_T^2}{m^2}}\biggr]
\,,\end{equation}
where for simplicity $\df\sigma$ denotes the cross section for $q_T\sim m$ in \eq{factorization_intermediate_qT} and $\df\sigma_{q_T\ll m}$ the cross section
for $q_T \ll m$ from \eq{factorization_small_qT}.

In the limit $q_T \ll m$, we can integrate out all massive modes, while a soft
or collinear emission at the scale $q_T$ cannot resolve the massive loop.
Therefore, the $(q_T, m)$-dependent NLP beam and soft functions must reduce
in this limit to an $m$-dependent factor times the standard massless $q_T$ beam
and soft functions,
\begin{align}\label{eq:BS_smallkT}
B_{n,\bar\chi\chi}(\xi,q_T,m,x_a)
&= F_n(\xi,m)\,B_g(q_T, x_a) \biggl[1 + \ORd{\frac{q_T^2}{m^2}}\biggr]
\,,\nn\\
S_{\bar\psi\psi}(\ell^+,\ell^-,q_T,m)
&= F_s(\ell^+,\ell^-,m)\,S_{gg}(q_T) \biggl[1 + \ORd{\frac{q_T^2}{m^2}}\biggr]
\,.\end{align}
This is analogous to the leading-power beam and soft functions with
massive secondary quarks in \refcite{Pietrulewicz:2017gxc}, which similarly factorize
in this limit into collinear and soft mass-mode functions times the standard massless
beam and soft functions.
The difference is that in that case, the cross section factorizes
multiplicatively and the mass effects only appear at higher orders.
In our case here, the mass effects already appear at the lowest nontrivial order,
leading to the additive NLP factorization structure and with the additional
complication of the endpoint divergences.

Here, the collinear mass-mode function $F_n(\xi,m)$ arises from integrating out
the massive collinear quarks in the $\cO_{bbg}$ operator. Likewise,
$F_s(\ell^+,\ell^-,m)$ arises from integrating
out the massive soft quarks in $\cO_{bbgg}$. Comparing \eq{BS_smallkT}
to the definitions in \eq{NLP_beam_soft}, we have the formal matching relation
\begin{align}
\bar\chi_{n,\omega\xi}\,\frac{\slashed{\nb}}{2}T^a\gamma_\perp^\mu\,\chi_{n,\omega(1-\xi)}
\to F_n(\xi,m)\, \cB_\perp^{a,\mu}
\,,\qquad
\cO^{ab,\mu\nu}_{s,\bar\psi\psi}
\to F_s(\ell^+,\ell^-,m)\, \cO_s^{(0)\,ab,\mu\nu}
\,.\end{align}
Color conservation and invariance under rotations and parity in the transverse plane allow only a single structure in each case, $\delta^{ab}\varepsilon_\perp^\mu$ and $\delta^{ab}g_\perp^{\mu\nu}$, respectively, so that $F_n$ and $F_s$ are indeed Lorentz-scalar functions.
Note that since we do not yet know the proper all-order renormalization and
cancellation of endpoint divergences at the level of the individual functions,
we will consider \eq{BS_smallkT} as the all-order
definition of $F_n$ and $F_s$ at both the bare and renormalized level.
At lowest order, $F_n$ and $F_s$ contain the massive $n$-collinear and soft bottom-quark loops
of \fig{modesLO0}, collected in \app{LO0-calc}. At higher orders, they contain the
virtual corrections to the massive loop as well as the effects of secondary massive quark
bubbles.

Now, using \eq{BS_smallkT} to take the $q_T\ll m$ limit of $\df\sigma$ and imposing
\eq{sigma_consistency} yields the consistency relation
\begin{align}\label{eq:consistency}
2\,\Re\big[C^*_{ggt}\,C_{ggb}(m,m_H)\big]
&= H_{gg}(m_H)+\int\!\df\xi\,H_{bbg}(\xi)\big[F_n(\xi,m)+F_{\nb}(\xi,m)\big]
\nn \\
& \quad
+ \int\!\df\ell^+\df\ell^-\,H_{bbgg}\,\cJ(\ell^+)\cJ(\ell^-)\,F_s(\ell^+,\ell^-,m)
\,.\end{align}
At lowest order, it corresponds to the decomposition of the massive form factor into hard, collinear, and soft contributions in \app{LO0-calc}. The terms on the right-hand side are individually endpoint-divergent
and \eq{consistency} holds with a common regulator.

As we have seen in \secs{collinearemission}{softemission},
in the sum of all diagrams with a single real
emission, the cancellation of
endpoint divergences in the $\xi$ and $\ell^\pm$ integrals
involves the lowest-order soft or collinear quark loops times a leading-power emission. The
lowest-order loops are fully contained in $F_n$ and $F_s$, which we can thus exploit
to isolate the LO$_0$-like endpoint divergences, effectively putting them in
a form such that their cancellation happens as on the
right-hand side of \eq{consistency}. To do so, we subtract the
$q_T \ll m$ limit from the NLP beam and soft functions and define
\begin{align} \label{eq:DeltaB_def}
\Delta B_{n,\bar\chi\chi}(\xi,q_T,m,x_a)
&= B_{n,\bar\chi\chi}(\xi,q_T,m,x_a) - F_n(\xi,m)\, B_g(q_T, x_a)
\,,\\ \label{eq:DeltaS_def}
\Delta S_{\bar\psi\psi}(\ell^+,\ell^-,q_T,m)
&= S_{\bar\psi\psi}(\ell^+,\ell^-,q_T,m)-F_s(\ell^+,\ell^-,m)\,S_{gg}(q_T)
\,.\end{align}
We will refer to $\Delta B_{n,\bar\chi\chi}$ and $\Delta S_{\bar\psi\psi}$ as the nonseparable
NLP beam and soft functions, as they contain the genuine, nontrivial dependence on $q_T/m$,
which in general will not separate into a product of a function of $m$ and a function of $q_T$.
By construction, they vanish for $q_T\ll m$ and at lowest order in $\alpha_s$.

Similarly, we can write the full cross section for $q_T \sim m$ as
\begin{align}
\df\sigma &= \df\sigma_{{q_T} \ll m} + \df\Delta \sigma
\,,\end{align}
which implicitly defines $\df\Delta\sigma = \df\sigma - \df\sigma_{{q_T} \ll m}$.
We can think of $\df\Delta\sigma$ as the ``nonsingular'' correction in $q_T/m$
which restores the full dependence on $q_T/m$ when transitioning from $q_T \ll m$
to $q_T \sim m$.
Using \eqs{DeltaB_def}{DeltaS_def}, we have
\begin{align}\label{eq:delta_sigma_def}
\df\Delta\sigma
&= \int\!\df \xi\, H_{bbg}(\xi) \Bigl[\Delta B_{n,\bar{\chi}\chi}(\xi,q_T,m,x_a)\otimes B_g(q_T, x_b)\otimes S_{gg}(q_T)+ (n, x_a \leftrightarrow \nb, x_b )\Bigr]
\nn \\ & \quad
+ \int \df\ell^+\df\ell^-\, H_{bbgg}\mathcal{J}(\ell^+)\mathcal{J}(\ell^-)\,B_g(q_T, x_a)\otimes B_g(q_T, x_b)\otimes \Delta S_{\bar{\psi}\psi}(\ell^+,\ell^-,q_T,m)
\,.\end{align}
The contribution to $\df\sigma$ proportional to $H_{gg}(m_H)$, where the $b$-quark loop is hard,
does not appear in \eq{delta_sigma_def} because it is already fully contained in $\df\sigma_{q_T\ll m}$.
While the subtractions in \eqs{DeltaB_def}{DeltaS_def} remove all endpoint divergences
in the $q_T \ll m$ limit, we have no reason to believe that this accounts
for all endpoint divergences for $q_T \sim m$.
A priori, the endpoint-divergent contributions could have
nontrivial $q_T/m$ dependence, such that there are endpoint divergences in each of
the integrals in \eq{delta_sigma_def}, which only cancel in the
sum of terms in $\df\Delta\sigma$ (since that must be the case separately
for $\df\sigma$ and $\df\sigma_{q_T\ll m}$).

As we will discuss in more detail in the next subsection, from our explicit
calculations we observe that this does not happen
at $\ord{\alpha_s}$. At this order, all endpoint divergences for $q_T \sim m$
are accounted for by the $q_T \ll m$ limit, which is an interesting and nontrivial
outcome of our analysis.
However, we stress that this \emph{does not} imply that $\df\Delta\sigma$ is free
of endpoint divergences to all orders.

Using the $q_T\ll m$ limit to subtract the LO$_0$-like massive loops
(as opposed to, e.g., just subtracting the lowest-order terms) has the further
advantage that the subtraction is a scale-independent cross section to all orders.
Hence, $\df\Delta\sigma$ itself must also be independent of the renormalization and
rapidity scales to all orders, i.e.,
\begin{align}\label{eq:delta_sigma_RGinv}
\mu\frac{\df}{\df\mu}\,\df\Delta\sigma = 0\,,\qquad \nu\frac{\df}{\df\nu}\,\df\Delta\sigma = 0
\,.\end{align}
This implies a nontrivial consistency relation between the anomalous dimensions of
$\Delta B_{n,\bar\chi\chi}$ and $\Delta S_{\bar\psi\psi}$.
In \refcite{Pietrulewicz:2017gxc}, the secondary quark-mass effects caused a
nontrivial $q_T/m$-dependent rapidity anomalous dimension at the order they first
appear.
Similarly here, $\Delta B_{n,\bar\chi\chi}$ and $\Delta S_{\bar\psi\psi}$ can
be expected to have rapidity divergences with a nontrivial dependence on $q_T/m$ and
associated $q_T/m$-dependent rapidity anomalous dimensions $\Delta\gamma_\nu(q_T/m)$,
which can only cancel between them, as the remaining
ingredients in \eq{delta_sigma_def}
do not depend on $m$.
A priori, $\Delta\gamma_\nu(q_T/m)$ could be nonzero starting at $\ord{\alpha_s}$,
which would cause NLL logarithms of $q_T/m_H$ with a nontrivial dependence on $q_T/m$.
However, we will see in the next subsection that
$\Delta\gamma_\nu(q_T/m)$
vanishes at $\ord{\alpha_s}$.

\subsection{Rapidity anomalous dimension at NLO}
\label{sec:rapidityadm}

At $\ord{\alpha_s}$, $\Delta B_{n,\bar\chi\chi}$ and $\Delta S_{\bar\psi\psi}$
only receive single real-emission contributions, while the
virtual corrections are fully removed by the subtractions in \eqs{DeltaB_def}{DeltaS_def}.
Hence, we can obtain them by interfering the one-gluon real-emission amplitudes computed in \secs{collinearemission}{softemission}
with the one-gluon real-emission amplitude of the standard leading-power operator generated by
the top-quark loop. For the beam functions, we need the amplitude for
a collinear emission from the collinear gluon field, while for the soft function we need
the soft-emission amplitude from the soft Wilson lines. These are given by
\begin{align}\label{eq:MLP_def}
\cA^{n}_{\rm LP} &= \bra{g_n(k)}\,g\,\mathcal{B}_{n\perp}\,\ket{g_n(p_1)}
\nn \\
&= \img g f^{abc}\bigg\{
-\frac{z(3-z)}{2\omega_n(1-z)}\,g_\perp^{\mu\nu}\nb^\rho
+\frac{\omega_n(1-z^2)}{2z\,k_T^2}\,g_\perp^{\mu\nu}n^\rho\nn \\
&\qquad\qquad
-\frac{1-z}{k_T^2}\Big[g_\perp^{\mu\nu}k_\perp^\rho+\frac{2}{z}\,g_\perp^{\mu\rho}k_\perp^\nu-2\,g_\perp^{\nu\rho}k_\perp^\mu-\frac{2}{\omega_n}\,k_\perp^\mu k_\perp^\nu\nb^\rho\Big]\bigg\}
\,,\nn \\
\cA^{s}_{\rm LP} &= \bra{g_s(k)}\,S_n^\dagger S_{\nb}\,\ket{0}\,g_\perp^{\mu\nu}
= \img g f^{abc}\Big(\frac{n^\sigma}{k^+}-\frac{\nb^\sigma}{k^-}\Big)g_\perp^{\mu\nu}
\,,\end{align}
and the mirror image of $\cA^{n}_{\rm LP}$ for an $\nb$-collinear gluon.

On the NLP side, we need the amplitude $\cA_{\bar\chi\chi}(\xi,m)$ for an
$n$-collinear emission from the $n$-collinear quark operator at fixed $\xi$ and
the amplitude $\cA_{\bar\psi\psi}(\ell^+,\ell^-,m)$ for a soft emission from
the soft-quark operator at fixed $\ell^\pm$ and without the respective hard Wilson
coefficients. Since the one-gluon matrix elements $\cM$ calculated in
\secs{collinearemission}{softemission} include the Wilson coefficients and are
integrated over $\xi$ and $\ell^\pm$, they are related by
\begin{align}
\cM_{\bar\chi\chi}(m) &= \int\df\xi\,C_{bbg}(\xi)\,\cA_{\bar\chi\chi}(\xi,m)
\,, \nn \\
\cM_{\bar\psi\psi}(m) &= \int\df\ell^+\df\ell^-\,C_{q\bar q}^S\,\cJ(\ell^+)\cJ(\ell^-)\,\cA_{\bar\psi\psi}(\ell^+,\ell^-,m)
\,.\end{align}
For the single real-emission contributions, we can perform the subtractions
in \eqs{DeltaB_def}{DeltaS_def} directly at the amplitude level, with the corresponding
subtracted one-gluon amplitudes given by
\begin{align}\label{eq:DeltaA}
\Delta\cA_{\bar\chi\chi}(\xi,m)
&= \cA_{\bar\chi\chi}(\xi,m) - F_n^{(0)}(\xi,m)\,\cA^{n}_{\rm LP}
\,,\nn\\
\Delta\cA_{\bar\psi\psi}(\ell^+,\ell^-,m)
&= \cA_{\bar\psi\psi}(\ell^+,\ell^-,m) - F_s^{(0)}(\ell^+,\ell^-,m)\,\cA^{s}_{\rm LP}
\,,\end{align}
where $F_{n,s}^{(0)}$ are the $\ord{\alpha_s^0}$ contributions given by the
LO$_0$ $n$-collinear and soft quark loops.

In terms of the diagram classes of \sec{collinearemission}, for
$\Delta\cA_{\bar\chi\chi}(\xi ,m)$ we have explicitly
\begin{align}\label{eq:DeltaMn_def}
\int\!\df \xi \, C_{bbg}(\xi)\, \Delta\cA_{\bar\chi\chi}(\xi ,m)
&= \Bigl(\cM_{d_{LE}}+\cM_{d_{VE}^{n}}+\cM_{d_{VE}^{qqgg}}+\cM_{d_{ggg}^{n}}\Bigr)
\nn \\ & \quad
-\cA^{n}_{\rm LP} \int\!\df \xi \,C_{bbg}(\xi)\, F^{(0)}_n(\xi,m)
= \Ord{\eps^0, \eta^0}
\,.\end{align}
From our calculation, we have verified explicitly that the right-hand side is free
of both $1/\eta$ and $1/\eps$ divergences.
As we saw in \sec{endpointdiv}, the $1/\eta$ endpoint divergences of the diagrams
involving an $n$-collinear loop in \eq{coll-LO1} cancel against those of the
LO$_0$ soft and $\nb$-collinear loops times $\cA^n_{\rm LP}$ in \eq{coll-LO0}.
The latter are in turn precisely
minus those of the subtraction term thanks to the LO$_0$ cancellation on the right-hand
side of \eq{consistency}. As a result, all endpoint divergences cancel on the right-hand side
of \eq{DeltaMn_def}. Thus, also the $\xi$ integral on the left-hand side is free of
endpoint divergences.%
\footnote{%
On the left-hand side this cancellation is nontrivial and not immediately manifest
at the level of the $\xi$ integrand, because the endpoint divergences from the central
region of the $\xi$ integral, $\cR^0$, cancel against contributions from the outer
regions, $\cR^\pm$.}
Similarly, $F_n(\xi,m)$ carries the local ultraviolet divergence of the collinear loop,
and the subtraction removes exactly this piece, so that \eq{DeltaMn_def} also has
no $1/\eps$ divergences.

For the case of a soft emission, in
terms of the diagrams in \sec{softemission} we have
\begin{align}\label{eq:DeltaMs_def}
& \int\df \ell^+ \df \ell^- \,C_{q\bar q}^S \cJ(\ell^+)\cJ(\ell^-)\Delta\cA_{\bar\psi\psi}(\ell^+, \ell^-,m)
\nn \\ & \quad
= \cM_{s,L}^{\mloone} + \cM_{s,V}^{\mloone}
-\cA^{s}_{\rm LP} \int\!\df \ell^+\df\ell^- \,C_{q\bar q}^S \cJ(\ell^+)\cJ(\ell^-) F^{(0)}_s(\ell^+,\ell^-,m)
\nn \\ & \quad
= -\frac{\img g^3 f^{abc} m}{8\pi^2}\,\, \frac{	\left( g_\perp^{\mu\sigma} k_\perp^{\nu} - g_\perp^{\nu\sigma} k_\perp^{\mu} \right)}{{k_T}\sqrt{k_T^2+4m^2}}
\Biggl[ \mathrm{Li}_2\Biggl( \frac{\sqrt{k_T^2+4m^2}-k_T}{\sqrt{k_T^2+4m^2}+k_T} \Biggr)
- \mathrm{Li}_2\Biggl( \frac{\sqrt{k_T^2+4m^2}+k_T}{\sqrt{k_T^2+4m^2}-k_T} \Biggr) \Biggr]
\,.\end{align}
Here, the subtraction term is precisely equal to $\cM_s^{\mlozero}$ times
the LP Wilson line in \eq{Msoft_cancellation}. Analogous to the collinear case and as written in the
last line of \eq{Msoft_cancellation}, the divergences in the soft-loop LO$_1$ diagrams
cancel against those of the collinear and hard contributions times the LP soft Wilson-line
emission, where the latter are equal to minus those of the subtraction term thanks to the
LO$_0$ cancellation.
As discussed in \sec{soft_intcancel}, the structure and cancellation of the endpoint divergences
at the level of the $\ell^\pm$ integrands is quite intricate. The $\ell^-\to0$ divergences
cancel between the two support regions of the loop-emission diagram. The divergences at $\ell^-\to-\infty$ cancel between the loop- and vertex-emission diagrams combined with the subtraction term.

The $\ord{\alpha_s}$ nonseparable beam and soft functions now follow by interfering the subtracted amplitudes with the leading-power amplitudes in \eq{MLP_def},
\begin{align}\label{eq:DeltaBn_amp}
\Delta B_{n,\bar\chi\chi/g}^{(1)}(\xi,q_T,m,z)
&= \left(\cA^{n}_{\rm LP}\right)^* \Delta\cA_{\bar\chi\chi}(\xi ,m)
\,,\nn \\
{\Delta S^{(1)}_{\bar\psi\psi}}(\ell^+,\ell^-,q_T,m)
&= {\frac{1}{2(2\pi)^{d-1}}\int_{-\infty}^{\infty}\df y_k}\left(\cA^{s}_{\rm LP}\right)^*\Delta\cA_{\bar\psi\psi}(\ell^+, \ell^- ,m)\Big|_{k_T=q_T}\,,
\end{align}
where the sum over the polarizations and colors of the emitted gluon is implied.
The beam function here is the partonic beam function, which is obtained from \eq{NLP_beam_soft} by replacing the hadron state $\ket{N}$ by an incoming on-shell gluon with momentum $p_1$. Thus, its last argument is the partonic momentum fraction $z = \omega_n/p_1^-$ relative to the incoming gluon momentum rather than the hadronic momentum fraction $x$. For the gluon-initiated channel considered here, the hadronic beam function entering \eq{delta_sigma_def} follows from the convolution with the gluon PDF $f_g$,
\begin{align}\label{eq:DeltaB_hadronic}
\Delta B_{n,\bar\chi\chi}^{(1)}(\xi,q_T,m,x)
= \int_{x}^1 \frac{\df z}{z}\, \Delta B^{(1)}_{n,\bar\chi\chi/g}(\xi,q_T,m,z)\, f_g\Big(\frac{x}{z}\Big) + \ord{\alpha_s^2}
\,.
\end{align}
Since $\Delta B^{(1)}_{n,\bar\chi\chi}$ is free of infrared divergences, see \eq{DeltaMn_def}, the partonic result at this order directly gives the matching coefficient onto the gluon PDF.
For the partonic beam function in \eq{DeltaBn_amp}, the measurement $\delta^{(2)}(\vec k_T+\vec q_T)$ together with the momentum fraction $z$ fixes the momentum of the emitted gluon completely.

For the soft function in \eq{DeltaBn_amp}, the integral over the rapidity $y_k$ of the emitted gluon remains, which in general requires a rapidity regulator.
However, \eq{DeltaMs_def} is antisymmetric under $\mu\leftrightarrow\nu$, whereas $\cA^{s}_{\rm LP}$ is proportional to $g_\perp^{\mu\nu}$. Hence,
$\Delta S_{\bar\psi\psi}$ vanishes at $\ord{\alpha_s}$,
\begin{equation}
\Delta S^{(1)}_{\bar\psi\psi}(\ell^+,\ell^-,q_T,m) = 0
\,.\end{equation}
So there are trivially no endpoint divergences at this order in the $\ell^\pm$ integrals
in \eq{delta_sigma_def}. There are also no rapidity divergences, which implies that the
rapidity anomalous dimension $\Delta\gamma_\nu(q_T/m)$ vanishes at $\ord{\alpha_s}$.

The same must then also hold for the beam function, which we can verify explicitly:
The $\xi$ integrals in \eq{delta_sigma_def} are free of endpoint divergences
at $\ord{\alpha_s}$ due to \eq{DeltaMn_def}.
Regarding the rapidity anomalous dimension, the rapidity divergence in the beam
function arises at the endpoint $z\to1$, where the emitted gluon becomes soft.
Expanding the partonic beam function $\Delta B^{(1)}_{n,\bar\chi\chi/g}$ in this limit we find that it vanishes,
\begin{align}\label{eq:DeltaB_softlimit}
\Delta B^{(1)}_{n,\bar\chi\chi/g}(\xi,q_T,m,z) = \ord{(1-z)^0} \quad\text{for } z\to 1
\,,\end{align}
i.e., there is no $1/(1-z)$ term and the $z$ integral of $\Delta B^{(1)}_{n,\bar\chi\chi/g}$ needs no rapidity regulator. Together with the absence of $1/\eps$ poles in \eq{DeltaMn_def}, this implies that
after renormalization $\Delta B^{(1)}_{n,\bar\chi\chi}$ will be $\mu$ and $\nu$ independent
at this order.

Therefore, at $\ord{\alpha_s}$, $\df\Delta\sigma$ in \eq{delta_sigma_def}
becomes
\begin{align}\label{eq:delta_sigma_NLO}
\df\Delta\sigma^{(1)}
= \int\!\df\xi\,H^{(0)}_{bbg}(\xi)
\Bigl[\Delta B^{(1)}_{n,\bar\chi\chi}(\xi,q_T,m,x_a)\otimes B_g^{(0)}(q_T, x_b)\otimes S^{(0)}_{gg}(q_T)
+ (n, x_a \leftrightarrow \bar n, x_b)\Bigr]
\,.\end{align}
At the renormalized level, its $\mu$ and $\nu$ independence, as required by
\eq{delta_sigma_RGinv}, is then trivially satisfied at this order since all
functions are $\mu$ and $\nu$ independent at the lowest order at which they
appear. Nevertheless, $\df\Delta\sigma^{(1)}$ is a nontrivial function of
$q_T/m$, which vanishes for $q_T\ll m$. In particular, $\df\Delta\sigma^{(1)}$
contains no logarithms of $m/m_H$ or $q_T/m_H$. This is a consequence of the
subtraction in \eq{DeltaMn_def}: we have verified with the explicit amplitudes
of \sec{collinearemission} that the subtraction of $F_n(\xi,m)\,\cA^n_{\rm LP}$
removes all terms proportional to $\log^2(-m_H^2/m^2)$ that appear in the
individual helicity amplitudes, so that $\Delta\cA_{\bar\chi\chi}$ depends
on the bottom-quark mass only through the ratio $k_T/m$.

Concerning the joint resummation of $q_T$ and $m$, this means that $\df\Delta\sigma^{(1)}$
effectively acts as a nonlogarithmic $\ord{\alpha_s}$ boundary term, which formally enters at NLL$'$.
This is analogous to the role played by the primary mass effects in \refcite{Pietrulewicz:2017gxc}, where the massive beam-function coefficient at one loop provides a nonlogarithmic boundary constant but with
a nontrivial $q_T/m$ dependence, which is free of rapidity divergences and decouples for $q_T\ll m$.
We leave the implementation of this joint resummation to future work.

\section{Conclusions}
\label{sec:ybytsummary}

In this paper, we studied finite bottom-quark mass effects in the Higgs $q_T$ spectrum arising from the interference between the top- and bottom-quark loops in gluon fusion.
Although formally suppressed by $m/m_H$, the bottom-quark loop contributes of $\cO(5\%)$ to the gluon-fusion cross section and is therefore relevant for precision predictions of the spectrum, e.g., when using
its shape to constrain the Higgs Yukawa couplings to the quarks.
In contrast to the mass-dependent $gg \to H$ form factor and $H\to\gamma\gamma$ amplitude studied previously, gluon emissions that the Higgs boson recoils against can resolve the bottom-quark loop, introducing
$q_T$ as another dynamical scale.

We formulated factorization theorems in SCET for the kinematic regimes $q_T \ll m \ll m_H$ and $q_T \sim m \ll m_H$.
In the first regime, the bottom quark can be integrated out and its mass dependence is entirely contained in the form factor $C_{ggb}(m, m_H)$ of \refcite{Liu:2022ajh}, multiplying the leading-power $q_T$ beam and soft functions.
In the phenomenologically most relevant regime $q_T \sim m$, where the gluon-fusion spectrum peaks, the mass dependence is carried in addition by new NLP $q_T$- and $m$-dependent beam and soft functions, containing a collinear or a soft bottom-quark loop, respectively.

We computed all amplitudes with a single collinear or soft emission that enter these functions at
the next-to-leading order in $\alpha_s$.
For the collinear emission, the endpoint divergences of the diagrams in which the gluon is emitted from the NLP beam function cancel against those of leading-power emissions multiplying the emission-free (LO$_0$) collinear and soft contributions.
Since the additional scale allows the coefficients of the endpoint divergences to depend on $m/k_T$, this cancellation point-by-point in $k_T$ is a nontrivial statement: the endpoint-divergent part factorizes into the LO$_0$ contributions times a leading-power emission, while the finite remainder retains a nontrivial dependence on $m/k_T$.
The remaining $\eps$ poles cancel once the hard contributions are included, and the Ward identities are satisfied for each operator separately.
Glauber operator insertions contribute nontrivially to the collinear emission, while for the soft dynamics we saw
that the Glauber insertions effectively
set the $\img 0$ prescriptions in various eikonal propagators.
Since the $\img 0$ prescriptions affect the regions $\ell^{\pm} \to 0$
which feature endpoint divergences,
we anticipate that these insights may also lead to broader lessons
regarding their field-theoretic treatment
and their connection to Glauber physics.
For the soft emission, each contributing diagram already satisfies the Ward identity individually.
Instead of regularizing each diagram separately, which is computationally challenging,
we combined all diagrams at loop integrand level to obtain a finite closed-form result, without the need of an explicit regulator.

We validated our results against the full QCD helicity amplitudes of \refcite{Baur:1989cm}.
We find full analytic agreement at leading power in both the collinear and the soft limit, where the agreement in the collinear limit requires the Glauber contribution.
The EFT amplitudes reproduce the singular behavior of the full theory, including the large logarithms of $m_H/m$, with a nonsingular remainder suppressed by at least one additional integer power of $\la$. In addition, the collinear amplitudes reduce to the soft ones for $z\to 1$, which completes a three-way consistency check.

We also discussed the implications of our results for the factorization theorem
and the consistency between the $q_T\ll m$ and $q_T\sim m$ limits.
Due to the structure of endpoint divergences we found at this order, it is possible
to arrange their cancellation in the factorization theorem in terms of the mass-dependent
form factor and a remainder, $\df\Delta\sigma^{(1)}$, that encodes the nontrivial
dependence on $q_T\sim m$ and is free of endpoint divergences.
An interesting open question for future study is whether
or not this will persist at higher orders. We further show that the structure
of the factorization theorem allows for a nontrivial $q_T/m$-dependent correction to the
standard $q_T$ rapidity anomalous dimension, which turns out to still vanish
at $\ord{\alpha_s}$, but presumably starts contributing at the next higher order.

Several steps remain towards a complete resummed prediction of the $y_b$-$y_t$-interference
contribution to the Higgs $q_T$ spectrum.
We already know that $\df\Delta\sigma$ will start contributing at least at NLL$'$
in the joint resummation. Its resummation
is however technically more involved than the usual leading-power $q_T$ resummation due to the additional convolutions in $\xi$ and $\ell^\pm$ in the factorization theorem. It also requires the renormalization of the NLP beam and soft functions beyond the order considered here, including the question whether an additive rapidity renormalization or further endpoint divergences arise at higher orders.
Furthermore, factorization theorems for the regimes $m \ll q_T \ll m_H$ and $m \ll q_T \sim m_H$ still need to be worked out for a complete description over the entire $q_T$ spectrum.
To that end, our results here provide an important constraint, as they must
be consistent with the description of the $m \ll q_T$ limit.
We leave the simultaneous resummation of mass and $q_T$ logarithms and the matching across the different kinematic regimes to future work.

\acknowledgments
We used generative AI tools (Claude Opus 5.5) to improve the language of the original manuscript,
to cross-check typeset equations against our results, and to simplify expressions and obtain
closed forms for the finite pieces in the auxiliary files from our own results.
All calculations were performed by the authors, and all AI-generated expressions
were verified against our original results.
We thank Aneesh V. Manohar  and Bruno Scheihing-Hitschfeld for helpful discussions.
The authors gratefully acknowledge the Erwin-Schrödinger International Institute for Mathematics and Physics at the University of Vienna for providing a wonderful space for discussions and partial financial support during the Programme "New Paradigms for Harnessing Quantum Field Theory at Colliders", July 27 -- August 28, 2026.
This project has received funding from the European Research Council (ERC) under
the European Union’s Horizon 2020 research and innovation programme (Grant agreement
101002090 COLORFREE).
I.S. was supported in part by the Office of Nuclear Physics of the U.S. Department of Energy under the Grant No.~DE-SC0011090.
Z.S. was supported in part by Department of Energy award DESC0025293 and National Science Foundation grant PHY-2515115.

\appendix
\addtocontents{toc}{\string\setcounter{tocdepth}{1}}
\bookmarksetup{depth=1}
\section{Leading-order results}
\label{app:LO0-calc}
\subsection{Leading-order collinear contribution}
\label{app:ybyt-collLO0}
\begin{figure}
	\begin{subfigure}{0.25\textwidth}
		\includegraphics[width=\textwidth]{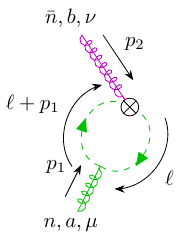}
	\end{subfigure}%
	\centering
	\caption{Leading-order $n$-collinear contribution.}
	\label{fig:collinearLO0}
\end{figure}

The $n$-collinear leading-order diagram is shown in \fig{collinearLO0}. We use massive SCET Feynman rules from \refcite{Leibovich:2003jd} and our Feynman rule for the effective vertex with the Higgs boson in \eq{LOcFR}.
We start by solving the $\omega_n$ and $\ell^-$ integrals that are fixed by $\delta$-functions
\begin{align}
	\mathcal{M}_c^{\mathrm{LO}}=&\int \df \xi \int \frac{\df \ell^+ \df^{d-2} \ell_\perp}{(2 \pi)^{d-1}}
	\frac{\img g^2 C_{bbg}(\xi) \omega_n\delta^{ab} m\mu^{2 \eps} g_\perp^{\mu\nu}}
	{4 \pi(\ell_\perp^2 +\omega_n \xi \ell^+-m^2 +\img0)\left[\ell_\perp^2+ \omega_n (\xi-1) \ell^+ -m^2 +\img0\right]}.
\end{align}
The $\xi$ integral will be divergent due to endpoint divergences.
These divergences only cancel once the collinear and soft contributions are combined.
We will regulate these divergences with two different regulators: the absolute-value or $\eta$-regulator~\cite{Chiu:2011qc,Chiu:2012ir,Echevarria:2011epo}
and the pure rapidity regulator~\cite{Ebert:2018gsn}. For the absolute-value regulator it is sufficient to include the regulator just before integrating over $\xi$. The pure rapidity regulator, on the other hand, depends on $\ell^-$ and $\ell^+$ and has to be included before performing the $\ell^+$ integral.
\subsubsection{Absolute-value regulator}

Then the final result for the $n$-collinear contribution is given by
\begin{align}
	\mathcal{M}_c^{\mathrm{LO}}
	=& -\frac{g^2}{(4\pi) ^{2}} \delta^{ab}  g_\perp^{\mu\nu}e^{\eps \gamma_E}\mu^{2 \eps} m^{1-2\eps} \Gamma(\eps)\int_0^1 \df \xi\left(\frac{1}{\xi}\Big|\frac{\omega_n\xi}{\nu}\Big|^{-\eta}+ \frac{1}{1 -\xi} \Big|\frac{\omega_n(1-\xi)}{\nu}\Big|^{-\eta}\right) \nn \\
	=&\frac{g^2}{(4\pi) ^{2}}  \delta^{ab}  g_\perp^{\mu\nu}e^{\eps \gamma_E}\mu^{2 \eps} m^{1-2\eps} \left| \frac{\omega_n}{\nu}\right| ^{-\eta} \frac{2\Gamma(\eps)}{\eta}.
\end{align}
For this regulator, the $\nb$-collinear diagram gives the same result as the $n$-collinear diagram. We therefore  simply multiply our final result by a factor of 2
\begin{align}
	\mathcal{M}_c^{\mathrm{LO}}+\mathcal{M}_{\bar{c}}^{\mathrm{LO}}=&
	\frac{g^2}{(4\pi) ^{2}} \delta^{ab}  g_\perp^{\mu\nu}e^{\eps \gamma_E}\mu^{2 \eps} m^{1-2\eps} \left| \frac{\omega_n}{\nu}\right| ^{-\eta} \frac{4\Gamma(\eps)}{\eta}.
\end{align}
\subsubsection{Pure rapidity regulator}
We use the pure rapidity regulator~\cite{Ebert:2018gsn} in the analytic form $w^2\upsilon^{\eta}\big[(\ell^-+\img0)/(\ell^++\img0)\big]^{-\eta/2}$, acting on the quark line that becomes soft at the endpoint, i.e.\ on $\ell$ for the $1/(1-\xi)$ term of $C_{bbg}(\xi)$; the $1/\xi$ term follows by charge conjugation and gives the same contribution.
Since the branch cut of the regulator in $\ell^+$ lies in the lower half plane, we close the $\ell^+$ contour in the upper half plane.
The $\ell^+$ integral then no longer vanishes for $\xi<0$, while it still vanishes for $\xi>1$.
Performing the $\ell^+$, $\ell_\perp$ and $\xi$ integrals we find
\begin{align}\label{eq:PRcollresLOn}
	\mathcal{M}_{c\text{ p}\eta}^{\mathrm{LO}} = \frac{g^2}{8\pi^2}\,\delta^{ab} g_\perp^{\mu\nu}\, e^{\eps\gamma_E}\mu^{2\eps} m^{1-2\eps}\,\upsilon^{\eta}
	\Big(\frac{\omega_n^2}{m^2}\Big)^{-\frac{\eta}{2}}\,\frac{\Gamma(\eps-\frac{\eta}{2})\,\Gamma(\eta)}{\Gamma(1+\frac{\eta}{2})}\,,
\end{align}
Here the subscript $\text{ p}\eta$ indicates that the results were obtained using the pure rapidity regulator.
Since the regulator is not invariant under $\ell\to-\ell$, the $\bar{n}$-collinear contribution cannot be obtained from the $n$-collinear result by $\eta \to -\eta$.
Instead it has to be computed with the same loop momentum, which gives
\begin{align}\label{eq:PRcollresLOnb}
	\mathcal{M}_{\bar{c}\text{ p}\eta}^{\mathrm{LO}} = -\frac{g^2}{8\pi^2}\,\delta^{ab} g_\perp^{\mu\nu}\, e^{\eps\gamma_E}\mu^{2\eps} m^{1-2\eps}\,\upsilon^{\eta}
	\Big(\frac{\omega_{\bar n}^2}{m^2}\Big)^{\frac{\eta}{2}}\,\Gamma\Big(\eps+\frac{\eta}{2}\Big)
	\bigg[\frac{e^{-\img\pi\eta/2}\,\Gamma(\frac{\eta}{2})}{\Gamma(1+\eta)}+\frac{\Gamma(-\eta)}{\Gamma(1-\frac{\eta}{2})}\bigg]\,,
\end{align}
In the sum of the $n$- and $\bar{n}$-collinear contributions the $1/\eta$ poles from the endpoint divergences cancel, and taking $\eta\to0$ we obtain
\begin{align}\label{eq:PRcollresLO}
	\mathcal{M}_{c\text{ p}\eta}^{\mathrm{LO}}+\mathcal{M}_{\bar{c}\text{ p}\eta}^{\mathrm{LO}}
	&= - \frac{g^2}{8\pi^{2}}\, \delta^{ab} g_\perp^{\mu\nu}\, e^{\eps \gamma_E} \mu^{2 \eps} m^{1-2\eps}\, \Gamma(\eps)
	\left[ \psi^{(0)}(\eps)+\gamma_E +\log \frac{-m_H^2-\img0}{m^2}\right] \nn \\
	&= \frac{g^2 m\,\delta^{ab} g_\perp^{\mu\nu}}{8\pi^2}\bigg[\frac{1}{\eps^2}-\frac{1}{\eps}\Big(\log\frac{m_H^2}{\mu^2}-\img\pi\Big)
	\nn \\ &\qquad
	+\frac12\log^2\frac{\mu^2}{m^2}-\log\frac{\mu^2}{m^2}\Big(\log\frac{m_H^2}{m^2}-\img\pi\Big)-\frac{\pi^2}{12}\bigg]\,,
\end{align}
where $\psi^{(0)}$ is the polygamma function of order 0 and $m_H^2=\omega_n\omega_{\bar n}$.
\subsection{Soft LO$_0$ }
\label{app:ybytsoftLO}
\subsubsection{Absolute-value regulator}
\begin{figure}
	\centering
	\includegraphics[width=0.2\textwidth]{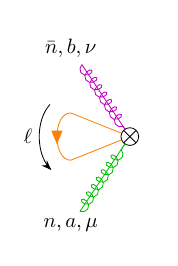}
	\caption{Leading-order soft contribution.}
	\label{fig:softLO0}
\end{figure}
\noindent
The soft leading-order diagram is shown in \fig{softLO0}.
Using the same Feynman rules as before, this diagram without any rapidity regulator is
\begin{align}
\label{eq:softLO0intnoreg}
\mathcal{M}^{s}_{\mathrm{LO}_{0}}
= - 2\,\delta^{ab} g_{\perp}^{\mu\nu}   \img g^{2} m \int \frac{\df^{d} \ell}{(2\pi)^d}
\frac{1} {(\ell^{+}+\img 0) (\ell^{-} - \img 0)} \frac{1}{\ell^{+}\ell^{-} + \ell_{\perp}^{2} - m^{2} + \img 0}
\,,\end{align}
where the factor of two accounts for the two orientations of the fermion loop, i.e.\ the two terms in \eq{LOsFR}, which give the same contribution.
To regulate the rapidity divergence with the absolute-value regulator,
we include a factor of $\big\vert \frac{2\ell_{z}}{\nu}\big\vert^{-\eta}$
and use integration variables $\df \ell^{0} \df \ell^{z} \df^{d-2} \ell_{\perp}$.
We first perform the analytic $\ell^{0}$ integral by contours,
then perform the $\ell^{z}$ integral, treating the $\pm \img 0$ prescriptions carefully.
Finally, after performing the $\ell_{\perp}$ integral, we get
\begin{align}
	\mathcal{M}^{s}_{\mathrm{LO}_{0}}
	=& - \delta^{ab} g_{\perp}^{\mu\nu} \frac{g^{2}m }{8\pi^{2}}
	\times
	\Bigg[
		\frac{ \Gamma(\frac12 - \frac\eta2) \Gamma(\frac\eta2) \Gamma(\frac\eta2 + \eps)}{\sqrt\pi \Gamma(1+\frac \eta2)}
		\bigg(\frac{2m}{\nu}  \bigg)^{-\eta} \bigg( \frac{m^{2} }{ \mu^{2} e^{\gamma_{E}}}\bigg)^{-\eps}
	\nn \\
	& \qquad\qquad\qquad
  - \Gamma(\eps) \bigg( \frac{m^{2} }{ \mu^{2} e^{\gamma_{E}}}\bigg)^{-\eps}  \bigg(\frac{\nu}{2}  \bigg)^{\eta} \pi \csc(\pi \eta) \Big[ (-\img 0)^{-\eta} - (\img 0)^{-\eta} \Big]
	\Bigg ]
	\label{eq:softLO0abs}
\,. \end{align}
Expanding in $\eta$ first, then $\eps$,
we get
\begin{empheq}{align}
	\mathcal{M}^{s}_{\mathrm{LO}_{0}}
	&= \delta^{ab} g_{\perp}^{\mu\nu} \frac{g^{2}m }{8\pi^{2}}
	\Bigg[ \frac{1}{\eps^{2}} - \frac{2}{\eps \eta}
	+ \frac{1}{\eps} \bigg[ \log \bigg( \frac{\mu^{2}}{\nu^{2}} \bigg) + \img \pi\bigg]
	\nn \\
	& \qquad
	+ \frac{2}{\eta} \log \bigg( \frac{m^{2}}{\mu^{2}} \bigg) + \frac12 \log^{2} \bigg( \frac{m^{2}}{\mu^{2}} \bigg)
	- \log \bigg( \frac{m^{2}}{\mu^{2}} \bigg) \bigg[ \log \bigg( \frac{m^{2}}{\nu^{2}} \bigg) + \img \pi \bigg]
	- \frac{\pi^{2}}{12}
	\Bigg].
\end{empheq}

\subsubsection{Glauber contribution}
At LO$_0$ there is also a contribution  where the quark exchanged between the two gluon vertices has Glauber scaling, $\ell\sim(\lambda^2,\lambda^2,\lambda)$.
At leading power the $n$-collinear propagator depends only on $\ell^+$ and the $\nb$-collinear one only on $\ell^-$,
\begin{align}
	\cM_G = 2\img g^2 m\, \omega_n^2\, \delta^{ab} g_\perp^{\mu\nu}
	\int\!\frac{\df^d\ell}{(2\pi)^d}\,
	\frac{1}{(\ell_\perp^2-m^2)\,(\ell_\perp^2-m^2-\omega_n\ell^+ +\img 0)\,(\ell_\perp^2-m^2+\omega_n\ell^- +\img 0)}\,,
\end{align}
where the factor of two accounts for the two orientations of the fermion loop.
Performing the $\ell^0$ integral by contours and the $\ell^z$ integral with the regulator $|2\ell^z/\nu|^{-\eta}$,
the eikonal propagators only enter through the signs of their $\img 0$ prescriptions~\cite{Rothstein:2016bsq} and we find for $\eta\to 0$
\begin{align}
	\cM_G = \delta^{ab} g_\perp^{\mu\nu}\frac{g^2 m}{8\pi^2}\, \img\pi\, e^{\eps\gamma_E}\,\Gamma(\eps)\left(\frac{\mu^2}{m^2}\right)^{\eps}
	= \delta^{ab} g_\perp^{\mu\nu}\frac{g^2 m}{8\pi^2}\left[\frac{\img\pi}{\eps} - \img\pi\log\frac{m^2}{\mu^2} + \ord{\eps}\right].
\end{align}
This is exactly the second term in \eq{softLO0abs}, which arises from the poles of $1/(\ell^+\ell^-)$ in the soft integrand, i.e.\ from the region $\ell^+\ell^-\ll\ell_\perp^2$.
The Glauber diagram therefore coincides with the Glauber zero-bin of the soft contribution, and since the soft integral is evaluated over the full $\ell$ space this region is already contained in $\cM_s$.
Adding $\cM_G$ separately would double count it, and the net Glauber contribution at LO$_0$ vanishes.
This is no longer the case for a collinear emission (\sec{glaubercoll}): there, the two $n$-collinear propagators in $d_G^{2a,b}$ have $\ell^+$ poles on opposite sides of the real axis,
the residue depends on the propagator offsets (and hence on $k_\perp$), and the result is not reproduced by the soft region.
\subsubsection{Pure rapidity regulator}
The soft contribution in  \fig{softLO0} can, of course, also be regulated with the pure rapidity regulator.
In this case, we start by evaluating the $\ell_\perp$-integral which can be straightforwardly done using known integral formul\ae.
The remaining integral is scaleless and the soft contribution using the pure rapidity regulator is zero.
This is expected, as the sum of the collinear contributions in this regulator is rapidity finite.
The Glauber region vanishes as well, since its $\ell^+$ integrand is analytic in the upper half plane.
\subsection{Leading-order hard contribution}
\label{app:ybytLO0-hard}
The leading-order hard contribution is given by the full QCD result minus the soft and collinear contributions.
However, we can also get this contribution from evaluating the full QCD diagram for $p_1, p_2 \gg m$ which amounts to evaluating this diagram using the method of regions \cite{Beneke:1997zp,Smirnov:1998vk,Smirnov:2002pj}.
Expanding the full QCD diagram in $m$ and performing the integrals, we find
\begin{align}
	\mathcal{M}_h= 2^{4 \eps -3} \pi^{\eps-\frac{1}{2}} g^2 m \delta^{ab} g_\perp^{\mu\nu} \frac{\csc (\pi \eps)}{\eps \Gamma(\frac{1}{2}-\eps)}\Big(\frac{\mu^2 e^{\gamma_E}}{4\pi}\Big)^{\eps}(-m_H^2 - \img 0)^{ -\eps}
	\left[\frac{2\eps^2 }{2 \eps ^2- 3 \eps+1}-1\right],
\end{align}
where we used that the incoming gluons can only have $\perp$ polarization and $p_1^-=p_2^+=\omega_n$, such that $m_H^2=(p_1+p_2)^2=\omega_n^2$. The hard contribution depends only on the boost-invariant combination $p_1^-p_2^+=m_H^2$. We work in the $\MSbar$ scheme, expand the result and define
\begin{align}\label{eq:defC1}
	\mathcal{M}_h= & \frac{g^2 m \delta^{ab}g_\perp^{\mu\nu}}{8 \pi^2} \left[-\frac{1}{\eps^2}+\frac{1}{\eps}\log\frac{-m_H^2 - \img 0}{\mu^2}-\frac{1}{2} \log^2 \frac{-m_H^2 - \img 0}{\mu^2}+2+\frac{\pi^2}{12}\right]\nn \\
	=& \frac{g^2 m \delta^{ab}g_\perp^{\mu\nu}}{8 \pi^2} C(m_H),
\end{align}
where $C(m_H)$ is a bare hard Wilson coefficient that we will use in the \loone~calculation.

\subsection{Cross checks}
We compared our results with the soft, collinear and hard contributions given in \refcite{Liu:2019oav} and find full agreement after taking into account the differences in our regulators.
Further, we checked that all endpoint divergences cancel between the soft and collinear contributions.
The remaining $\eps$ poles cancel against the hard contribution and we arrive at a finite result
\begin{align}
	\mathcal{M}_h+\mathcal{M}_c+\mathcal{M}_{\bar{c}}+\mathcal{M}_s
	=&\frac{g^2 m\delta^{ab}  g_\perp^{\mu\nu}}{8 \pi^2} \left[-\frac{1}{2}\left(\log\frac{m_H^2}{m^2}-\img \pi\right)^2 +2\right].
	\label{eq:full_LO0}
\end{align}
which agrees with the leading term in the full QCD result given in \refcite{Spira:1995rr}.

\section{Ward identities for collinear emission}
\label{app:wardcoll}
The Ward identities provide a powerful cross-check on our results.
They should be satisfied for the sum of all diagrams for each operator. We check the Ward identity for the emitted gluon, $\eps_\mu(p_1)\,\eps_\nu(p_2)\,k_\rho\,\cM^{\mu\nu\rho}=0$, i.e.\ we replace the polarization vector of the emitted gluon by its momentum $k^\rho$, while the incoming gluons with momenta $p_1$ and $p_2$ are contracted with physical, transverse polarization vectors.
This contraction with $\eps(p_1)$ and $\eps(p_2)$ is implied in all diagrammatic equations below.
The Ward identities for the incoming gluons $p_1^\mu$ and $p_2^\nu$ are satisfied by construction, since we are assuming transverse polarization of the incoming gluons.

We begin with the simpler contributions, namely the diagrams in which the quarks in the loop have hard, soft or anti-collinear scaling,
\begin{align}
	\left(\raisebox{-0.45\height}{\includegraphics[height=3cm]{figs/diagrams/LO1/LP-emissions/anticollLO1-d1-nolabel}}
	+\raisebox{-0.45\height}{\includegraphics[height=3.3cm]{figs/diagrams/LO1/LP-emissions/anticollLO1-d2-nolabel}} \right) k^\rho &=0,\\
	\left( \raisebox{-0.45\height}{\includegraphics[height=3.cm]{figs/diagrams/LO1/LP-emissions/softLO1-d1-nolabel}}
	+  \raisebox{-0.45\height}{\includegraphics[height=3.5cm]{figs/diagrams/LO1/LP-emissions/softLO1-d2-nolabel}} \right) k^\rho  &=0\\
	C(m_H) \times	\left( \raisebox{-0.45\height}{\includegraphics[height=2cm]{figs/diagrams/LO1/LP-emissions/hardLO1-d1-nolabel}}
	+  \raisebox{-0.45\height}{\includegraphics[height=2cm]{figs/diagrams/LO1/LP-emissions/hardLO1-d2-nolabel}} \right) k^\rho  &=0
\end{align}
so that the Ward identities are satisfied separately for each of these operators.

The diagrams with a collinear loop all contribute to the same operator and must therefore be considered together,
\begin{align}
	&\left(
	\raisebox{-0.45\height}{\includegraphics[height=2cm]{figs/diagrams/LO1/NLP-emission/collLO1-d1-nolabel}}
	+ \raisebox{-0.45\height}{\includegraphics[height=2.4cm]{figs/diagrams/LO1/NLP-emission/collLO1-d2-nolabel}}
	+  \raisebox{-0.45\height}{\includegraphics[height=2cm]{figs/diagrams/LO1/NLP-emission/collLO1-d3-FFGG-nolabel}}
	+ \raisebox{-0.45\height}{\includegraphics[height=2cm]{figs/diagrams/LO1/NLP-emission/collLO1-d3-LR-nolabel}}
	+ \raisebox{-0.45\height}{\includegraphics[height=2cm]{figs/diagrams/LO1/NLP-emission/collLO1-d3-LL-nolabel}}
	\right) k^\rho
	= 0.
	\end{align}
Here it is essential to include the finite parts, which we did not give explicitly in \secs{VEsection}{LEsection}. Once all these contributions are taken into account, the identity is fulfilled.

Finally, we have the Glauber contribution which has to satisfy the Ward identities on its own and we find
\begin{align}
	&\left( \raisebox{-0.45\height}{\includegraphics[height=2.4cm]{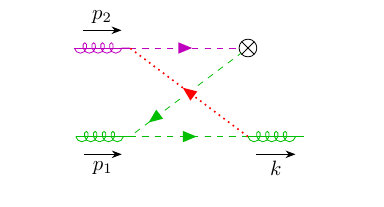}}
	+ \raisebox{-0.45\height}{\includegraphics[height=2.4cm]{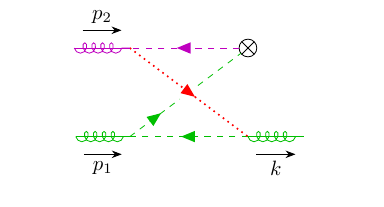}}
	\right) k^\rho
=0.
\end{align}

\section{Feynman rules}
\label{app:feynman_rules}
For the derivation of these Feynman rules, we wrote down the full QCD diagram and expanded the momenta in the appropriate soft or $n_i$-collinear limit.

Springs with lines denote collinear gluons, regular springs soft gluons, dashed fermion lines collinear quarks and regular fermion lines soft quarks. We use the same color coding as before where $n$-collinear gluons are shown in green and $\nb$-collinear gluons are shown in magenta. Then the collinear Feynman rules are given by
\begin{align}
	\label{eq:LOcFR}
	\raisebox{-0.45\height}{\includegraphics{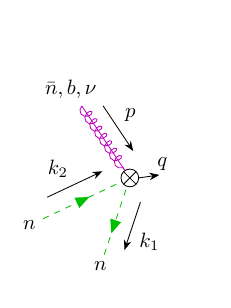}}
	=& g T^b \Big( \frac{1}{k_1^-} - \frac{1}{k_2^-}\Big)
	\frac{\slashed{\bar{n}}}{2} \gamma_\perp^\nu \delta(\omega_1-k_1^-)\delta(\omega_2+k_2^-)\delta(p^+-q^+) \nn \\[-4em]
	=& - g T^b C_{bbg}(\xi)\frac{\slashed{\bar{n}}}{2} \gamma_\perp^\nu
	\delta(\omega_n-k_2^-+k_1^-)\delta(\xi \omega-k_2^-)\delta(p^+-q^+) , \\
	\raisebox{-0.45\height}{\includegraphics{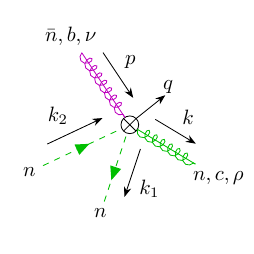}} =
	&g^2  C_{bbg}(\xi)\frac{\nbar^\rho}{\omega_n} \frac{\slashed{\bar{n}}}{2}\gamma_\perp^\nu \delta(p^+-q^+)  \nn \\[-4em]
	&\times \left[ T^b T^c\delta(\omega_n-k_2^-+k_1^-+k^-)\delta(\xi \omega-k_2^-+k^-)\right.
	\nn \\
	& \left.\phantom{\times \big [} - T^c T^b\delta(\omega_n-k_2^-+k_1^-+k^-)\delta(\xi \omega-k_2^-)\right], \label{eq:d1cFR}
\end{align}
In the second lines of \eqs{LOcFR}{LOacFR}, the Jacobian $\df\omega_1\df\omega_2=\omega_n\,\df\omega_n\df\xi$ has been absorbed, since $1/k_1^--1/k_2^-=-C_{bbg}(\xi)/\omega_n$ (and analogously for the $\nb$-collinear rule).
For \eq{d1cFR}, we also checked that the derivation from the full QCD diagram and using collinear gauge invariance in SCET yield the same result. For the $\nb$-collinear contribution, the leading-order Feynman rule is obtained from symmetry: interchanging ($n \to \nb$) switches $n$- and $\nb$-collinear modes.
\begin{align}
	\label{eq:LOacFR}
	\raisebox{-0.45\height}{\includegraphics{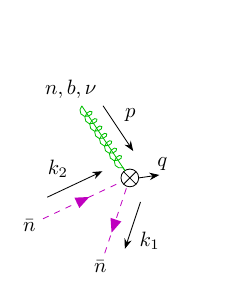}}
	=& g T^b \Big( \frac{1}{k_1^+} - \frac{1}{k_2^+}\Big)
	\frac{\slashed{{n}}}{2}\gamma_\perp^\nu \delta(\omega_1-k_1^+)\delta(\omega_2+k_2^+)\delta(p^--q^-) \nn \\[-5em]
	=& - g T^b C_{bbg}(\xi)\frac{\slashed{{n}}}{2} \gamma_\perp^\nu
	\delta(\omega_n-k_2^++k_1^+)\delta(\xi \omega-k_2^+)\delta(p^--q^-) ,
\end{align}
\begin{align}\label{eq:d1acFR}
	\raisebox{-0.45\height}{\includegraphics{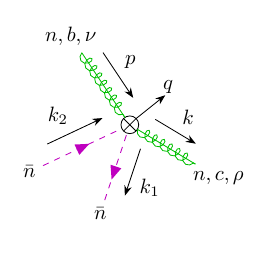}} =
	&- \img g^2 T^d f^{dcb}\frac{\nb^{\rho}}{k^-} C_{bbg}(\xi)\frac{\slashed{{n}}}{2}
	\gamma_\perp^\nu \delta(p^--q^-)
	\delta(\omega_n-k_2^++k_1^+)\delta(\xi \omega-k_2^+).
	\nn\\[-4em]
\end{align}
The soft Feynman rules are given by
\begin{align}
	\label{eq:LOsFR}
	\raisebox{-0.45\height}{\includegraphics{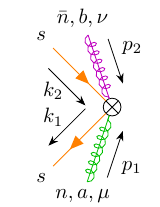}} &=\hspace{5pt}g^2 \Big(\frac{T^a T^b}{(k_2^- +\img 0) (k_1^+ - \img 0)} \gamma_\perp^\mu \frac{\slashed{n}\slashed{\nbar}}{4}\gamma_\perp^\nu  +\frac{T^b T^a}{(k_2^+ +\img 0) (k_1^- -\img 0)} \gamma_\perp^\nu \frac{\slashed{\nbar}\slashed{n}}{4}\gamma_\perp^\mu \Big) \,,
\end{align}

\begin{align}\label{eq:svertFR}
	\raisebox{-0.45\height}{\includegraphics{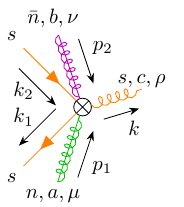}} =
	& \hspace{5pt} g^3  \gamma_\perp^\mu \frac{\slashed{n}\slashed{\nbar}}{4}\gamma_\perp^\nu
	\bigg[  \frac{\nbar^{\sigma}}{k^{-}+\img0} \Big( \frac{T^{a} T^{c} T^{b}}{(k_{1}^{+} -\img0) (k_{2}^{-}+\img0)} -  \frac{T^{a} T^{b} T^{c}}{(k_{1}^{+} - \img0) [(k_{2}-k)^{-} +\img0]}\Big)
	\nn \\[-3em]
	& \qquad\qquad+ \frac{n^{\sigma}}{k^{+}-\img0} \Big( \frac{T^{c} T^{a} T^{b}}{(k_{2}^{-}+\img0) [(k_{1} + k)^{+} - \img0]} -  \frac{T^{a} T^{c} T^{b}}{(k_{2}^{-} +\img0) (k_{1}^{+}-\img0)} \Big) \bigg ]
\nn \\ &+ (n, \mu, a) \leftrightarrow (\nbar, \nu,b)
\,,
\end{align}

\begin{align}\label{eq:d1sFR}
	\raisebox{-0.45\height}{\includegraphics{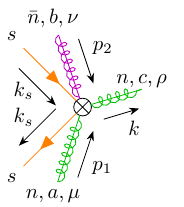}} = &-\frac{\img g^3}{2}   \frac{f^{abc}}{k_s^+ k_s^-} \frac{\nb^\rho}{k^-}\gamma_\perp^\nu\frac{\slashed{n}\slashed{\nbar}}{4}\gamma_\perp^\mu \,,
\end{align}
where in \eq{d1sFR} we have already taken the color trace and set the momenta of the two soft quarks equal, $k_1=k_2=k_s$, as is the case when the soft loop is closed with a single soft-quark propagator, which is the only way this vertex contributes in our calculation.
Inside this trace, only the mass term of the propagator contributes, and the two terms of \eq{LOsFR} give the same Dirac trace, so that they can be combined into a single structure, exactly like the LO$_{0}$ case in \eq{softLO0intnoreg}.

\addcontentsline{toc}{section}{References}
\bibliographystyle{jhep}
\bibliography{../paper/refs}

\providecommand{\href}[2]{#2}\begingroup\raggedright\begin{thebibliography}{100}

\bibitem{ATLAS:2012yve}
{\scshape ATLAS} collaboration, \emph{{Observation of a new particle in the
  search for the Standard Model Higgs boson with the ATLAS detector at the
  LHC}}, \href{https://doi.org/10.1016/j.physletb.2012.08.020}{\emph{Phys.
  Lett. B} {\bfseries 716} (2012) 1}
  [\href{https://arxiv.org/abs/1207.7214}{{\ttfamily 1207.7214}}].

\bibitem{CMS:2012qbp}
{\scshape CMS} collaboration, \emph{{Observation of a New Boson at a Mass of
  125 GeV with the CMS Experiment at the LHC}},
  \href{https://doi.org/10.1016/j.physletb.2012.08.021}{\emph{Phys. Lett. B}
  {\bfseries 716} (2012) 30} [\href{https://arxiv.org/abs/1207.7235}{{\ttfamily
  1207.7235}}].

\bibitem{Cepeda:2019klc}
M.~Cepeda et~al., \emph{{Report from Working Group 2}: {Higgs Physics at the
  HL-LHC and HE-LHC}},
  \href{https://doi.org/10.23731/CYRM-2019-007.221}{\emph{CERN Yellow Rep.
  Monogr.} {\bfseries 7} (2019) 221}
  [\href{https://arxiv.org/abs/1902.00134}{{\ttfamily 1902.00134}}].

\bibitem{Wilczek:1977zn}
F.~Wilczek, \emph{{Decays of Heavy Vector Mesons Into Higgs Particles}},
  \href{https://doi.org/10.1103/PhysRevLett.39.1304}{\emph{Phys. Rev. Lett.}
  {\bfseries 39} (1977) 1304}.

\bibitem{Shifman:1979eb}
M.~A. Shifman, A.~I. Vainshtein, M.~B. Voloshin and V.~I. Zakharov,
  \emph{{Low-Energy Theorems for Higgs Boson Couplings to Photons}},
  {\emph{Sov. J. Nucl. Phys.} {\bfseries 30} (1979) 711}.

\bibitem{Inami:1982xt}
T.~Inami, T.~Kubota and Y.~Okada, \emph{{Effective Gauge Theory and the Effect
  of Heavy Quarks in Higgs Boson Decays}},
  \href{https://doi.org/10.1007/BF01571710}{\emph{Z. Phys. C} {\bfseries 18}
  (1983) 69}.

\bibitem{DAWSON1991283}
S.~Dawson, \emph{Radiative corrections to higgs boson production},
  \href{https://doi.org/https://doi.org/10.1016/0550-3213(91)90061-2}{\emph{Nuclear
  Physics B} {\bfseries 359} (1991) 283}.

\bibitem{Djouadi:1991tka}
A.~Djouadi, M.~Spira and P.~M. Zerwas, \emph{{Production of Higgs bosons in
  proton colliders: QCD corrections}},
  \href{https://doi.org/10.1016/0370-2693(91)90375-Z}{\emph{Phys. Lett. B}
  {\bfseries 264} (1991) 440}.

\bibitem{Spira:1995rr}
M.~Spira, A.~Djouadi, D.~Graudenz and P.~M. Zerwas, \emph{{Higgs boson
  production at the LHC}},
  \href{https://doi.org/10.1016/0550-3213(95)00379-7}{\emph{Nucl. Phys. B}
  {\bfseries 453} (1995) 17}
  [\href{https://arxiv.org/abs/hep-ph/9504378}{{\ttfamily hep-ph/9504378}}].

\bibitem{Harlander:2002wh}
R.~V. Harlander and W.~B. Kilgore, \emph{{Next-to-next-to-leading order Higgs
  production at hadron colliders}},
  \href{https://doi.org/10.1103/PhysRevLett.88.201801}{\emph{Phys. Rev. Lett.}
  {\bfseries 88} (2002) 201801}
  [\href{https://arxiv.org/abs/hep-ph/0201206}{{\ttfamily hep-ph/0201206}}].

\bibitem{Ravindran:2003um}
V.~Ravindran, J.~Smith and W.~L. van Neerven, \emph{{NNLO corrections to the
  total cross-section for Higgs boson production in hadron hadron collisions}},
  \href{https://doi.org/10.1016/S0550-3213(03)00457-7}{\emph{Nucl. Phys. B}
  {\bfseries 665} (2003) 325}
  [\href{https://arxiv.org/abs/hep-ph/0302135}{{\ttfamily hep-ph/0302135}}].

\bibitem{Anastasiou:2015ema}
C.~Anastasiou, C.~Duhr, F.~Dulat, F.~Herzog and B.~Mistlberger, \emph{{Higgs
  Boson Gluon-Fusion Production in QCD at Three Loops}},
  \href{https://doi.org/10.1103/PhysRevLett.114.212001}{\emph{Phys. Rev. Lett.}
  {\bfseries 114} (2015) 212001}
  [\href{https://arxiv.org/abs/1503.06056}{{\ttfamily 1503.06056}}].

\bibitem{Anastasiou:2016cez}
C.~Anastasiou, C.~Duhr, F.~Dulat, E.~Furlan, T.~Gehrmann, F.~Herzog et~al.,
  \emph{{High precision determination of the gluon fusion Higgs boson
  cross-section at the LHC}},
  \href{https://doi.org/10.1007/JHEP05(2016)058}{\emph{JHEP} {\bfseries 05}
  (2016) 058} [\href{https://arxiv.org/abs/1602.00695}{{\ttfamily
  1602.00695}}].

\bibitem{Mistlberger:2018etf}
B.~Mistlberger, \emph{{Higgs boson production at hadron colliders at N$^{3}$LO
  in QCD}}, \href{https://doi.org/10.1007/JHEP05(2018)028}{\emph{JHEP}
  {\bfseries 05} (2018) 028}
  [\href{https://arxiv.org/abs/1802.00833}{{\ttfamily 1802.00833}}].

\bibitem{deFlorian:1999zd}
D.~de~Florian, M.~Grazzini and Z.~Kunszt, \emph{{Higgs production with large
  transverse momentum in hadronic collisions at next-to-leading order}},
  \href{https://doi.org/10.1103/PhysRevLett.82.5209}{\emph{Phys. Rev. Lett.}
  {\bfseries 82} (1999) 5209}
  [\href{https://arxiv.org/abs/hep-ph/9902483}{{\ttfamily hep-ph/9902483}}].

\bibitem{Ravindran:2002dc}
V.~Ravindran, J.~Smith and W.~L. Van~Neerven, \emph{{Next-to-leading order QCD
  corrections to differential distributions of Higgs boson production in hadron
  hadron collisions}},
  \href{https://doi.org/10.1016/S0550-3213(02)00333-4}{\emph{Nucl. Phys. B}
  {\bfseries 634} (2002) 247}
  [\href{https://arxiv.org/abs/hep-ph/0201114}{{\ttfamily hep-ph/0201114}}].

\bibitem{Glosser:2002gm}
C.~J. Glosser and C.~R. Schmidt, \emph{{Next-to-leading corrections to the
  Higgs boson transverse momentum spectrum in gluon fusion}},
  \href{https://doi.org/10.1088/1126-6708/2002/12/016}{\emph{JHEP} {\bfseries
  12} (2002) 016} [\href{https://arxiv.org/abs/hep-ph/0209248}{{\ttfamily
  hep-ph/0209248}}].

\bibitem{Chen:2014gva}
X.~Chen, T.~Gehrmann, E.~W.~N. Glover and M.~Jaquier, \emph{{Precise QCD
  predictions for the production of Higgs + jet final states}},
  \href{https://doi.org/10.1016/j.physletb.2014.11.021}{\emph{Phys. Lett. B}
  {\bfseries 740} (2015) 147}
  [\href{https://arxiv.org/abs/1408.5325}{{\ttfamily 1408.5325}}].

\bibitem{Boughezal:2015aha}
R.~Boughezal, C.~Focke, W.~Giele, X.~Liu and F.~Petriello, \emph{{Higgs boson
  production in association with a jet at NNLO using jettiness subtraction}},
  \href{https://doi.org/10.1016/j.physletb.2015.06.055}{\emph{Phys. Lett. B}
  {\bfseries 748} (2015) 5} [\href{https://arxiv.org/abs/1505.03893}{{\ttfamily
  1505.03893}}].

\bibitem{Boughezal:2015dra}
R.~Boughezal, F.~Caola, K.~Melnikov, F.~Petriello and M.~Schulze, \emph{{Higgs
  boson production in association with a jet at next-to-next-to-leading
  order}}, \href{https://doi.org/10.1103/PhysRevLett.115.082003}{\emph{Phys.
  Rev. Lett.} {\bfseries 115} (2015) 082003}
  [\href{https://arxiv.org/abs/1504.07922}{{\ttfamily 1504.07922}}].

\bibitem{Caola:2015wna}
F.~Caola, K.~Melnikov and M.~Schulze, \emph{{Fiducial cross sections for Higgs
  boson production in association with a jet at next-to-next-to-leading order
  in QCD}}, \href{https://doi.org/10.1103/PhysRevD.92.074032}{\emph{Phys. Rev.
  D} {\bfseries 92} (2015) 074032}
  [\href{https://arxiv.org/abs/1508.02684}{{\ttfamily 1508.02684}}].

\bibitem{Chen:2016zka}
X.~Chen, J.~Cruz-Martinez, T.~Gehrmann, E.~W.~N. Glover and M.~Jaquier,
  \emph{{NNLO QCD corrections to Higgs boson production at large transverse
  momentum}}, \href{https://doi.org/10.1007/JHEP10(2016)066}{\emph{JHEP}
  {\bfseries 10} (2016) 066}
  [\href{https://arxiv.org/abs/1607.08817}{{\ttfamily 1607.08817}}].

\bibitem{Campbell:2019gmd}
J.~M. Campbell, R.~K. Ellis and S.~Seth, \emph{{H + 1 jet production
  revisited}}, \href{https://doi.org/10.1007/JHEP10(2019)136}{\emph{JHEP}
  {\bfseries 10} (2019) 136}
  [\href{https://arxiv.org/abs/1906.01020}{{\ttfamily 1906.01020}}].

\bibitem{Chen:2019wxf}
X.~Chen, T.~Gehrmann, E.~W.~N. Glover and A.~Huss, \emph{{Fiducial cross
  sections for the four-lepton decay mode in Higgs-plus-jet production up to
  NNLO QCD}}, \href{https://doi.org/10.1007/JHEP07(2019)052}{\emph{JHEP}
  {\bfseries 07} (2019) 052}
  [\href{https://arxiv.org/abs/1905.13738}{{\ttfamily 1905.13738}}].

\bibitem{Bozzi:2005wk}
G.~Bozzi, S.~Catani, D.~de~Florian and M.~Grazzini, \emph{{Transverse-momentum
  resummation and the spectrum of the Higgs boson at the LHC}},
  \href{https://doi.org/10.1016/j.nuclphysb.2005.12.022}{\emph{Nucl. Phys. B}
  {\bfseries 737} (2006) 73}
  [\href{https://arxiv.org/abs/hep-ph/0508068}{{\ttfamily hep-ph/0508068}}].

\bibitem{Becher:2012yn}
T.~Becher, M.~Neubert and D.~Wilhelm, \emph{{Higgs-Boson Production at Small
  Transverse Momentum}},
  \href{https://doi.org/10.1007/JHEP05(2013)110}{\emph{JHEP} {\bfseries 05}
  (2013) 110} [\href{https://arxiv.org/abs/1212.2621}{{\ttfamily 1212.2621}}].

\bibitem{Neill:2015roa}
D.~Neill, I.~Z. Rothstein and V.~Vaidya, \emph{{The Higgs Transverse Momentum
  Distribution at NNLL and its Theoretical Errors}},
  \href{https://doi.org/10.1007/JHEP12(2015)097}{\emph{JHEP} {\bfseries 12}
  (2015) 097} [\href{https://arxiv.org/abs/1503.00005}{{\ttfamily
  1503.00005}}].

\bibitem{Bizon:2017rah}
W.~Bizon, P.~F. Monni, E.~Re, L.~Rottoli and P.~Torrielli,
  \emph{{Momentum-space resummation for transverse observables and the Higgs
  p$_{\perp}$ at N$^{3}$LL+NNLO}},
  \href{https://doi.org/10.1007/JHEP02(2018)108}{\emph{JHEP} {\bfseries 02}
  (2018) 108} [\href{https://arxiv.org/abs/1705.09127}{{\ttfamily
  1705.09127}}].

\bibitem{Chen:2018pzu}
X.~Chen, T.~Gehrmann, E.~W.~N. Glover, A.~Huss, Y.~Li, D.~Neill et~al.,
  \emph{{Precise QCD Description of the Higgs Boson Transverse Momentum
  Spectrum}}, \href{https://doi.org/10.1016/j.physletb.2018.11.037}{\emph{Phys.
  Lett. B} {\bfseries 788} (2019) 425}
  [\href{https://arxiv.org/abs/1805.00736}{{\ttfamily 1805.00736}}].

\bibitem{Bizon:2018foh}
W.~Bizo{\'n}, X.~Chen, A.~Gehrmann-De~Ridder, T.~Gehrmann, N.~Glover, A.~Huss
  et~al., \emph{{Fiducial distributions in Higgs and Drell-Yan production at
  N$^{3}$LL+NNLO}}, \href{https://doi.org/10.1007/JHEP12(2018)132}{\emph{JHEP}
  {\bfseries 12} (2018) 132}
  [\href{https://arxiv.org/abs/1805.05916}{{\ttfamily 1805.05916}}].

\bibitem{Gutierrez-Reyes:2019rug}
D.~Gutierrez-Reyes, S.~Leal-Gomez, I.~Scimemi and A.~Vladimirov,
  \emph{{Linearly polarized gluons at next-to-next-to leading order and the
  Higgs transverse momentum distribution}},
  \href{https://doi.org/10.1007/JHEP11(2019)121}{\emph{JHEP} {\bfseries 11}
  (2019) 121} [\href{https://arxiv.org/abs/1907.03780}{{\ttfamily
  1907.03780}}].

\bibitem{Becher:2020ugp}
T.~Becher and T.~Neumann, \emph{{Fiducial $q_T$ resummation of color-singlet
  processes at N$^3$LL+NNLO}},
  \href{https://doi.org/10.1007/JHEP03(2021)199}{\emph{JHEP} {\bfseries 03}
  (2021) 199} [\href{https://arxiv.org/abs/2009.11437}{{\ttfamily
  2009.11437}}].

\bibitem{Billis:2021ecs}
G.~Billis, B.~Dehnadi, M.~A. Ebert, J.~K.~L. Michel and F.~J. Tackmann,
  \emph{{Higgs pT Spectrum and Total Cross Section with Fiducial Cuts at Third
  Resummed and Fixed Order in QCD}},
  \href{https://doi.org/10.1103/PhysRevLett.127.072001}{\emph{Phys. Rev. Lett.}
  {\bfseries 127} (2021) 072001}
  [\href{https://arxiv.org/abs/2102.08039}{{\ttfamily 2102.08039}}].

\bibitem{Ebert:2016idf}
M.~A. Ebert, S.~Liebler, I.~Moult, I.~W. Stewart, F.~J. Tackmann, K.~Tackmann
  et~al., \emph{{Exploiting jet binning to identify the initial state of
  high-mass resonances}},
  \href{https://doi.org/10.1103/PhysRevD.94.051901}{\emph{Phys. Rev. D}
  {\bfseries 94} (2016) 051901}
  [\href{https://arxiv.org/abs/1605.06114}{{\ttfamily 1605.06114}}].

\bibitem{Bishara:2016jga}
F.~Bishara, U.~Haisch, P.~F. Monni and E.~Re, \emph{{Constraining Light-Quark
  Yukawa Couplings from Higgs Distributions}},
  \href{https://doi.org/10.1103/PhysRevLett.118.121801}{\emph{Phys. Rev. Lett.}
  {\bfseries 118} (2017) 121801}
  [\href{https://arxiv.org/abs/1606.09253}{{\ttfamily 1606.09253}}].

\bibitem{Soreq:2016rae}
Y.~Soreq, H.~X. Zhu and J.~Zupan, \emph{{Light quark Yukawa couplings from
  Higgs kinematics}},
  \href{https://doi.org/10.1007/JHEP12(2016)045}{\emph{JHEP} {\bfseries 12}
  (2016) 045} [\href{https://arxiv.org/abs/1606.09621}{{\ttfamily
  1606.09621}}].

\bibitem{Cal:2023mib}
P.~Cal, R.~von Kuk, M.~A. Lim and F.~J. Tackmann, \emph{{qT spectrum for Higgs
  boson production via heavy quark annihilation at N3LL'+aN3LO}},
  \href{https://doi.org/10.1103/PhysRevD.110.076005}{\emph{Phys. Rev. D}
  {\bfseries 110} (2024) 076005}
  [\href{https://arxiv.org/abs/2306.16458}{{\ttfamily 2306.16458}}].

\bibitem{LHCHiggsCrossSectionWorkingGroup:2016ypw}
{\scshape LHC Higgs Cross Section Working Group} collaboration, D.~de~Florian
  et~al., \emph{{Handbook of LHC Higgs Cross Sections: 4. Deciphering the
  Nature of the Higgs Sector}},
  \href{https://doi.org/10.23731/CYRM-2017-002}{\emph{CERN Yellow Rep. Monogr.}
  {\bfseries 2} (2017) 1} [\href{https://arxiv.org/abs/1610.07922}{{\ttfamily
  1610.07922}}].

\bibitem{Czakon:2023kqm}
M.~Czakon, F.~Eschment, M.~Niggetiedt, R.~Poncelet and T.~Schellenberger,
  \emph{{Top-Bottom Interference Contribution to Fully Inclusive Higgs
  Production}},
  \href{https://doi.org/10.1103/PhysRevLett.132.211902}{\emph{Phys. Rev. Lett.}
  {\bfseries 132} (2024) 211902}
  [\href{https://arxiv.org/abs/2312.09896}{{\ttfamily 2312.09896}}].

\bibitem{Lindert:2017pky}
J.~M. Lindert, K.~Melnikov, L.~Tancredi and C.~Wever, \emph{{Top-bottom
  interference effects in Higgs plus jet production at the LHC}},
  \href{https://doi.org/10.1103/PhysRevLett.118.252002}{\emph{Phys. Rev. Lett.}
  {\bfseries 118} (2017) 252002}
  [\href{https://arxiv.org/abs/1703.03886}{{\ttfamily 1703.03886}}].

\bibitem{Melnikov:2016qoc}
K.~Melnikov, L.~Tancredi and C.~Wever, \emph{{Two-loop $gg \to Hg$ amplitude
  mediated by a nearly massless quark}},
  \href{https://doi.org/10.1007/JHEP11(2016)104}{\emph{JHEP} {\bfseries 11}
  (2016) 104} [\href{https://arxiv.org/abs/1610.03747}{{\ttfamily
  1610.03747}}].

\bibitem{Melnikov:2017pgf}
K.~Melnikov, L.~Tancredi and C.~Wever, \emph{{Two-loop amplitudes for $q g \to
  H q$ and $q \bar{q} \to H g$ mediated by a nearly massless quark}},
  \href{https://doi.org/10.1103/PhysRevD.95.054012}{\emph{Phys. Rev. D}
  {\bfseries 95} (2017) 054012}
  [\href{https://arxiv.org/abs/1702.00426}{{\ttfamily 1702.00426}}].

\bibitem{Grazzini:2013mca}
M.~Grazzini and H.~Sargsyan, \emph{{Heavy-quark mass effects in Higgs boson
  production at the LHC}},
  \href{https://doi.org/10.1007/JHEP09(2013)129}{\emph{JHEP} {\bfseries 09}
  (2013) 129} [\href{https://arxiv.org/abs/1306.4581}{{\ttfamily 1306.4581}}].

\bibitem{Banfi:2013eda}
A.~Banfi, P.~F. Monni and G.~Zanderighi, \emph{{Quark masses in Higgs
  production with a jet veto}},
  \href{https://doi.org/10.1007/JHEP01(2014)097}{\emph{JHEP} {\bfseries 01}
  (2014) 097} [\href{https://arxiv.org/abs/1308.4634}{{\ttfamily 1308.4634}}].

\bibitem{Caola:2018zye}
F.~Caola, J.~M. Lindert, K.~Melnikov, P.~F. Monni, L.~Tancredi and C.~Wever,
  \emph{{Bottom-quark effects in Higgs production at intermediate transverse
  momentum}}, \href{https://doi.org/10.1007/JHEP09(2018)035}{\emph{JHEP}
  {\bfseries 09} (2018) 035}
  [\href{https://arxiv.org/abs/1804.07632}{{\ttfamily 1804.07632}}].

\bibitem{Kotsky:1997rq}
M.~I. Kotsky and O.~I. Yakovlev, \emph{{On the resummation of double logarithms
  in the process Higgs ---{\ensuremath{>}} gamma gamma}},
  \href{https://doi.org/10.1016/S0370-2693(97)01260-4}{\emph{Phys. Lett. B}
  {\bfseries 418} (1998) 335}
  [\href{https://arxiv.org/abs/hep-ph/9708485}{{\ttfamily hep-ph/9708485}}].

\bibitem{Akhoury:2001mz}
R.~Akhoury, H.~Wang and O.~I. Yakovlev, \emph{{On the Resummation of large QCD
  logarithms in Higgs ---{\ensuremath{>}} gamma gamma decay}},
  \href{https://doi.org/10.1103/PhysRevD.64.113008}{\emph{Phys. Rev. D}
  {\bfseries 64} (2001) 113008}
  [\href{https://arxiv.org/abs/hep-ph/0102105}{{\ttfamily hep-ph/0102105}}].

\bibitem{Liu:2017vkm}
T.~Liu and A.~A. Penin, \emph{{High-Energy Limit of QCD beyond the Sudakov
  Approximation}},
  \href{https://doi.org/10.1103/PhysRevLett.119.262001}{\emph{Phys. Rev. Lett.}
  {\bfseries 119} (2017) 262001}
  [\href{https://arxiv.org/abs/1709.01092}{{\ttfamily 1709.01092}}].

\bibitem{Liu:2018czl}
T.~Liu and A.~Penin, \emph{{High-Energy Limit of Mass-Suppressed Amplitudes in
  Gauge Theories}}, \href{https://doi.org/10.1007/JHEP11(2018)158}{\emph{JHEP}
  {\bfseries 11} (2018) 158}
  [\href{https://arxiv.org/abs/1809.04950}{{\ttfamily 1809.04950}}].

\bibitem{Penin:2014msa}
A.~A. Penin, \emph{{High-Energy Limit of Quantum Electrodynamics beyond Sudakov
  Approximation}},
  \href{https://doi.org/10.1016/j.physletb.2015.04.036}{\emph{Phys. Lett. B}
  {\bfseries 745} (2015) 69} [\href{https://arxiv.org/abs/1412.0671}{{\ttfamily
  1412.0671}}].

\bibitem{Anastasiou:2020vkr}
C.~Anastasiou and A.~Penin, \emph{{Light Quark Mediated Higgs Boson Threshold
  Production in the Next-to-Leading Logarithmic Approximation}},
  \href{https://doi.org/10.1007/JHEP07(2020)195}{\emph{JHEP} {\bfseries 07}
  (2020) 195} [\href{https://arxiv.org/abs/2004.03602}{{\ttfamily
  2004.03602}}].

\bibitem{Liu:2024tkc}
T.~Liu, A.~A. Penin and A.~Rehman, \emph{{Light quark mediated Higgs boson
  production in association with a jet at the next-to-next-to-leading order and
  beyond}}, \href{https://doi.org/10.1007/JHEP04(2024)031}{\emph{JHEP}
  {\bfseries 04} (2024) 031}
  [\href{https://arxiv.org/abs/2402.18625}{{\ttfamily 2402.18625}}].

\bibitem{Liu:2019oav}
Z.~L. Liu and M.~Neubert, \emph{{Factorization at subleading power and
  endpoint-divergent convolutions in $h\to\gamma\gamma$ decay}},
  \href{https://doi.org/10.1007/JHEP04(2020)033}{\emph{JHEP} {\bfseries 04}
  (2020) 033} [\href{https://arxiv.org/abs/1912.08818}{{\ttfamily
  1912.08818}}].

\bibitem{Liu:2020wbn}
Z.~L. Liu, B.~Mecaj, M.~Neubert and X.~Wang, \emph{{Factorization at subleading
  power and endpoint divergences in $h\to\gamma\gamma$ decay. Part II.
  Renormalization and scale evolution}},
  \href{https://doi.org/10.1007/JHEP01(2021)077}{\emph{JHEP} {\bfseries 01}
  (2021) 077} [\href{https://arxiv.org/abs/2009.06779}{{\ttfamily
  2009.06779}}].

\bibitem{Liu:2022ajh}
Z.~L. Liu, M.~Neubert, M.~Schnubel and X.~Wang, \emph{{Factorization at
  next-to-leading power and endpoint divergences in gg \textrightarrow{} h
  production}}, \href{https://doi.org/10.1007/JHEP06(2023)183}{\emph{JHEP}
  {\bfseries 06} (2023) 183}
  [\href{https://arxiv.org/abs/2212.10447}{{\ttfamily 2212.10447}}].

\bibitem{Bauer:2000ew}
C.~W. Bauer, S.~Fleming and M.~E. Luke, \emph{{Summing Sudakov logarithms in $B
  \to X_s \gamma $in effective field theory.}},
  \href{https://doi.org/10.1103/PhysRevD.63.014006}{\emph{Phys. Rev. D}
  {\bfseries 63} (2000) 014006}
  [\href{https://arxiv.org/abs/hep-ph/0005275}{{\ttfamily hep-ph/0005275}}].

\bibitem{Bauer:2000yr}
C.~W. Bauer, S.~Fleming, D.~Pirjol and I.~W. Stewart, \emph{{An Effective field
  theory for collinear and soft gluons: Heavy to light decays}},
  \href{https://doi.org/10.1103/PhysRevD.63.114020}{\emph{Phys. Rev. D}
  {\bfseries 63} (2001) 114020}
  [\href{https://arxiv.org/abs/hep-ph/0011336}{{\ttfamily hep-ph/0011336}}].

\bibitem{Bauer:2001ct}
C.~W. Bauer and I.~W. Stewart, \emph{{Invariant operators in collinear
  effective theory}},
  \href{https://doi.org/10.1016/S0370-2693(01)00902-9}{\emph{Phys. Lett. B}
  {\bfseries 516} (2001) 134}
  [\href{https://arxiv.org/abs/hep-ph/0107001}{{\ttfamily hep-ph/0107001}}].

\bibitem{Bauer:2001yt}
C.~W. Bauer, D.~Pirjol and I.~W. Stewart, \emph{{Soft collinear factorization
  in effective field theory}},
  \href{https://doi.org/10.1103/PhysRevD.65.054022}{\emph{Phys. Rev. D}
  {\bfseries 65} (2002) 054022}
  [\href{https://arxiv.org/abs/hep-ph/0109045}{{\ttfamily hep-ph/0109045}}].

\bibitem{Beneke:2002ph}
M.~Beneke, A.~P. Chapovsky, M.~Diehl and T.~Feldmann, \emph{{Soft collinear
  effective theory and heavy to light currents beyond leading power}},
  \href{https://doi.org/10.1016/S0550-3213(02)00687-9}{\emph{Nucl. Phys. B}
  {\bfseries 643} (2002) 431}
  [\href{https://arxiv.org/abs/hep-ph/0206152}{{\ttfamily hep-ph/0206152}}].

\bibitem{Beneke:2002ni}
M.~Beneke and T.~Feldmann, \emph{{Multipole expanded soft collinear effective
  theory with nonAbelian gauge symmetry}},
  \href{https://doi.org/10.1016/S0370-2693(02)03204-5}{\emph{Phys. Lett. B}
  {\bfseries 553} (2003) 267}
  [\href{https://arxiv.org/abs/hep-ph/0211358}{{\ttfamily hep-ph/0211358}}].

\bibitem{Kolodrubetz:2016uim}
D.~W. Kolodrubetz, I.~Moult and I.~W. Stewart, \emph{{Building Blocks for
  Subleading Helicity Operators}},
  \href{https://doi.org/10.1007/JHEP05(2016)139}{\emph{JHEP} {\bfseries 05}
  (2016) 139} [\href{https://arxiv.org/abs/1601.02607}{{\ttfamily
  1601.02607}}].

\bibitem{Moult:2016fqy}
I.~Moult, L.~Rothen, I.~W. Stewart, F.~J. Tackmann and H.~X. Zhu,
  \emph{{Subleading Power Corrections for N-Jettiness Subtractions}},
  \href{https://doi.org/10.1103/PhysRevD.95.074023}{\emph{Phys. Rev. D}
  {\bfseries 95} (2017) 074023}
  [\href{https://arxiv.org/abs/1612.00450}{{\ttfamily 1612.00450}}].

\bibitem{Feige:2017zci}
I.~Feige, D.~W. Kolodrubetz, I.~Moult and I.~W. Stewart, \emph{{A Complete
  Basis of Helicity Operators for Subleading Factorization}},
  \href{https://doi.org/10.1007/JHEP11(2017)142}{\emph{JHEP} {\bfseries 11}
  (2017) 142} [\href{https://arxiv.org/abs/1703.03411}{{\ttfamily
  1703.03411}}].

\bibitem{Moult:2018jjd}
I.~Moult, I.~W. Stewart, G.~Vita and H.~X. Zhu, \emph{{First Subleading Power
  Resummation for Event Shapes}},
  \href{https://doi.org/10.1007/JHEP08(2018)013}{\emph{JHEP} {\bfseries 08}
  (2018) 013} [\href{https://arxiv.org/abs/1804.04665}{{\ttfamily
  1804.04665}}].

\bibitem{Ebert:2018gsn}
M.~A. Ebert, I.~Moult, I.~W. Stewart, F.~J. Tackmann, G.~Vita and H.~X. Zhu,
  \emph{{Subleading power rapidity divergences and power corrections for
  q$_{T}$}}, \href{https://doi.org/10.1007/JHEP04(2019)123}{\emph{JHEP}
  {\bfseries 04} (2019) 123}
  [\href{https://arxiv.org/abs/1812.08189}{{\ttfamily 1812.08189}}].

\bibitem{Beneke:2018gvs}
M.~Beneke, A.~Broggio, M.~Garny, S.~Jaskiewicz, R.~Szafron, L.~Vernazza et~al.,
  \emph{{Leading-logarithmic threshold resummation of the Drell-Yan process at
  next-to-leading power}},
  \href{https://doi.org/10.1007/JHEP03(2019)043}{\emph{JHEP} {\bfseries 03}
  (2019) 043} [\href{https://arxiv.org/abs/1809.10631}{{\ttfamily
  1809.10631}}].

\bibitem{Beneke:2019mua}
M.~Beneke, M.~Garny, S.~Jaskiewicz, R.~Szafron, L.~Vernazza and J.~Wang,
  \emph{{Leading-logarithmic threshold resummation of Higgs production in gluon
  fusion at next-to-leading power}},
  \href{https://doi.org/10.1007/JHEP01(2020)094}{\emph{JHEP} {\bfseries 01}
  (2020) 094} [\href{https://arxiv.org/abs/1910.12685}{{\ttfamily
  1910.12685}}].

\bibitem{Moult:2019uhz}
I.~Moult, I.~W. Stewart, G.~Vita and H.~X. Zhu, \emph{{The Soft Quark
  Sudakov}}, \href{https://doi.org/10.1007/JHEP05(2020)089}{\emph{JHEP}
  {\bfseries 05} (2020) 089}
  [\href{https://arxiv.org/abs/1910.14038}{{\ttfamily 1910.14038}}].

\bibitem{Leibovich:2003jd}
A.~K. Leibovich, Z.~Ligeti and M.~B. Wise, \emph{{Comment on quark masses in
  SCET}}, \href{https://doi.org/10.1016/S0370-2693(03)00565-3}{\emph{Phys.
  Lett. B} {\bfseries 564} (2003) 231}
  [\href{https://arxiv.org/abs/hep-ph/0303099}{{\ttfamily hep-ph/0303099}}].

\bibitem{Rothstein:2003wh}
I.~Z. Rothstein, \emph{{Factorization, power corrections, and the pion
  form-factor}}, \href{https://doi.org/10.1103/PhysRevD.70.054024}{\emph{Phys.
  Rev. D} {\bfseries 70} (2004) 054024}
  [\href{https://arxiv.org/abs/hep-ph/0301240}{{\ttfamily hep-ph/0301240}}].

\bibitem{Chay:2005ck}
J.~Chay, C.~Kim and A.~K. Leibovich, \emph{{Quark mass effects in the
  soft-collinear effective theory and anti-B ---> X(s) gamma in the endpoint
  region}}, \href{https://doi.org/10.1103/PhysRevD.72.014010}{\emph{Phys. Rev.
  D} {\bfseries 72} (2005) 014010}
  [\href{https://arxiv.org/abs/hep-ph/0505030}{{\ttfamily hep-ph/0505030}}].

\bibitem{Gritschacher:2013pha}
S.~Gritschacher, A.~H. Hoang, I.~Jemos and P.~Pietrulewicz, \emph{{Secondary
  Heavy Quark Production in Jets through Mass Modes}},
  \href{https://doi.org/10.1103/PhysRevD.88.034021}{\emph{Phys. Rev. D}
  {\bfseries 88} (2013) 034021}
  [\href{https://arxiv.org/abs/1302.4743}{{\ttfamily 1302.4743}}].

\bibitem{Pietrulewicz:2017gxc}
P.~Pietrulewicz, D.~Samitz, A.~Spiering and F.~J. Tackmann,
  \emph{{Factorization and Resummation for Massive Quark Effects in Exclusive
  Drell-Yan}}, \href{https://doi.org/10.1007/JHEP08(2017)114}{\emph{JHEP}
  {\bfseries 08} (2017) 114}
  [\href{https://arxiv.org/abs/1703.09702}{{\ttfamily 1703.09702}}].

\bibitem{Liu:2020tzd}
Z.~L. Liu, B.~Mecaj, M.~Neubert and X.~Wang, \emph{{Factorization at subleading
  power, Sudakov resummation, and endpoint divergences in soft-collinear
  effective theory}},
  \href{https://doi.org/10.1103/PhysRevD.104.014004}{\emph{Phys. Rev. D}
  {\bfseries 104} (2021) 014004}
  [\href{https://arxiv.org/abs/2009.04456}{{\ttfamily 2009.04456}}].

\bibitem{Beneke:2020ibj}
M.~Beneke, M.~Garny, S.~Jaskiewicz, R.~Szafron, L.~Vernazza and J.~Wang,
  \emph{{Large-x resummation of off-diagonal deep-inelastic parton scattering
  from d-dimensional refactorization}},
  \href{https://doi.org/10.1007/JHEP10(2020)196}{\emph{JHEP} {\bfseries 10}
  (2020) 196} [\href{https://arxiv.org/abs/2008.04943}{{\ttfamily
  2008.04943}}].

\bibitem{Bodwin:2021cpx}
G.~T. Bodwin, J.-H. Ee, J.~Lee and X.-P. Wang, \emph{{Analyticity,
  renormalization, and evolution of the soft-quark function}},
  \href{https://doi.org/10.1103/PhysRevD.104.016010}{\emph{Phys. Rev. D}
  {\bfseries 104} (2021) 016010}
  [\href{https://arxiv.org/abs/2101.04872}{{\ttfamily 2101.04872}}].

\bibitem{Beneke:2022obx}
M.~Beneke, M.~Garny, S.~Jaskiewicz, J.~Strohm, R.~Szafron, L.~Vernazza et~al.,
  \emph{{Next-to-leading power endpoint factorization and resummation for
  off-diagonal ``gluon'' thrust}},
  \href{https://doi.org/10.1007/JHEP07(2022)144}{\emph{JHEP} {\bfseries 07}
  (2022) 144} [\href{https://arxiv.org/abs/2205.04479}{{\ttfamily
  2205.04479}}].

\bibitem{Bell:2022ott}
G.~Bell, P.~B{\"o}er and T.~Feldmann, \emph{{Muon-electron backward scattering:
  a prime example for endpoint singularities in SCET}},
  \href{https://doi.org/10.1007/JHEP09(2022)183}{\emph{JHEP} {\bfseries 09}
  (2022) 183} [\href{https://arxiv.org/abs/2205.06021}{{\ttfamily
  2205.06021}}].

\bibitem{Feldmann:2022ixt}
T.~Feldmann, N.~Gubernari, T.~Huber and N.~Seitz, \emph{{Contribution of the
  electromagnetic dipole operator $O_7$ to the $\bar{B}_s \to \mu^+\mu^-$ decay
  amplitude}}, \href{https://doi.org/10.1103/PhysRevD.107.013007}{\emph{Phys.
  Rev. D} {\bfseries 107} (2023) 013007}
  [\href{https://arxiv.org/abs/2211.04209}{{\ttfamily 2211.04209}}].

\bibitem{Cornella:2022ubo}
C.~Cornella, M.~K{\"o}nig and M.~Neubert, \emph{{Structure-dependent QED
  effects in exclusive $B$ decays at subleading power}},
  \href{https://doi.org/10.1103/PhysRevD.108.L031502}{\emph{Phys. Rev. D}
  {\bfseries 108} (2023) L031502}
  [\href{https://arxiv.org/abs/2212.14430}{{\ttfamily 2212.14430}}].

\bibitem{Hurth:2023paz}
T.~Hurth and R.~Szafron, \emph{{Refactorisation in subleading $\bar{B} \to X_s
  \gamma$}}, \href{https://doi.org/10.1016/j.nuclphysb.2023.116200}{\emph{Nucl.
  Phys. B} {\bfseries 991} (2023) 116200}
  [\href{https://arxiv.org/abs/2301.01739}{{\ttfamily 2301.01739}}].

\bibitem{Beneke:2024cpq}
M.~Beneke, Y.~Ji and X.~Wang, \emph{{Renormalization of the
  next-to-leading-power $\gamma\gamma \to h $ and $gg\to h$ soft quark
  functions}},  \href{https://arxiv.org/abs/2403.17738}{{\ttfamily
  2403.17738}}.

\bibitem{Boussarie:2023izj}
R.~Boussarie et~al., \emph{{TMD Handbook}},
  \href{https://arxiv.org/abs/2304.03302}{{\ttfamily 2304.03302}}.

\bibitem{Baur:1989cm}
U.~Baur and E.~W.~N. Glover, \emph{{Higgs Boson Production at Large Transverse
  Momentum in Hadronic Collisions}},
  \href{https://doi.org/10.1016/0550-3213(90)90532-I}{\emph{Nucl. Phys. B}
  {\bfseries 339} (1990) 38}.

\bibitem{Chetyrkin:1997iv}
K.~G. Chetyrkin, B.~A. Kniehl and M.~Steinhauser, \emph{{Hadronic Higgs decay
  to order $\alpha_s^4$}},
  \href{https://doi.org/10.1103/PhysRevLett.79.353}{\emph{Phys. Rev. Lett.}
  {\bfseries 79} (1997) 353}
  [\href{https://arxiv.org/abs/hep-ph/9705240}{{\ttfamily hep-ph/9705240}}].

\bibitem{vanRitbergen:1997va}
T.~van Ritbergen, J.~A.~M. Vermaseren and S.~A. Larin, \emph{{The Four loop
  beta function in quantum chromodynamics}},
  \href{https://doi.org/10.1016/S0370-2693(97)00370-5}{\emph{Phys. Lett. B}
  {\bfseries 400} (1997) 379}
  [\href{https://arxiv.org/abs/hep-ph/9701390}{{\ttfamily hep-ph/9701390}}].

\bibitem{Czakon:2004bu}
M.~Czakon, \emph{{The four-loop QCD beta-function and anomalous dimensions}},
  \href{https://doi.org/10.1016/j.nuclphysb.2005.01.012}{\emph{Nucl. Phys. B}
  {\bfseries 710} (2005) 485}
  [\href{https://arxiv.org/abs/hep-ph/0411261}{{\ttfamily hep-ph/0411261}}].

\bibitem{Chetyrkin:2005ia}
K.~G. Chetyrkin, J.~H. Kuhn and C.~Sturm, \emph{{QCD decoupling at four
  loops}}, \href{https://doi.org/10.1016/j.nuclphysb.2006.03.020}{\emph{Nucl.
  Phys. B} {\bfseries 744} (2006) 121}
  [\href{https://arxiv.org/abs/hep-ph/0512060}{{\ttfamily hep-ph/0512060}}].

\bibitem{Schroder:2005hy}
Y.~Schroder and M.~Steinhauser, \emph{{Four-loop decoupling relations for the
  strong coupling}},
  \href{https://doi.org/10.1088/1126-6708/2006/01/051}{\emph{JHEP} {\bfseries
  01} (2006) 051} [\href{https://arxiv.org/abs/hep-ph/0512058}{{\ttfamily
  hep-ph/0512058}}].

\bibitem{Gerlach:2018hen}
M.~Gerlach, F.~Herren and M.~Steinhauser, \emph{{Wilson coefficients for Higgs
  boson production and decoupling relations to $
  \mathcal{O}\left({\alpha}_s^4\right) $}},
  \href{https://doi.org/10.1007/JHEP11(2018)141}{\emph{JHEP} {\bfseries 11}
  (2018) 141} [\href{https://arxiv.org/abs/1809.06787}{{\ttfamily
  1809.06787}}].

\bibitem{Liu:2021mac}
Z.~L. Liu, M.~Neubert, M.~Schnubel and X.~Wang, \emph{{Radiative quark jet
  function with an external gluon}},
  \href{https://doi.org/10.1007/JHEP02(2022)075}{\emph{JHEP} {\bfseries 02}
  (2022) 075} [\href{https://arxiv.org/abs/2112.00018}{{\ttfamily
  2112.00018}}].

\bibitem{Chang:2017atu}
C.-H. Chang, I.~W. Stewart and G.~Vita, \emph{{A Subleading Power Operator
  Basis for the Scalar Quark Current}},
  \href{https://doi.org/10.1007/JHEP04(2018)041}{\emph{JHEP} {\bfseries 04}
  (2018) 041} [\href{https://arxiv.org/abs/1712.04343}{{\ttfamily
  1712.04343}}].

\bibitem{Gehrmann:2014vha}
T.~Gehrmann and D.~Kara, \emph{{The $Hb\bar{b}$ form factor to three loops in
  QCD}}, \href{https://doi.org/10.1007/JHEP09(2014)174}{\emph{JHEP} {\bfseries
  09} (2014) 174} [\href{https://arxiv.org/abs/1407.8114}{{\ttfamily
  1407.8114}}].

\bibitem{Chakraborty:2022yan}
A.~Chakraborty, T.~Huber, R.~N. Lee, A.~von Manteuffel, R.~M. Schabinger, A.~V.
  Smirnov et~al., \emph{{Hbb vertex at four loops and hard matching
  coefficients in SCET for various currents}},
  \href{https://doi.org/10.1103/PhysRevD.106.074009}{\emph{Phys. Rev. D}
  {\bfseries 106} (2022) 074009}
  [\href{https://arxiv.org/abs/2204.02422}{{\ttfamily 2204.02422}}].

\bibitem{Ebert:2017uel}
M.~A. Ebert, J.~K.~L. Michel and F.~J. Tackmann, \emph{{Resummation Improved
  Rapidity Spectrum for Gluon Fusion Higgs Production}},
  \href{https://doi.org/10.1007/JHEP05(2017)088}{\emph{JHEP} {\bfseries 05}
  (2017) 088} [\href{https://arxiv.org/abs/1702.00794}{{\ttfamily
  1702.00794}}].

\bibitem{Moult:2017xpp}
I.~Moult, M.~P. Solon, I.~W. Stewart and G.~Vita, \emph{{Fermionic Glauber
  Operators and Quark Reggeization}},
  \href{https://doi.org/10.1007/JHEP02(2018)134}{\emph{JHEP} {\bfseries 02}
  (2018) 134} [\href{https://arxiv.org/abs/1709.09174}{{\ttfamily
  1709.09174}}].

\bibitem{Chiu:2011qc}
J.-y. Chiu, A.~Jain, D.~Neill and I.~Z. Rothstein, \emph{{The Rapidity
  Renormalization Group}},
  \href{https://doi.org/10.1103/PhysRevLett.108.151601}{\emph{Phys. Rev. Lett.}
  {\bfseries 108} (2012) 151601}
  [\href{https://arxiv.org/abs/1104.0881}{{\ttfamily 1104.0881}}].

\bibitem{Chiu:2012ir}
J.-Y. Chiu, A.~Jain, D.~Neill and I.~Z. Rothstein, \emph{{A Formalism for the
  Systematic Treatment of Rapidity Logarithms in Quantum Field Theory}},
  \href{https://doi.org/10.1007/JHEP05(2012)084}{\emph{JHEP} {\bfseries 05}
  (2012) 084} [\href{https://arxiv.org/abs/1202.0814}{{\ttfamily 1202.0814}}].

\bibitem{Echevarria:2011epo}
M.~G. Echevarria, A.~Idilbi and I.~Scimemi, \emph{{Factorization Theorem For
  Drell-Yan At Low $q_T$ And Transverse Momentum Distributions
  On-The-Light-Cone}},
  \href{https://doi.org/10.1007/JHEP07(2012)002}{\emph{JHEP} {\bfseries 07}
  (2012) 002} [\href{https://arxiv.org/abs/1111.4996}{{\ttfamily 1111.4996}}].

\bibitem{ParticleDataGroup:2024cfk}
{\scshape Particle Data Group} collaboration, S.~Navas et~al., \emph{{Review of
  particle physics}},
  \href{https://doi.org/10.1103/PhysRevD.110.030001}{\emph{Phys. Rev. D}
  {\bfseries 110} (2024) 030001}.

\bibitem{Ellis:1987xu}
R.~K. Ellis, I.~Hinchliffe, M.~Soldate and J.~J. van~der Bij, \emph{{Higgs
  Decay to $\tau^+ \tau^-$: A Possible Signature of Intermediate Mass Higgs
  Bosons at SSC}},
  \href{https://doi.org/10.1016/0550-3213(88)90019-3}{\emph{Nucl. Phys. B}
  {\bfseries 297} (1988) 221}.

\bibitem{Mertig:1990an}
R.~Mertig, M.~Bohm and A.~Denner, \emph{{FEYN CALC: Computer algebraic
  calculation of Feynman amplitudes}},
  \href{https://doi.org/10.1016/0010-4655(91)90130-D}{\emph{Comput. Phys.
  Commun.} {\bfseries 64} (1991) 345}.

\bibitem{Shtabovenko:2016sxi}
V.~Shtabovenko, R.~Mertig and F.~Orellana, \emph{{New Developments in FeynCalc
  9.0}}, \href{https://doi.org/10.1016/j.cpc.2016.06.008}{\emph{Comput. Phys.
  Commun.} {\bfseries 207} (2016) 432}
  [\href{https://arxiv.org/abs/1601.01167}{{\ttfamily 1601.01167}}].

\bibitem{Shtabovenko:2020gxv}
V.~Shtabovenko, R.~Mertig and F.~Orellana, \emph{{FeynCalc 9.3: New features
  and improvements}},
  \href{https://doi.org/10.1016/j.cpc.2020.107478}{\emph{Comput. Phys. Commun.}
  {\bfseries 256} (2020) 107478}
  [\href{https://arxiv.org/abs/2001.04407}{{\ttfamily 2001.04407}}].

\bibitem{Shtabovenko:2023idz}
V.~Shtabovenko, R.~Mertig and F.~Orellana, \emph{{FeynCalc 10: Do multiloop
  integrals dream of computer codes?}},
  \href{https://doi.org/10.1016/j.cpc.2024.109357}{\emph{Comput. Phys. Commun.}
  {\bfseries 306} (2025) 109357}
  [\href{https://arxiv.org/abs/2312.14089}{{\ttfamily 2312.14089}}].

\bibitem{Berger:2010xi}
C.~F. Berger, C.~Marcantonini, I.~W. Stewart, F.~J. Tackmann and W.~J.
  Waalewijn, \emph{{Higgs Production with a Central Jet Veto at NNLL+NNLO}},
  \href{https://doi.org/10.1007/JHEP04(2011)092}{\emph{JHEP} {\bfseries 04}
  (2011) 092} [\href{https://arxiv.org/abs/1012.4480}{{\ttfamily 1012.4480}}].

\bibitem{Rothstein:2016bsq}
I.~Z. Rothstein and I.~W. Stewart, \emph{{An Effective Field Theory for Forward
  Scattering and Factorization Violation}},
  \href{https://doi.org/10.1007/JHEP08(2016)025}{\emph{JHEP} {\bfseries 08}
  (2016) 025} [\href{https://arxiv.org/abs/1601.04695}{{\ttfamily
  1601.04695}}].

\bibitem{Jacob:1959at}
M.~Jacob and G.~C. Wick, \emph{{On the general theory of collisions for
  particles with spin}},
  \href{https://doi.org/10.1016/0003-4916(59)90051-X}{\emph{Annals Phys.}
  {\bfseries 7} (1959) 404}.

\bibitem{Haber:1994pe}
H.~E. Haber, \emph{{Spin formalism and applications to new physics searches}},
  in \emph{{21st Annual SLAC Summer Institute on Particle Physics: Spin
  Structure in High-energy Processes}}, pp.~231--272, 1994,
  \href{https://arxiv.org/abs/hep-ph/9405376}{{\ttfamily hep-ph/9405376}}.

\bibitem{Beneke:1997zp}
M.~Beneke and V.~A. Smirnov, \emph{{Asymptotic expansion of Feynman integrals
  near threshold}},
  \href{https://doi.org/10.1016/S0550-3213(98)00138-2}{\emph{Nucl. Phys. B}
  {\bfseries 522} (1998) 321}
  [\href{https://arxiv.org/abs/hep-ph/9711391}{{\ttfamily hep-ph/9711391}}].

\bibitem{Smirnov:1998vk}
V.~A. Smirnov and E.~R. Rakhmetov, \emph{{The Strategy of regions for
  asymptotic expansion of two loop vertex Feynman diagrams}},
  \href{https://doi.org/10.1007/BF02557396}{\emph{Theor. Math. Phys.}
  {\bfseries 120} (1999) 870}
  [\href{https://arxiv.org/abs/hep-ph/9812529}{{\ttfamily hep-ph/9812529}}].

\bibitem{Smirnov:2002pj}
V.~A. Smirnov, \emph{{Applied asymptotic expansions in momenta and masses}},
  {\emph{Springer Tracts Mod. Phys.} {\bfseries 177} (2002) 1}.

\end{thebibliography}\endgroup

\end{document}